\documentclass[sigconf]{acmart}
\AtBeginDocument{%
  \providecommand\BibTeX{{%
    \normalfont B\kern-0.5em{\scshape i\kern-0.25em b}\kern-0.8em\TeX}}}

\setcopyright{rightsretained}
\copyrightyear{2024}
\acmYear{2024}
\acmConference[CHI '24]{Proceedings of the CHI Conference on Human Factors in Computing Systems}{May 11--16, 2024}{Honolulu, HI, USA}
\acmBooktitle{Proceedings of the CHI Conference on Human Factors in Computing Systems (CHI '24), May 11--16, 2024, Honolulu, HI, USA}
\acmDOI{10.1145/3613904.3642834}
\acmISBN{979-8-4007-0330-0/24/05}

\ccsdesc[500]{Information systems~Crowdsourcing}
\ccsdesc[500]{Applied computing~Annotation}
\ccsdesc[500]{Computing methodologies~Natural language processing}
\ccsdesc[500]{Human-centered computing~Human computer interaction (HCI)}

\title[If in a Crowdsourced Data Annotation Pipeline, a GPT-4]{If in a Crowdsourced Data Annotation Pipeline, a GPT-4}

\usepackage{xspace}

\newcommand{\ie}{{\it i.e.}\xspace}

\usepackage{multirow}
\usepackage{subcaption}

\usepackage{pdflscape}

\begin{document}



\author{Zeyu He}
\affiliation{%
  \institution{The Pennsylvania State University}
  \city{University Park}
  \state{PA}
  \country{USA}
}
\email{zmh5268@psu.edu}

\author{Chieh-Yang Huang}
\affiliation{%
  \institution{The Pennsylvania State University}
  \city{University Park}
  \state{PA}
  \country{USA}
}
\email{chiehyang@alumni.psu.edu}

\author{Chien-Kuang Cornelia Ding}
\affiliation{%
  \institution{University of California, San Francisco}
  \city{San Francisco}
  \state{CA}
  \country{USA}
}
\email{cornelia.ding@ucsf.edu}

\author{Shaurya Rohatgi}
\affiliation{%
  \institution{The Pennsylvania State University}
  \city{University Park}
  \state{PA}
  \country{USA}
}
\email{szr207@psu.edu}

\author{Ting-Hao `Kenneth' Huang}
\affiliation{%
  \institution{The Pennsylvania State University}
  \city{University Park}
  \state{PA}
  \country{USA}
}
\email{txh710@psu.edu}

\renewcommand{\shortauthors}{He, et al.}

\begin{abstract}
Recent studies indicated GPT-4 outperforms online crowd workers in data labeling accuracy, notably workers from Amazon Mechanical Turk (MTurk). 
However, these studies were criticized for deviating from standard crowdsourcing practices and emphasizing individual workers' performances over the whole data-annotation process. 
This paper compared GPT-4 and an ethical and well-executed MTurk pipeline, with 415 workers labeling 3,177 sentence segments from 200 scholarly articles using the CODA-19 scheme.
Two worker interfaces yielded 127,080 labels, which were then used to infer the final labels through eight label-aggregation algorithms.
Our evaluation showed that despite best practices, MTurk pipeline's highest accuracy was 81.5\%, whereas GPT-4 achieved 83.6\%. 
Interestingly, when combining GPT-4's labels with crowd labels collected via an advanced worker interface for aggregation, 
2 out of the 8 algorithms achieved an even higher accuracy (87.5\%, 87.0\%). 
Further analysis suggested that, when the crowd's and GPT-4's labeling strengths are complementary,
aggregating them could increase labeling accuracy.\footnote{Minor errors were corrected on June 26, 2024. These errors did not impact the conclusion of the paper. The information in this arXiv paper has been corrected, and the corrigendum is attached at the end (Appendix~\ref{sec:corrigendum}). }

\end{abstract}



\keywords{Crowdsourcing; Data Annotation; GPT-4; Large Language Model}



\maketitle


\section{Introduction}







With GPT-4 demonstrating its impressive ability to follow written instructions and respond to questions, a series of new studies emerged, stating that GPT-4's ability to label data has surpassed online crowd workers, notably Amazon Mechanical Turk (MTurk) workers~\cite{tornberg2023chatgpt, gilardi2023chatgpt, alizadeh2023open, zhu2023can}.
While these studies could be encouraging-- the ultimate goal of collecting human annotations at scale is to train advanced AI models-- several criticisms have emerged questioning the credibility of these studies~\cite{Xiang_2023,Casilli_2023,Turkopticon_2023}.
Aside from obvious issues like testing GPT-4's ability with datasets existing before GPT-4's knowledge cut-off date,\footnote{https://ediscoverychannel.com/2023/08/29/chatgpt-a-legaltech-perspective-part-two-limitations/}
criticisms fall into two categories:

\begin{itemize}
    \item 
First, 
\textbf{many studies deviated from standard crowdsourcing practices for annotating data.}
For example, several studies solely used MTurk's Master Qualification to select workers, which is known to be ineffective~\cite{rouse2019reliability}. 
Furthermore, in many of these studies, requesters did not monitor the quality of workers' labels and removed underperforming workers in the midst of the annotation process, which is a common practice when constructing a new dataset.

\item
Second, and more fundamentally, 
\textbf{these studies focused primarily on individual workers' performances instead of holistic data-annotation process.}
Literature in collective intelligence suggests that aggregating individual judgments, despite their potential inaccuracy, can lead to a final decision superior to any single person's judgment~\cite{raykar10a,10.14778/3055540.3055547}.
Furthermore, in real-world data annotation efforts, 
in addition to individual workers' performances,
many more factors contribute to the quality of the final resulting labels, including interface design, requesters' monitoring and communication effort, payment, and label aggregation techniques.
Some of these studies, with the goal of isolating the variable of worker performance, adopted approaches that were impractical or unfeasible for real-world crowdsourcing, {\em e.g.},
using the entire dataset's gold-standard labels to eliminate underperforming workers,
or collecting only one label per data instance.

\end{itemize}

This paper presents a \textbf{holistic investigation that compares the data labeling abilities of GPT-4 and a well-executed, ethical MTurk data-annotation process}.
We had 415 MTurk workers label 3,177 sentence segments in 200 scholarly papers published in 2022 or later, where each segment collected 20 labels.
We used the CODA-19 dataset's 5-class label scheme~\cite{huang-etal-2020-coda}, categorizing sentence segments by their research aspects: Background, Purpose, Method, Finding/Contribution, and Other.
We further experimented with two distinctly designed worker interfaces, recognizing the potential biases of any particular annotation interface~\cite{rahmanian2014user,toomim2011utility}.
A total of 127,080 labels was gathered in the study (3,177 sentence segments $\times$ 2 interface variations $\times$ 20 labels.)\footnote{We will keep the dataset offline for a minimum of one year to avoid data contamination concerns of large language models pre-trained using data on the public Internet.
Access to the dataset can be obtained by submitting a request to the first author. Before sharing the data, we will hash the worker IDs to ensure they are unique within the dataset but unrecognizable as real worker IDs.}
We then applied 8 label aggregation algorithms, including Majority Voting and DawidSkene, to determine the final labels and compared the labeling accuracies with GPT-4.
We found that, even with the best crowdsourcing practices, \textbf{MTurk's top-performing pipeline' accuracy of 81.5\% did not surpass GPT-4's 83.6\% (p<0.05).}
Interestingly, \textbf{when combining GPT-4's labels with crowd labels collected via an advanced worker interface for aggregation, 
2 out of the 8 algorithms achieved a higher accuracy (87.5\% with p<0.01 and 87.0\% with p<0.01) compared to GPT-4's standalone performance (83.6\%)}.
Further analysis suggested that, when the crowd's and GPT-4's labeling strengths are complementary-- with crowds better at labeling the Finding/Contribution class and GPT-4 excelling in all other classes-- aggregating them could further increase labeling accuracy.

The contribution of this paper is three-fold.
First, it responds to recent speculations about GPT-4's labeling ability surpassing online crowd workers by focusing on the performance of holistic crowdsourcing pipelines, an area previously overlooked. 
Second, our study highlights the value of crowdsourced labels in scenarios where GPT-4's accuracy generally outperforms yet complements crowd efforts, demonstrating that adding crowd labels can further enhance accuracy. 
Third, and most importantly, this study sheds light on the evolving role and best practices for crowdsourcing in the era of Large Language Models (LLMs), particularly when LLMs often exhibit superior labeling accuracy compared to crowd workers.

\section{Related Work}


\subsection{Crowd Workers vs. GPT}

Recent advancements in Large Language Models (LLMs), especially GPT-3 and GPT-4, have spurred a flurry of studies examining their data labeling abilities compared to human annotators.
While a subset of these studies was centered on expert annotators like English teachers~\cite{chiang2023can}, most of them targeted online crowd workers.
Often, these studies indicated that LLMs, particularly GPT-4, can surpass crowd workers in annotation quality, suggesting a potential to replace them~\cite{tornberg2023chatgpt,gilardi2023chatgpt,he2023annollm,alizadeh2023open}.
A summary of these studies can be found in Table~\ref{tab:related-work}.
In certain extended explorations, there have been attempts to replace human crowd workers with LLMs within human computation workflows~\cite{wu2023llms}.
\begin{table*}[t]
\centering
\small
\begin{tabular}{@{}cccccccccc@{}}
\toprule
\multirow{2}{*}{\textbf{}} & \multicolumn{2}{c}{\textbf{Overview}} & \multicolumn{2}{c}{\textbf{\begin{tabular}[c]{@{}c@{}}Available\\ Online\\ Pre-Sep 2021?\end{tabular}}} & \multicolumn{2}{c}{\textbf{\begin{tabular}[c]{@{}c@{}}Crowd\\ Worker\\ Selection\end{tabular}}} & \multicolumn{3}{c}{\textbf{\begin{tabular}[c]{@{}c@{}}Crowdsourced\\ Annotation\\ Practice\end{tabular}}} \\ \cmidrule(l){2-10} 
 & \textbf{Platform} & \textbf{Task} & \textbf{Data} & \textbf{\begin{tabular}[c]{@{}c@{}}Gold\\ Label\end{tabular}} & \textbf{\begin{tabular}[c]{@{}c@{}}Worker\\ Qualification\end{tabular}} & \textbf{\begin{tabular}[c]{@{}c@{}}Manual\\ Filtering\end{tabular}} & \textbf{\begin{tabular}[c]{@{}c@{}}\#Item\\ Labeled\end{tabular}} & \textbf{\begin{tabular}[c]{@{}c@{}}\#Label\\ Per\\ Item\end{tabular}} & \textbf{\begin{tabular}[c]{@{}c@{}}Label\\ Aggregation\end{tabular}} \\ \midrule
\cite{tornberg2023chatgpt} & MTurk & \begin{tabular}[c]{@{}c@{}}Text\\ Classification\end{tabular} & Yes & Inferenced & \begin{tabular}[c]{@{}c@{}}Master,\\ US-Only\end{tabular} & No & 500 & 10 & MV \\ \midrule
\cite{gilardi2023chatgpt,alizadeh2023open} & MTurk & \begin{tabular}[c]{@{}c@{}}Text\\ Classification\end{tabular} & Partial & No & \begin{tabular}[c]{@{}c@{}}Master,\\ US-Only,\\ HIT Approval \%,\\ \#HITs Approved\end{tabular} & No & 6,183 & 2 & No \\ \midrule
\cite{he2023annollm} & UHRS & \begin{tabular}[c]{@{}c@{}}Text\\ Classification\end{tabular} & Partial\footnotemark & Partial & No & No & 9,903 & 3+ & \begin{tabular}[c]{@{}c@{}}Collect until\\ 3 agree.\end{tabular} \\ \midrule
\cite{cegin2023chatgpt} & - & Paraphrases & Yes & - & \multicolumn{5}{c}{No new crowdsourced annotations were collected.} \\ \midrule
\cite{zhu2023can} & - & \begin{tabular}[c]{@{}c@{}}Text\\ Classification\end{tabular} & Partial & Partial & \multicolumn{5}{c}{No new crowdsourced annotations were collected.} \\ \midrule
\cite{zhao2021lmturk} & - & \begin{tabular}[c]{@{}c@{}}Text\\ Classification\end{tabular} & Yes & Yes & \multicolumn{5}{c}{No new crowdsourced annotations were collected.} \\ \midrule
\cite{huang2023chatgpt} & - & \begin{tabular}[c]{@{}c@{}}Writing,\\ Text\\ Classification\end{tabular} & \begin{tabular}[c]{@{}c@{}}Yes,\\ Yes\end{tabular} & \begin{tabular}[c]{@{}c@{}}-,\\ Yes\end{tabular} & \multicolumn{5}{c}{No new crowdsourced annotations were collected.} \\ \midrule
\cite{wang2021want} & - & \begin{tabular}[c]{@{}c@{}}Writing,\\ Text\\ Classification\end{tabular} & \begin{tabular}[c]{@{}c@{}}Yes,\\ Yes\end{tabular} & \begin{tabular}[c]{@{}c@{}}-,\\ Yes\end{tabular} & \multicolumn{5}{c}{No new crowdsourced annotations were collected.} \\ \midrule
\cite{bansal2023large} & - & \begin{tabular}[c]{@{}c@{}}Text\\ Classification\end{tabular} & Yes & Yes & \multicolumn{5}{c}{No new crowdsourced annotations were collected.} \\ \midrule
\cite{Ding2022IsGA} & - & \begin{tabular}[c]{@{}c@{}}Text\\ Classification,\\ NER, ASTE\end{tabular} & \begin{tabular}[c]{@{}c@{}}Yes,\\ Yes,\\ Yes\end{tabular} & \begin{tabular}[c]{@{}c@{}}Yes,\\ Yes,\\ Yes\end{tabular} & \multicolumn{5}{c}{No new crowdsourced annotations were collected.} \\ \midrule
\textbf{Ours} & \textbf{MTurk} & \textbf{\begin{tabular}[c]{@{}c@{}}Text\\ Classification\end{tabular}} & \textbf{No} & \textbf{No} & \textbf{\begin{tabular}[c]{@{}c@{}}US-Only,\\ HIT Approval \%,\\ \#HITs Approved,\\ Adult Content\end{tabular}} & \textbf{Yes} & \textbf{3,177} & \textbf{20} & \textbf{\begin{tabular}[c]{@{}c@{}}MV,\\ Dawid-Skene,\\ One-coin \\ Dawid-Skene,\\ M-MSR, \\ Wawa, ZBS,\\ GLAD, \\ MACE\end{tabular}} \\ \bottomrule
\end{tabular}
\caption{Recent studies have conducted comparisons between LLMs and Crowd Workers. However, many of these studies utilized data predating ChatGPT's cutoff date, potentially resulting in data contamination where GPT-4 was evaluated using its own training data. Notably, none of these studies incorporated manual filtering, and they often did not employ multiple aggregation methods.
}
\label{tab:related-work}
\end{table*}

This progress in AI can be exciting, especially given that the motivation for collecting human annotations has long been to train powerful AI models.
However, this research direction is not without its skeptics.
A primary critique, evident from Table~\ref{tab:related-work}, is that many studies leveraged existing datasets with gold labels already available online before ChatGPT's knowledge cutoff date (September 2021)~\cite{tornberg2023chatgpt, zhu2023can, he2023annollm, alizadeh2023open}.
Given that ChatGPT utilized vast swaths of online data for training, testing it with datasets available before this cutoff raises concerns about data contamination~\cite{magar-schwartz-2022-data, deng2023investigating, jiang2024investigating}-- essentially, testing GPT-4 with its training data. 
While it is convenient to use existent datasets for initial GPT-4 benchmarks, it is crucial for an unbiased assessment in which new datasets are curated and used.

Beyond the data contamination concern, the crowdsourcing research community has highlighted additional issues.
Many of these studies deviated from standard MTurk data annotation practices.
For instance, they have depended on the frequently criticized MTurk Master Qualification~\cite{rouse2019reliability}, either underpaid workers or failed to disclose pay rates, and lacked thoroughness in filtering out underperforming workers during the annotation process~\cite{tornberg2023chatgpt, he2023annollm, alizadeh2023open, huang2023chatgpt}.
Additionally, these studies often overlooked the collective nature of crowdsourcing, focusing instead on individual performance.
This focus gave rise to impractical testing methods, like using the entire dataset's gold labels to filter out underperforming workers~\cite{tornberg2023chatgpt}.
Consequently, these studies did not explore label aggregation techniques, which, when implemented, could yield high-quality results even if individual worker contributions were mediocre.

One of the few studies that took a more holistic view of crowdsourced labels rather than solely focusing on individual workers' performance was conducted by~\citet{li2024comparative}.
They compared LLMs' labeling performances to both individual and aggregated labels in crowdsourced NLP datasets, including RTE (Recognizing Textual Entailment)~\cite{snow2008cheap} and QUIZ~\cite{li2017hyper}.
However, the datasets used in this study-- from 2008 and 2017-- predate ChatGPT's knowledge cutoff.

\subsection{Crowd Label Aggregation and Quality Control}

In crowdsourcing, human computation, and database literature, there has been a significant focus on addressing the issue of unreliable quality in crowdsourced labels. 
This has led to two main research areas: {\em (i)} improving the quality of labels collected via crowdsourcing~\cite{snow2008cheap,daniel2018quality,allahbakhsh2013quality} and {\em (ii)} developing label aggregation techniques~\cite{10.14778/3055540.3055547}. 
Both strategies, which can be used simultaneously for better final label quality~\cite{dumitrache2021empirical,kazai2011search}, involve different approaches.
Quality control in crowdsourcing examines how factors like fair payment~\cite{mao2013volunteering,toomim2011utility}, 
instructions and training~\cite{liu2016effective,gillier2018effects}, 
task design~\cite{10.1145/2499149.2499168}, and interface design~\cite{toomim2011utility} affect the quality of crowd work. 
Common practices~\cite{allahbakhsh2013quality} include inserting gold labels for quality checks, using reputation systems, time locks, attention checks~\cite{pei2020attention}, and recruiting workers with specific qualifications~\cite{kummerfeld2021quantifying}.
Meanwhile, label aggregation research aims to derive high-quality labels from numerous unreliable ones~\cite{10.14778/3055540.3055547,10.1145/3447548.3467411}. 
Techniques include EM-based methods, weighted voting, bidding, and, more recently, neural aggregation methods.
(We describe the details of all the aggregation algorithms used in this work in Section~\ref{sec:aggregation-models}.)

However, most of these prior works were based on the assumption that AI models could not yet perform these labeling tasks effectively. 
Our research operates under a new paradigm where large language models can already perform these tasks fairly well in a zero-shot manner, leading to new research questions and challenges.


\section{Comparative Study Procedure}
\footnotetext{The QK dataset mentioned in this paper is internal and not publicly available.
}



In this paper, we followed the best practice of crowdsourcing to use MTurk to label new data, applied a variety of label aggregation techniques to induce final labels, and compared results with GPT-4. 
This section details the procedure.
Previous research indicates worker interface design can influence performance~\cite{rahmanian2014user,toomim2011utility}; thus, we tested two different interfaces in our study.
Among several available guidelines for reporting crowdsourcing experiments~\cite{10.1145/3479531,10.1145/3531146.3534647}, we adhered to the guidelines and checklist developed by~\citet{10.1145/3479531} to ensure our experiments are detailed enough for reliable replication by the community.




\subsection{Annotation Scheme and Data\label{subsec:scheme-and-data}}
\paragraph{Annotation Scheme and Instruction.}
We aimed to compare MTurk and GPT-4's ability to label text items, as GPT-4 currently performs best with text rather than with video or images.
For our study, we chose the CODA-19 label scheme~\cite{huang-etal-2020-coda}, which categorizes sentence segments in paper abstracts into research aspects, \ie, Background, Purpose, Method, Finding/Contribution, and Other.
We obtained the detailed annotation instructions via CODA-19's GitHub repository\footnote{CODA-19: COVID-19 Research Aspect Dataset: https://github.com/windx0303/CODA-19} and used it in our study.

This task was picked for its balanced difficulty: 
It demands reading scholarly articles, making it more challenging than basic sentiment labeling.
However, it is not as hard as expert-only labeling tasks like disease mentions, and MTurk workers have successfully completed it before~\cite{huang-etal-2020-coda,chan2018solvent}.

\paragraph{Data.}
The original CODA-19 dataset~\cite{huang-etal-2020-coda} contains biomedical papers published before April 2020 extracted from the COVID-19 Open Research Dataset (CORD-19) dataset~\cite{wang-etal-2020-cord}.
In this study, we sampled papers from the most recent release of the CORD-19 dataset, dated June 2, 2022, which housed around one million documentations.
To prevent our test data from overlapping with OpenAI GPT's training data, we limited our study to documents published after ChatGPT's last knowledge update in September 2021, focusing on 2022 publications or later.
Using \texttt{langdetect},\footnote{Language detection library in Python: https://github.com/fedelopez77/langdetect} we identified and retained only English papers.
After this process, our dataset comprised 123,881 papers with full text and metadata.

For our main study, we randomly sampled 200 papers from this dataset as the test set.
We segmented the abstracts of these papers into 3,177 sentence segments, averaging 15.89 segments per abstract, following CODA-19's approach~\cite{huang-etal-2020-coda}.

For developing worker interfaces (Section~\ref{subsec:mturk-procedure}), we also randomly sampled 200 different papers from the dataset as the interface development set.
During the interface design phase of the Basic Worker Interface (Section~\ref{subsec:mturk-procedure}), the papers in the interface development set were used as prototyping materials, such as placeholder texts for layout adjustments and texts for MTurk tasks to test interface functionalities.
We deliberately separated the test set from the interface development set to prevent bias, ensuring the interface does not unfairly favor papers in the test set.





\subsection{Collecting Labels via Amazon Mechanical Turk\label{subsec:mturk-procedure}}
\paragraph{Worker Interfaces}

\begin{figure*}[t]
    \centering
    \includegraphics[width=\textwidth]{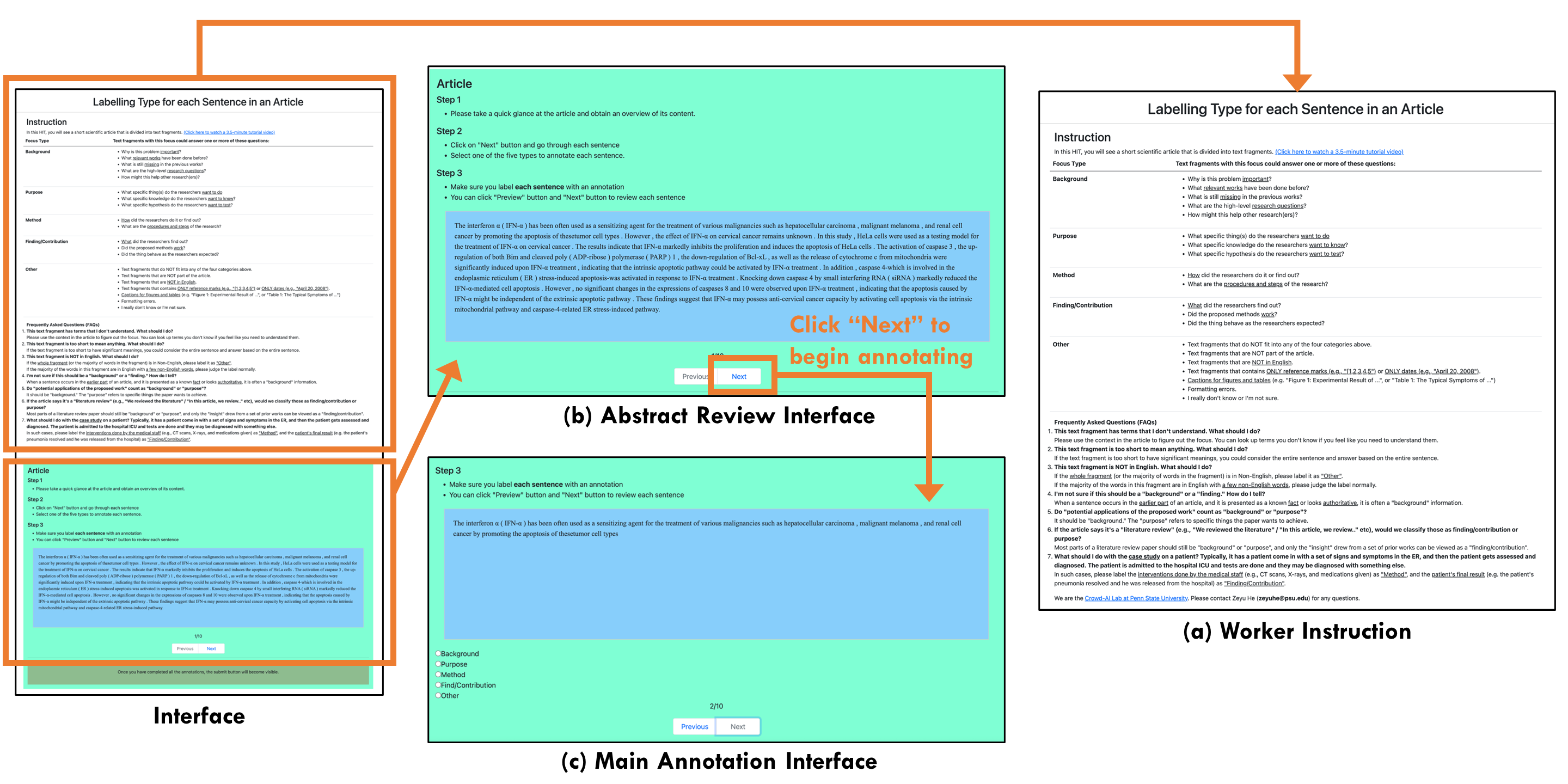}
    \caption{The basic worker interface, individually designed by one of the authors, has a focus on prioritizing task simplicity and user-friendliness.}
    \label{fig:basic_ui}
    \Description{The basic worker interface. It has an overview of the worker interface. The interface contains a worker instruction and an abstract review interface. By clicking the next and previous button on the abstract review interface, the worker can go to the next/previous text segment to make annotations or make changes.}
\end{figure*}
\begin{figure*}[t]
    \centering
    \includegraphics[width=\textwidth]{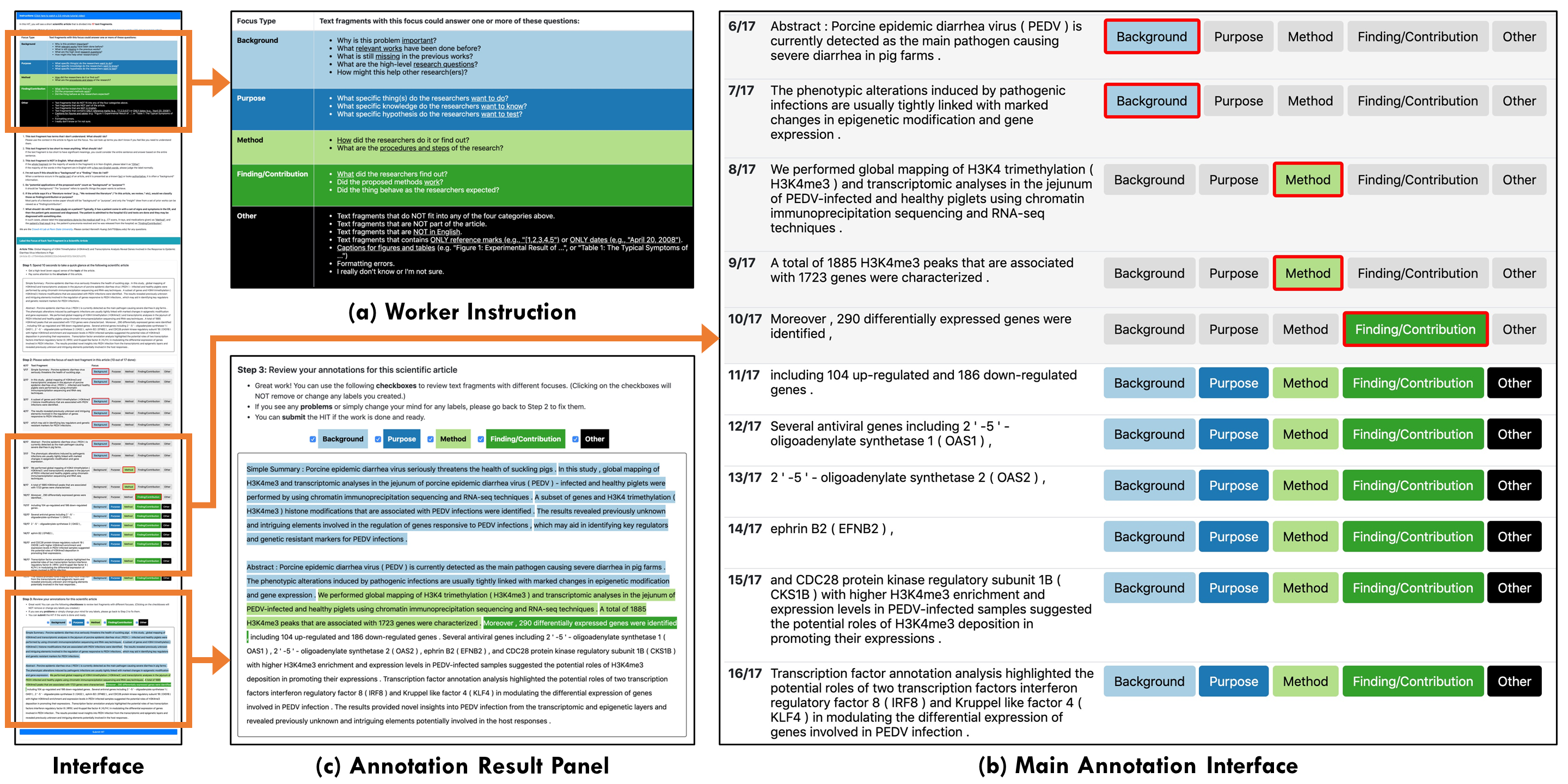}
    \caption{The advanced worker interface, adopted from CODA-19~\cite{huang-etal-2020-coda}, incorporates advanced features such as a visual feedback button, color-coded annotation view, and a time lock mechanism to deter hasty spam submissions.}
    \label{fig:pretty_ui}
    \Description{The advanced worker interface. It has an overview of the worker interface. The interface contains a worker instruction, a main annotation interface, and an annotation result panel.}
\end{figure*}

Prior studies suggested that the design of the worker interface would impact annotation performances on MTurk~\cite{rahmanian2014user, toomim2011utility}.
To address potential biases, we tested two interfaces in our study.
Both displayed the original \textsc{CODA-19} instructions but were independently designed by different individuals:

\begin{itemize}

    \item \textbf{Basic Worker Interface (Figure~\ref{fig:basic_ui}):}
    An author of this paper, unfamiliar with designing interfaces for MTurk tasks, was tasked with creating a worker interface using the original \textsc{CODA-19} annotation instructions, including examples and FAQs. We emphasized simplicity and usability in the design. The interface had instructions at the top (Figure~\ref{fig:basic_ui}a) and an annotation section below. Workers skimmed the abstract first (Figure~\ref{fig:basic_ui}b), labeled text segments, and then reviewed and corrected their labels using the ``prev'' and ``next'' buttons (Figure~\ref{fig:basic_ui}c).
    
    \item \textbf{Advanced Worker Interface (Figure~\ref{fig:pretty_ui}):} 
It is the original interface that was used for constructing the \textsc{CODA-19} dataset~\cite{huang-etal-2020-coda}, designed by a crowdsourcing expert with extensive experience in designing MTurk task interfaces.
Although this interface had a similar layout to the basic worker interface, it has several advanced features, such as visual feedback on button clicks, a color-coded annotation overview, and a time lock to prevent hasty spam submissions.

\end{itemize}

Both interfaces show the original \textsc{CODA-19} annotation instructions.
We did not explicitly tell workers that they were part of an experiment comparing MTurk pipelines with GPT; we simply stated it was a data labeling task.
This approach was chosen to replicate a typical data labeling scenario.

\textsc{CODA-19}'s successful collection with MTurk workers~\cite{huang-etal-2020-coda} confirmed the efficacy of the labeling scheme and task instructions, so we did not perform a pilot study for the identical Advanced Worker Interface. 
For the Basic Worker Interface, which was newly created for this work, we conducted a small set of pilot studies on MTurk to verify its functionality.

\paragraph{Worker Recruitment and Grouping.}
We first set up a \$1 qualification Human Intelligence Task (HIT) on MTurk with the basic interface, in which workers needed to watch a tutorial video, review instructions, and annotate a research abstract to qualify.
Of those who passed, we divided 800 workers into two groups: 
400 for the basic and 400 for the advanced interface.
Workers could only access the HITs of their qualified group.
Additionally, we applied four MTurk built-in qualifications: Locale (US Only), HIT Approval Rate ($\geq$98\%), Approved HITs ($\geq$3000), and Adult Content Qualification.\footnote{Our study used scholarly articles from the biomedical domain, which might include mentions of body parts or explicit medical details that could be uncomfortable for some workers. To prevent discomfort, we applied Adult Content Qualification, allowing only workers willing to see potentially offensive content to participate in our task.}

\paragraph{Posting Tasks in Batches and Monitoring Label Quality.}
We follow the best crowdsourcing practice to annotate the data using MTurk. 
When experienced requesters use MTurk to label unseen data, they rarely post all the data at once.
It is more common to post data in batches and monitor label quality carefully between each.
The original \textsc{CODA-19} dataset was constructed using this approach, and we adopted the same strategy in our study.

We divided 200 abstracts (see Section~\ref{subsec:scheme-and-data}) into four batches of 50, posting one at a time.
For each abstract, we created two HITs: one with the basic interface and the other with the advanced interface.
We recruited 20 workers via 20 assignments (from the qualified pool of 400) for each HIT.

Once a batch was completed, we first approved all submitted work in that batch.
Next, we assessed label quality and removed qualifications from underperforming workers to prevent them from accessing our future HITs.
In a practical label collection scenario, we would not have gold-standard labels for the entire dataset and could not afford to check every label manually.
Therefore, the first author, who is a Ph.D. student in Informatics (the ``CS Expert'' in Section~\ref{subsec:expert-procedure}), manually labeled only 10 abstracts per batch.
We then used these labels to compute three worker quality control statistics:
{\em (i)} label accuracy, based on only 10 manually labeled abstracts per batch, 
{\em (ii)} probability of agreeing with the majority label, and 
{\em (iii)} probability of labeling ``Other,'' a rare label.
For {\em (i)} and {\em (ii)}, we reviewed the bottom 30 workers' labels, and for {\em (iii)}, the top 30's.
We developed an interface allowing for rapid label inspection (details in the Appendix Figure~\ref{fig:worker-result-comparison-visualization}). 
If we observed a worker consistently providing incorrect labels or seemingly spamming our task, we revoked their qualification.
When a few removed workers contacted us to complain and request reinstatement, we justified our exclusion decision using the three statistics.

The crowdsourced data collection study took place between August 22nd, 2023, and September 7th, 2023.
The first HIT batch was posted on August 22nd, 2023; 
the second batch was posted on August 30th, 2023; 
the third batch was posted on September 5th, 2023; 
and the fourth batch was posted on September 7th, 2023.
Each batch was completed within one day after posting. 
With two different interfaces, we collected a total of 127,080 labels (3,177 sentence segments $\times$ 2 interface variations $\times$ 20 workers.)\footnote{The leading authors' institute has a guideline specifying that Institutional Review Board (IRB) approval is not necessary if a study on Amazon Mechanical Turk is focused only on asking workers about third parties and does not gather their personal opinions, beliefs, or experiences. Following this guideline, we did not seek IRB approval for the study.}

\paragraph{Worker Wage.} 
To target an hourly wage of \$10, we referenced literature which states the average reading speed on computer monitors is 250 words per minute~\cite{siegenthaler2012reading}. 
We determined working minutes by dividing the number of tokens by 250, then rounding up.
The task payment was set using the formula: \$0.05 + (Estimated Working Minutes $\times$ \$0.17).
Consequently, 53\% of our HITs were priced at \$0.22, 46\% at \$0.39, 0.5\% at \$0.56, and 0.5\% at \$0.73. 
For each HIT, we posted 40 assignments.
Factoring in the 40\% MTurk fee, the average cost to code each abstract with 40 workers was \$16.94. 

Beyond our initial calculations, multiple indicators suggest our workers earned over \$10 hourly. 
In this project, each worker received an average of (\$16.94/40)/1.4=\$0.303 for annotating one abstract. 
Our biomedical expert typically spent 96 seconds per abstract, leading to an estimated hourly wage of \$0.303$\times$(3600/96)=\$11.36.
Meanwhile, according to SOLVENT, MTurk workers took a median time of 1.3 minutes per abstract~\cite{chan2018solvent}.
This corresponds to an hourly rate of \$0.303$\times$(60/1.3)=\$13.98. 
Based on these figures, we are confident our workers' hourly rate exceeded \$10.

\subsection{Collecting Gold-Standard Labels Using Experts\label{subsec:expert-procedure}}
Similar to CODA-19~\cite{huang-etal-2020-coda}, we worked with two experts, a biomedical expert (Bio Expert), and a computer science expert (CS Expert). 
Both of these experts, who are also co-authors of this paper, manually annotated the entire test set of 200 abstracts from our MTurk study using the advanced interface.
The inter-annotator agreement (Cohen's kappa) between the two was 0.788.

\paragraph{Bio Expert provided gold-standard labels.}
Dr. Chien-Kuang Cornelia Ding is the biomedical expert, referred to as the ``Bio Expert'' in Table~\ref{table:all-worker-aggregation}.
She is a faculty member in the Department of Pathology at the University of California, San Francisco.
Dr. Ding possesses an M.D. and a Ph.D. in Genetics and Genomics. 
Notably, Dr. Ding played a critical role in the creation of the original CODA-19 label scheme, spending considerable time on manual data annotation to pinpoint corner cases and shape the initial CODA-19 instructions~\cite{huang-etal-2020-coda}.
Therefore, we trusted the Bio Expert's judgment throughout this study and treated her labels as the gold standard.

\paragraph{CS Expert labeled data for quality control in the annotation process and benchmarking non-experts' performance limits.}
The ``CS Expert'' in Table~\ref{table:all-worker-aggregation}, the first author of the paper, is a Ph.D. student in Informatics and well-acquainted with our annotation scheme. 
The purpose of collecting the CS Expert's labels is two-fold.
First, we used a subset of the CS Expert's labels to remove underperforming workers in the annotation process (Section~\ref{subsec:mturk-procedure}), simulating situations where Bio Expert's gold-standard labels are partially unavailable, requiring requesters to label some data to evaluate workers' performance. 
These situations are typical when computer science researchers develop datasets in specialized fields like biomedicine~\cite{sazzed-2022-banglabiomed,zong-etal-2022-extracting,mohr-etal-2022-covert,macherla2023mddial, parmar-etal-2022-boxbart, luo2022biotabqa}.\footnote{We confirmed with a co-author of the papers \citet{macherla2023mddial}, \citet{parmar-etal-2022-boxbart}, and \citet{luo2022biotabqa} that CS graduate students annotated data for quality control purposes in these projects, a detail not clearly stated in the paper.}
In such instances, experts within the domain often are unable to label large datasets quickly, so out-of-domain experts-- frequently CS graduate students-- sometime need to label portions of the data to assess the crowd labels' quality.
Second, the accuracy of the CS Expert's labels sets an estimated upper limit for non-expert performance, given their familiarity with the task and focused attention.
This benchmark helps in understanding the potential improvements in MTurk workers' performance through interface enhancements and attention checks.


\subsection{Annotating Data Using GPT-4\label{subsec:gpt-procedure}}

We used the full worker instruction from the original CODA-19 dataset as GPT-4's prompt for our data labeling~\cite{huang-etal-2020-coda}. 
Our initial perception was that GPT-4 underperformed in this specific task, given that it was reported to have inferior performance compared to the SciBERT model fine-tuned on the CODA-19 dataset~\cite{chandrasekhar-etal-2023-good}. 
However, we noticed that the prompt used in the said study
did not contain the entire abstract~\cite{chandrasekhar-etal-2023-good}, which might have led GPT-4 to rely on partial context for predictions.
So, we modified the prompt to include the full abstract for a zero-shot approach.
See Table~\ref{table:zero-shot-table} in the Appendix for our prompt details. 

Following prior studies that compared GPT-4's zero-shot capabilities with crowd workers~\cite{tornberg2023chatgpt}, we tested GPT-4 using both high (1.0) and low (0.2) temperature settings. 
For each setting, we executed the model five times and employed majority voting to determine the final label for every sentence segment.
We employed OpenAI's GPT-4 8K context model, priced at \$0.03 per 1,000 input tokens and \$0.06 per 1,000 output tokens.\footnote{OpenAI's Pricing: https://openai.com/pricing\#language-models}
The zero-shot GPT-4 experiment was conducted after collecting all crowd worker's data, on September 8th and 9th, 2023.
Post annotation, we recorded 2,507,240 input tokens and 780,979 output tokens, bringing the total GPT-4 cost to \$122.08.
This amounted to an average cost of \$0.61 per abstract.

\subsection{Label Cleaning Strategies\label{sec:label-cleaning}}
\paragraph{Label Cleaning Strategies.}
As described in Section~\ref{subsec:mturk-procedure}, we removed underperforming workers' qualifications after each data batch so they can not participate in future batches.
This raises an interesting question: How should we treat the labels these removed workers submit? 
We explored three strategies in this paper:


\begin{itemize}
    
    \item \textbf{All:} Retain every collected label without any exclusions.
    
    \item \textbf{Exclude-By-Worker:} Exclude labels from any MTurk worker who was ever removed.
    
    \item \textbf{Exclude-By-Batch:} Only exclude a label if its annotator was removed during that specific data batch. This means if a worker was removed from a given batch, we only exclude their labels from that batch but retain those from prior batches.
    
\end{itemize}

Only the selected labels will proceed to the follow-up label aggregation step.


\subsection{Label Aggregation Methods\label{sec:aggregation-models}}
In our study, we explored a range of label aggregation algorithms.
First, we adopted the majority voting method, including its tie-breaker approach, directly from CODA-19~\cite{huang-etal-2020-coda}.
Second,
we utilized a series of aggregation algorithms 
provided by Crowd-Kit~\cite{CrowdKit,ustalov2021general},
such as 
Dawid-Skene~\cite{dawid1979maximum},
One-coin Dawid-Skene~\cite{zhang2014spectral},
M-MSR (Matrix Mean-Subsequence-Reduced Algorithm)~\cite{ma2020adversarial},
Worker Agreement with Aggregate (Wawa)~\cite{Appen_2021},
Zero-Based Skill (ZBS)~\cite{ZeroBasedSkill}
and GLAD (Generative model of Labels, Abilities, and Difficulties)~\cite{whitehill2009whose}.
Finally, 
we also experimented with MACE (Multi-Annotator Competence Estimation) implemented by~\citet{hovy2013learning}.
This subsection delves into the details of these aggregation algorithms, setups, and implementations.


\subsubsection{Choosing Crowd-Kit}
In our study, we implemented all the algorithms using Crowd-Kit,\footnote{Crowd-Kit: https://github.com/Toloka/crowd-kit/} except for Majority Voting and MACE.
This choice was driven by our aim to mimic a realistic crowdsourced data annotation process, leading us to select publicly available and easy-to-use toolkits for label aggregation.
We selected Crowd-Kit for its regular updates and better maintenance over other tools like Active Crowd Toolkit~\cite{venanzi2015activecrowdtoolkit} and the CrowdTruth~\cite{CrowdTruth2} framework's GitHub repository.\footnote{Crowd-Kit received an update just a few days before we submitted the final version of this paper to CHI in February 2024.}
Additionally, Crowd-Kit, with over 180 stars, is GitHub's most popular label aggregation toolkit.
It is likely to be the preferred choice for many users seeking to aggregate algorithms for crowd labels.
\citet{evtikhiev2023out} employed the M-MSR algorithm from Crowd-Kit to derive human ``ground truth'' grades;
\citet{hiippala-etal-2022-developing} utilized its Dawid-Skene algorithm for identifying the most probable answers from three responses.



\subsubsection{Comparison between each algorithm}
We chose these algorithms for their representativeness and common use in label aggregation.
This sub-subsection describes and compares all the algorithms used in our work.
\textbf{Majority Vote}, a basic method, determines the final label based on the most common answer among workers.
Its primary drawback is the equal weighting it assigns to all workers, regardless of their expertise level.
The \textbf{Dawid-Skene} algorithm~\cite{dawid1979maximum} assesses the expertise level of each worker by representing them with a confusion matrix and utilizes the Expectation-Maximization (EM) framework to formulate two iterative steps for this process; 
the \textbf{One-Coin Dawid-Skene}~\cite{zhang2014spectral} variant follows the same principles as the original Dawid-Skene model, but it differs in the way it calculates worker errors during the M-step of the algorithm. 
The one-coin model, due to its simplicity, is easier to estimate and has better convergence properties.
The \textbf{M-MSR}~\cite{ma2020adversarial} model operates under the assumption that workers possess varying levels of expertise and represent these workers through vectors denoting their skills.
It then estimates each worker's probability distribution by solving a rank-one matrix completion problem.
The \textbf{GLAD}~\cite{whitehill2009whose} algorithm models the difficulty of each task and then analyzes each worker's response to determine the optimal values through the application of Gradient Descent~\cite{10.14778/3055540.3055547}. 
The \textbf{MACE}~\cite{hovy2013learning} approach focuses on determining if a worker is spamming and creates a probability model that associates each worker with a specific label probability distribution.
For more detailed insights, we would like to point to the paper by \citet{10.14778/3055540.3055547}, which surveys 17 aggregation algorithms and finds the Dawid-Skene algorithm to generally provide reliable results.


Moreover, we engaged in discussions with the developers of Crowd-Kit (who are not co-authors of this paper)
about the \textbf{Wawa}~\cite{Appen_2021} and \textbf{ZBS}~\cite{ZeroBasedSkill} methods, which were not based on published academic papers.
They explained that both Wawa and ZBS were developed in-house at Toloka,\footnote{Toloka: https://toloka.ai/} the data labeling platform that develops and maintains Crowd-Kit.
Wawa offers a modest yet intuitive enhancement over the Majority Vote method.
ZBS, on the other hand, was initially designed to manage streaming responses, particularly when new annotations are introduced.
It can rapidly and efficiently update the skills of annotators, maintaining a modest quality improvement over Majority Voting.
Importantly, it achieves this without the need to reprocess the entire dataset to infer the final label.

\subsubsection{Implementation Details}
First, we used the majority voting method, including its tie-breaker approach, directly from CODA-19~\cite{huang-etal-2020-coda}.
In cases of a tie, we prioritize the labels in the following sequence: Finding, Method, Purpose, Background, and Other.
Second,
we used the algorithms of Dawid-Skene, One-Coin Dawid-Skene, M-MSR, GLAD, Wawa, and ZBS provided by Crowd-Kit.
For all these models, we adhered to the default parameters specified by Crowd-Kit.
M-MSR occasionally encountered failures when the number of crowd workers was less than 10. We chose to ignore these failures in our simulation phase. In cases where a failure occurred, we reinitiated the simulation using a newly shuffled group of workers. 
Finally, we selected the MACE implementation by~\citet{hovy2013learning} due to its extensive use and high speed.
Additionally, its aggregation results proved to be more stable compared to those of Crowd-Kit's MACE.


\section{Experimental Results}

In this section, we first overview the comparative results of GPT-4 and MTurk pipeline with a variety of settings (Section~\ref{sec:mturk-vs-gpt}) and then show the results of incorporating GPT-4 into MTurk pipelines (Section~\ref{sec:mturk-plus-gpt}.)

Notably, we experimented with two interfaces, the basic interface and the advanced interface, so all the experiments have two sets of results.

\subsection{GPT-4 vs. MTurk Pipelines\label{sec:mturk-vs-gpt}}

\begin{table*}[t]
\centering
\setlength{\tabcolsep}{2.8pt}
\begin{tabular}{llrrrrrrrrrrrrrrrrr}
\toprule
\multicolumn{1}{l}{\textbf{Eval}} & \multicolumn{3}{c}{\textbf{Background}} & \multicolumn{3}{c}{\textbf{Purpose}} & \multicolumn{3}{c}{\textbf{Method}} & \multicolumn{3}{c}{\textbf{Finding}} & \multicolumn{3}{c}{\textbf{Other}} & \multicolumn{1}{c}{\multirow{2}{*}{\textbf{Acc}}} & \multicolumn{1}{c}{\multirow{2}{*}{\textbf{Kappa}}} \\ \cmidrule(lr){2-4} \cmidrule(lr){5-7} \cmidrule(lr){8-10} \cmidrule(lr){11-13} \cmidrule(lr){14-16}
\textbf{Label} & \multicolumn{1}{c}{P} & \multicolumn{1}{c}{R} & \multicolumn{1}{c}{F1} & \multicolumn{1}{c}{P} & \multicolumn{1}{c}{R} & \multicolumn{1}{c}{F1} & \multicolumn{1}{c}{P} & \multicolumn{1}{c}{R} & \multicolumn{1}{c}{F1} & \multicolumn{1}{c}{P} & \multicolumn{1}{c}{R} & \multicolumn{1}{c}{F1} & \multicolumn{1}{c}{P} & \multicolumn{1}{c}{R} & \multicolumn{1}{c}{F1} & \multicolumn{1}{c}{} & \multicolumn{1}{c}{} \\ \midrule
Basic UI MV & .713 & .281 & .403 & .149 & .525 & .232 & .368 & .599 & .456 & .772 & .507 & .612 & 1.000 & .286 & .444 & .477 & .285 \\
Advanced UI MV & .598 & .307 & .405 & .112 & .438 & .179 & .373 & .634 & .469 & .815 & .425 & .559 & \multicolumn{1}{c}{-} & 0 & \multicolumn{1}{c}{-} & .442 & .259 \\
GPT-4 (t=.2) & .860 & .913 & .885 & .499 & .843 & .627 & .775 & .871 & .820 & .982 & .784 & .872 & .322 & .905 & .475 & .836 & .764 \\
GPT-4 (t=1.0) & .859 & .903 & .881 & .493 & .829 & .619 & .766 & .876 & .818 & .978 & .783 & .870 & .346 & .857 & .493 & .833 & .760 \\
CS Expert & .900 & .801 & .848 & .541 & .853 & .662 & .856 & .801 & .828 & .913 & .915 & .914 & 1.000 & .619 & .765 & \textbf{.859} & .788 \\ \bottomrule
\end{tabular}
\caption{Performance using Bio Expert as the gold standard.  The CS Expert achieves the highest accuracy of 85.9\% over all models. GPT-4 at temperature of .2 and 1.0 have accuracies of 83.6\% and 83.3\%. Basic and Advanced Majority Vote had accuracy of 47.7\% and 44.2\%.}
\label{table:all-worker-stats-summary}
\end{table*}

\begin{figure*}[t]
    \centering
    \includegraphics[width=.99\textwidth]{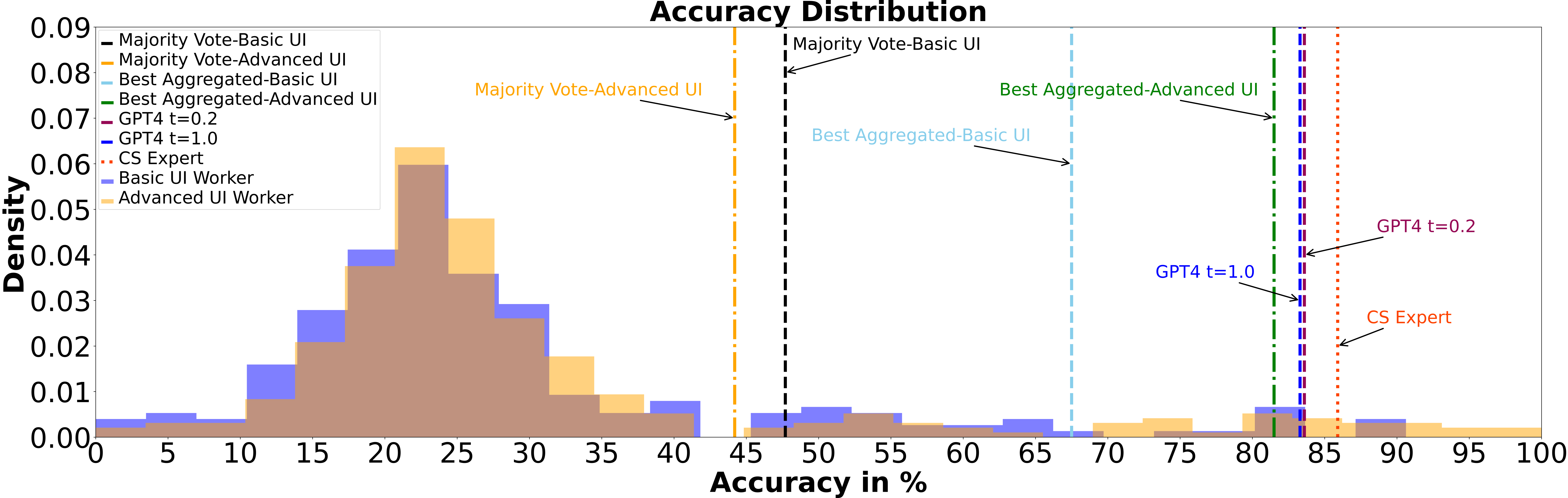}
    \caption{A density histogram of the response of all individual crowd workers' accuracy (filtered by the Exclude-By-Worker strategy.) The dash lines show majority vote accuracy and the best aggregation algorithm accuracy on both basic and advanced interface response, the accuracy of GPT-4 (t=0.2 and t=1.0), and the accuracy of CS Expert. The accuracy uses the Bio Expert as the gold standard. 
    }
    \label{fig:crow-histogram}
    \Description{It is a density histogram of response of all individual crowd worker's accuracy. The dash lines show majority vote accuracy and the best aggregation accuracy for both basic and advanced interface response, the accuracy of GPT-4 and the accuracy of the CS Expert. The accuracy uses the Bio Expert as the gold standard.}
\end{figure*}

To evaluate the labeling accuracy, we used the Bio Expert's labels as the gold-standard labels.
We employed the majority vote as our baseline model, as it was used in the original CODA-19 paper.
Table~\ref{table:all-worker-stats-summary} shows the results.
\textbf{GPT-4 exhibited accuracies of 83.6\% and 83.3\% at low (0.2) and high (1.0) temperatures, respectively.}
Using Majority Voting (MV), labels provided by MTurk workers from the Basic and Advanced interface groups achieved accuracies of 47.7\% and 44.2\%, respectively.
The CS Expert achieved the highest accuracy of 85.9\%.

Figure~\ref{fig:crow-histogram} shows the accuracy distribution of MTurk workers within the basic and advanced interface groups.
We adopted the histogram format from \citet{tornberg2023chatgpt}'s work for easier comparison. 
As highlighted in our introduction, while most workers reached accuracies of 20\% to 30\%, the aggregated result is notably more reliable.
We also noticed that workers in the advanced interface group exhibited marginally better performance, with their histogram leaning more to the right.





\subsubsection{Exclude-By-Worker is the best label-cleaning strategy.}
Next, we experimented with the three different label-cleaning strategies mentioned in Section~\ref{sec:aggregation-models}.
Table~\ref{table:basic-acc-result} and Table~\ref{table:advanced-acc-result} show the accuracy of three strategies paired with different aggregation models. 
Both the Exclude-By-Worker and Exclude-By-Batch strategies enhanced the aggregation accuracy, but Exclude-By-Worker produced superior results.

When pairing the One-Coin Dawid-Skene aggregation method with MTurk workers in the Advanced Interface group using the Exclude-By-Worker approach, we achieved \textbf{the best accuracy of 81.5\%}.
While this surpassed other aggregation models with different strategies, it \textbf{did not exceed the performance of GPT-4 (83.6\%)}.




\begin{table*}[t]
\centering
\begin{tabular}{lccccccccc}
\multicolumn{10}{c}{\textbf{Basic Interface}}\\
\toprule
\tiny Acc of GPT-4 (t=0.2) = \textbf{.836} & \multicolumn{3}{c}{\textbf{All Workers}} & \multicolumn{3}{c}{\textbf{Exclude-By-Worker}} & \multicolumn{3}{c}{\textbf{Exclude-By-Batch}} \\ \cmidrule(lr){2-4} \cmidrule(lr){5-7} \cmidrule(lr){8-10}
\textbf{Method} & \textbf{Acc} & \textbf{P-Value} & \textbf{95\% CI} & \textbf{Acc} & \textbf{P-Value} & \textbf{95\% CI} & \textbf{Acc} & \textbf{P-Value} & \textbf{95\% CI} \\ \midrule
\textbf{MV} & .477 & <.001 & {[}.459, .494{]} & .522 & <.001 & {[}.504, .539{]} & .484 & <.001 & {[}.466, .501{]} \\
\textbf{DawidSkene} & .549 & <.001 & {[}.532, .567{]} & .604 & <.001 & {[}.587, .621{]} & .592 & <.001 & {[}.575, .609{]} \\
\textbf{OneCoin} & .663 & <.001 & {[}.646, .679{]} & .671 & <.001 & {[}.655, .688{]} & .649 & <.001 & {[}.633, .666{]} \\
\textbf{GLAD} & .608 & <.001 & {[}.591, .625{]} & .616 & <.001 & {[}.599, .633{]} & .597 & <.001 & {[}.580, .614{]} \\
\textbf{M-MSR} & .436 & <.001 & {[}.419, .453{]} & .443 & <.001 & {[}.425, .460{]} & .414 & <.001 & {[}.396, .431{]} \\
\textbf{MACE} & \textbf{.675} & <.001 & {[}.659, .691{]} & \textbf{.675} & <.001 & {[}.659, .691{]} & \textbf{\underline{.675}} & <.001 & {[}.660, .692{]} \\
\textbf{Wawa} & .527 & <.001 & {[}.509, .544{]} & .555 & <.001 & {[}.537, .572{]} & .524 & <.001 & {[}.506, .541{]} \\
\textbf{ZBS} & .547 & <.001 & {[}.530, .564{]} & .580 & <.001 & {[}.563, .597{]} & .553 & <.001 & {[}.536, .570{]} \\ \midrule
\textbf{Avg. Acc} & .560 & - & - & .583 & -  & - & .561 & - & - \\ \midrule
\textbf{\#workers} & \multicolumn{3}{c}{216} & \multicolumn{3}{c}{134} & \multicolumn{3}{c}{176}
\\ \bottomrule
\end{tabular}
\caption{Aggregation Accuracy Results of the Basic Interface Group. \textbf{Bold} and \underline{underline} highlight the highest score within the column and across the table, respectively. P-value is obtained by comparing with GPT-4 over the article-level accuracy. 
(\textsuperscript{**}: p<0.01; \textsuperscript{***}: p<0.001. Paired t-test. N=200)
}
\label{table:basic-acc-result}
\end{table*}

\begin{table*}[t]
\centering
\begin{tabular}{lccccccccc}
\multicolumn{10}{c}{\textbf{Advanced Interface}}\\
\toprule
\tiny Acc of GPT-4 (t=0.2) =\textbf{.836} & \multicolumn{3}{c}{\textbf{All Workers}} & \multicolumn{3}{c}{\textbf{Exclude-By-Worker}} & \multicolumn{3}{c}{\textbf{Exclude-By-Batch}} \\ \cmidrule(lr){2-4} \cmidrule(lr){5-7} \cmidrule(lr){8-10}
\textbf{Method} & \textbf{Acc} & \textbf{P-Value} & \textbf{95\% CI} & \textbf{Acc} & \textbf{P-Value} & \textbf{95\% CI} & \textbf{Acc} & \textbf{P-Value} & \textbf{95\% CI} \\ \midrule
\textbf{MV} & .442 & <.001 & {[}.425, .459{]} & .490 & <.001 & {[}.473, .507{]} & .458 & <.001 & {[}.440, .475{]} \\
\textbf{DawidSkene} & .555 & <.001  & {[}.537, .572{]} & .637 & <.001 & {[}.620, .654{]} & .583 & <.001 & {[}.565, .600{]} \\
\textbf{OneCoin} & .517 & <.001  & {[}.500, .535{]} & \textbf{\underline{.815}} & .018 & {[}.801, .828{]} & .536 & <.001 & {[}.519, .553{]} \\
\textbf{GLAD} & .631 & <.001 & {[}.614, .648{]} & .692 & <.001 & {[}.675, .708{]} & .639 & <.001 & {[}.621, .655{]} \\
\textbf{M-MSR} & .428 & <.001  & {[}.411, .445{]} & .467 & <.001 & {[}.450, .485{]} & .438 & <.001 & {[}.421, .455{]} \\
\textbf{MACE} & \textbf{.798} & <.001  & {[}.784, .812{]} & .807 & .003 & {[}.793, .821{]} &\textbf{ .787} & <.001 & {[}.773, .801{]} \\
\textbf{Wawa} & .470 & <.001  & {[}.452, .487{]} & .545 & <.001 & {[}.528, .562{]} & .484 & <.001 & {[}.466, .501{]} \\
\textbf{ZBS} & .528 & <.001  & {[}.510, .545{]} & .596 & <.001 & {[}.579, .614{]} & .542 & <.001 & {[}.524, .559{]} \\ \midrule
\textbf{Avg. Acc} & .546 & - & - & .631 & - & - & .558 & - & -\\ \midrule
\textbf{\#workers} & \multicolumn{3}{c}{199} & \multicolumn{3}{c}{129} & \multicolumn{3}{c}{162}
\\ \bottomrule
\end{tabular}
\caption{Aggregation Accuracy Results of the Advanced Interface Group. \textbf{Bold} and \underline{underline} highlight the highest score within the column and across the table, respectively.  P-value is obtained by comparing with GPT-4 over the article-level accuracy. 
(\textsuperscript{**}: p<0.01; \textsuperscript{***}: p<0.001. Paired t-test. N=200)
}
\label{table:advanced-acc-result}
\end{table*}


\subsubsection{While more crowd labels enhanced accuracy, even 20 could not surpass GPT-4.}
\begin{figure*}
    \centering
    \begin{subfigure}{0.34\textheight}
        \centering
        \includegraphics[width=\linewidth]{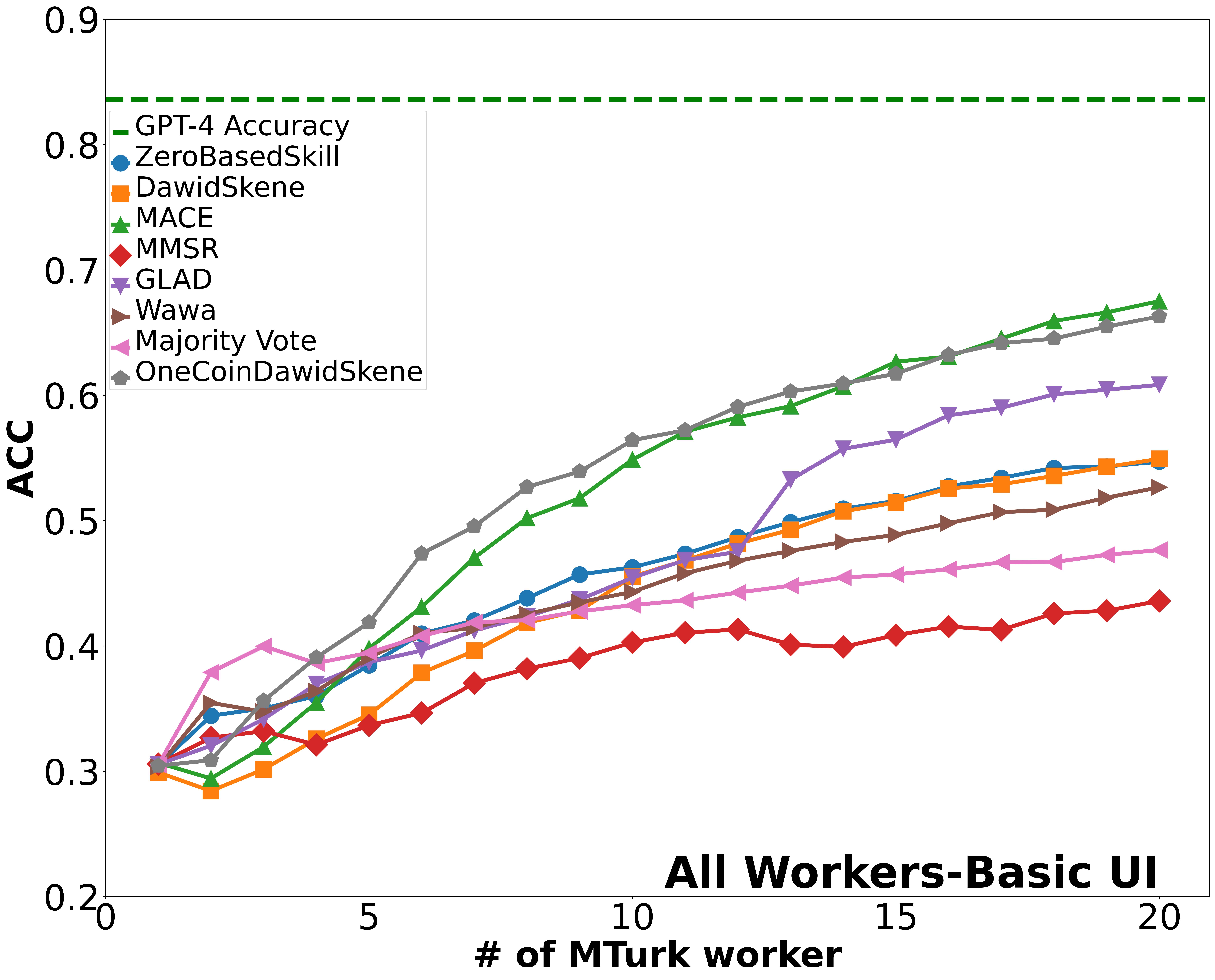}
        \caption{All Workers-Basic UI}
        \label{fig:all-aggregation-together-basic-all-worker}
    \end{subfigure}
    \hfill
    \begin{subfigure}{0.34\textheight}
        \centering
        \includegraphics[width=\linewidth]{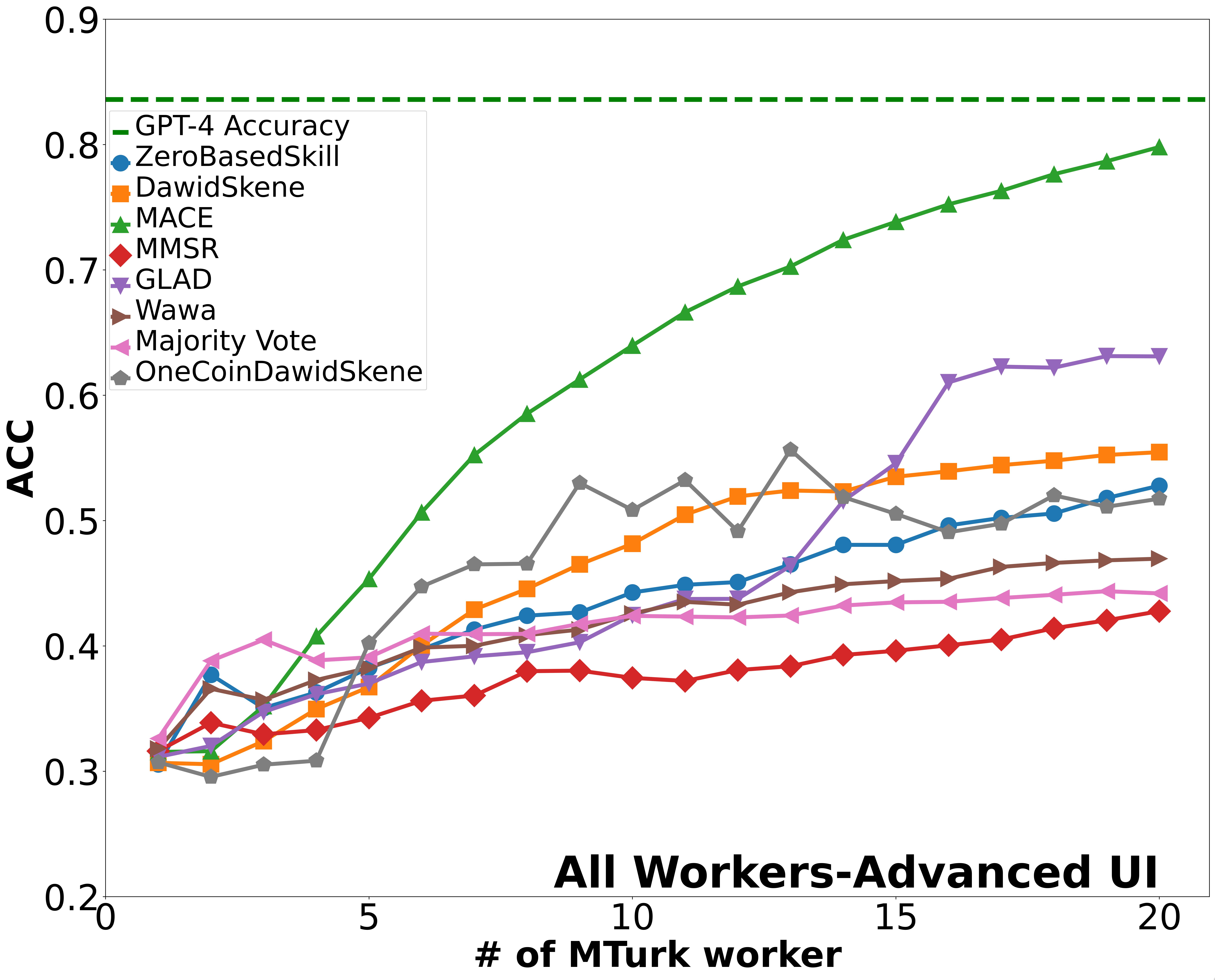}
        \caption{All Workers-Advanced UI}
        \label{fig:all-aggregation-together-advance-all-worker}
    \end{subfigure}
    
    \begin{subfigure}{0.34\textheight}
        \centering
        \includegraphics[width=\linewidth]{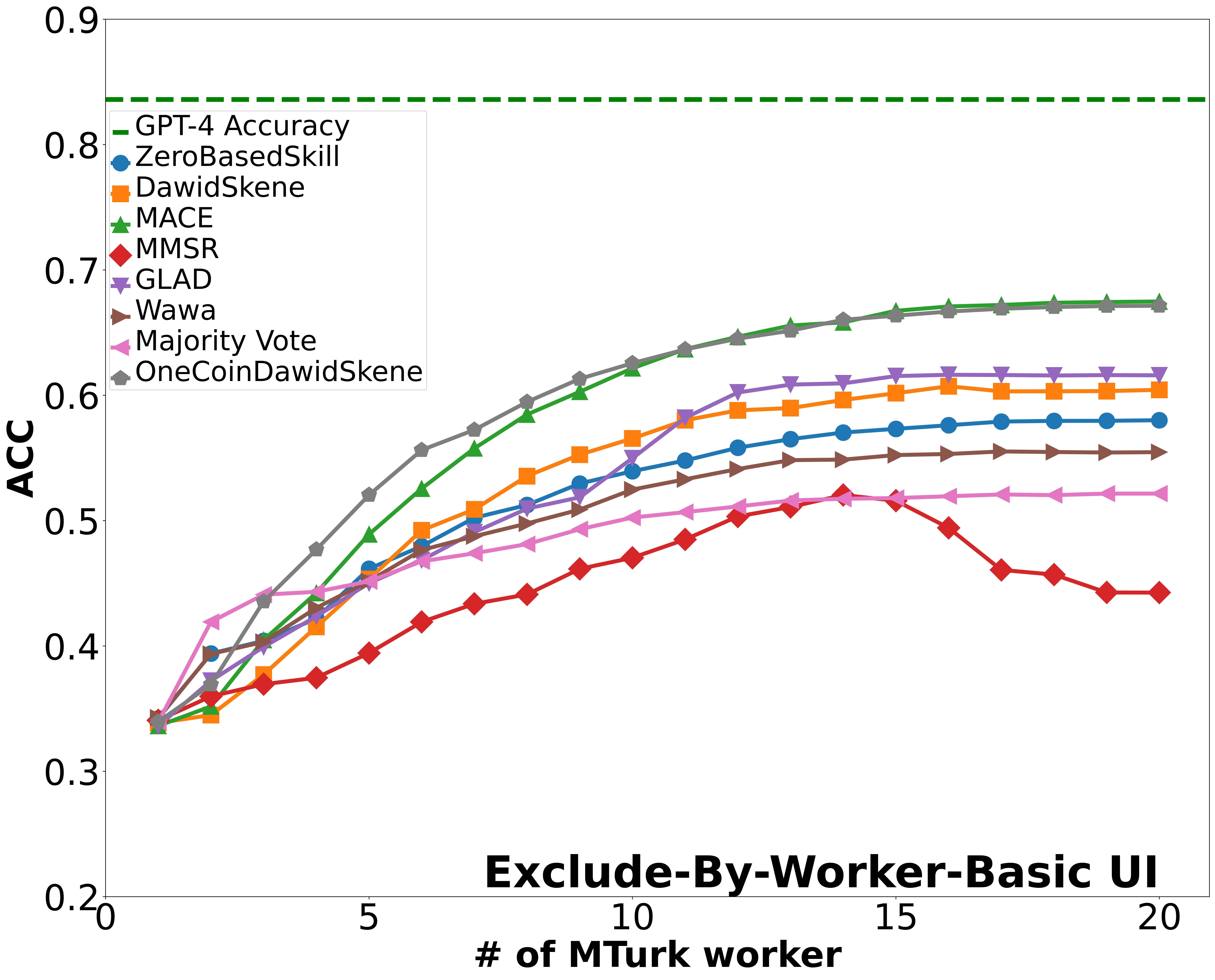}
        \caption{Exclude-By-Worker-Basic UI}
        \label{fig:all-aggregation-together-basic-no-bad-worker}
    \end{subfigure}
    \hfill
    \begin{subfigure}{0.34\textheight}
        \centering
        \includegraphics[width=\linewidth]{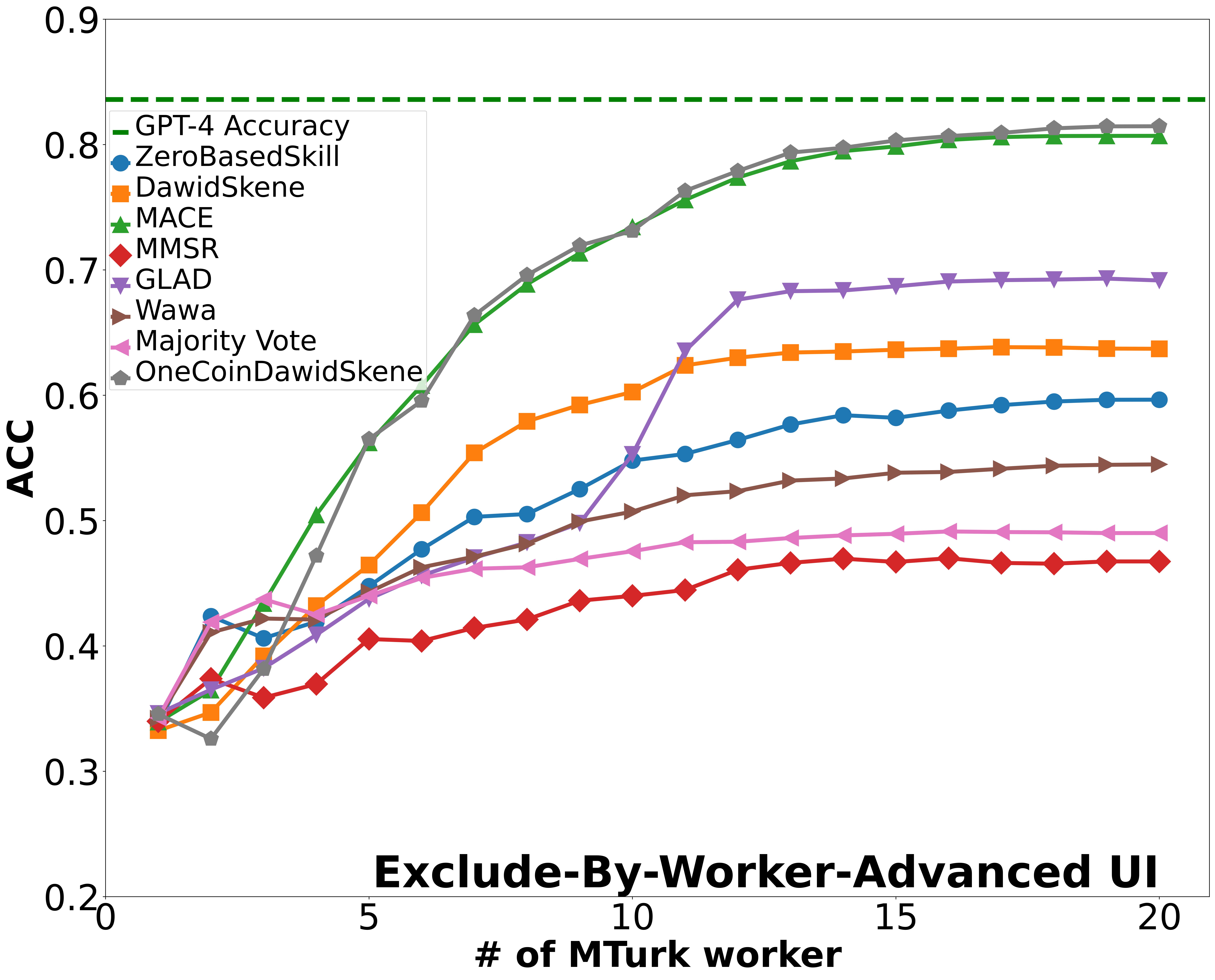}
        \caption{Exclude-By-Worker-Advanced UI}
        \label{fig:all-aggregation-together-advance-no-bad-worker}
    \end{subfigure}
    
    \begin{subfigure}{0.34\textheight}
        \centering
        \includegraphics[width=\linewidth]{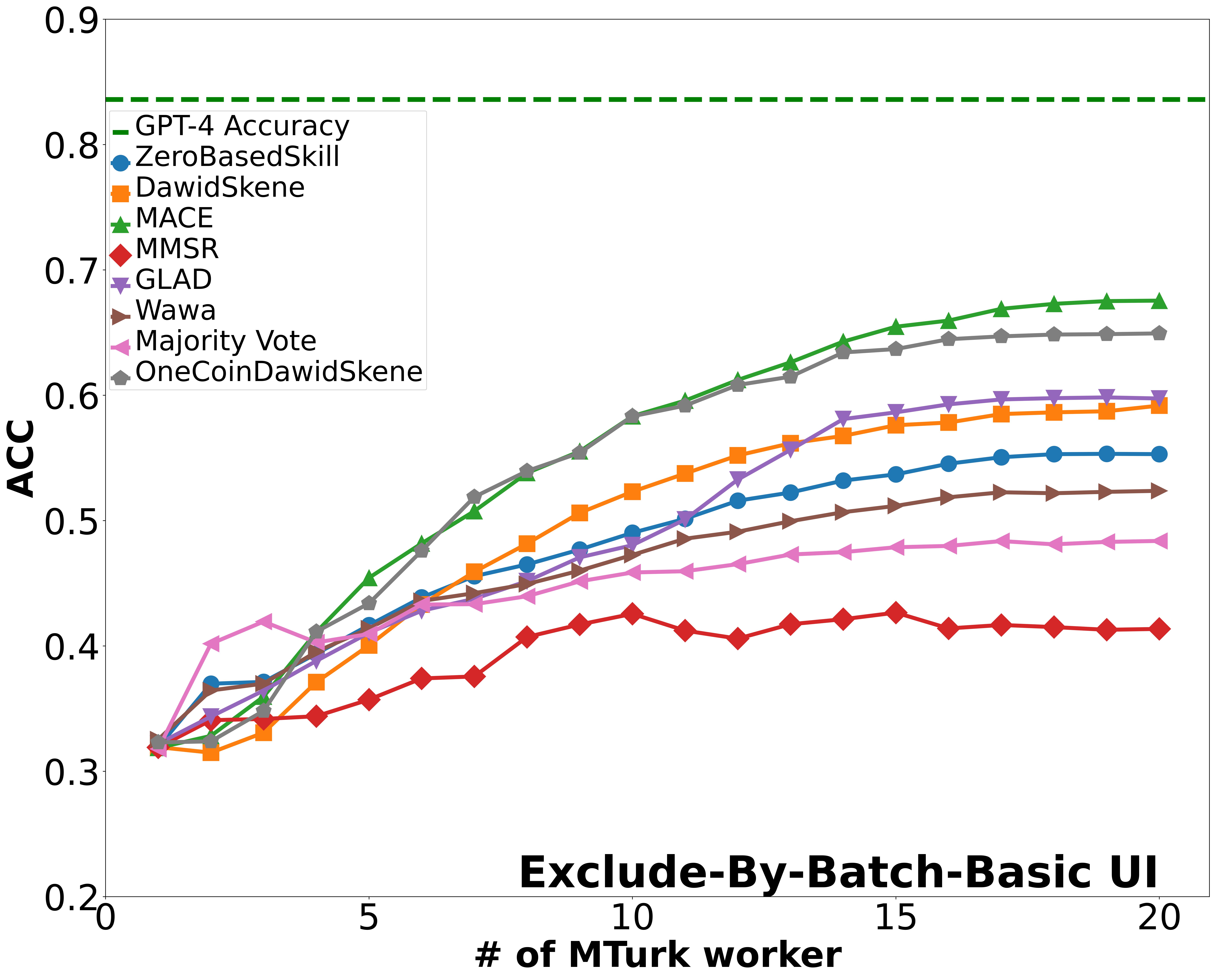}
        \caption{Exclude-By-Batch-Basic UI}
        \label{fig:all-aggregation-together-basic-no-bad-label}
    \end{subfigure}
    \hfill
    \begin{subfigure}{0.34\textheight}
        \centering
        \includegraphics[width=\linewidth]{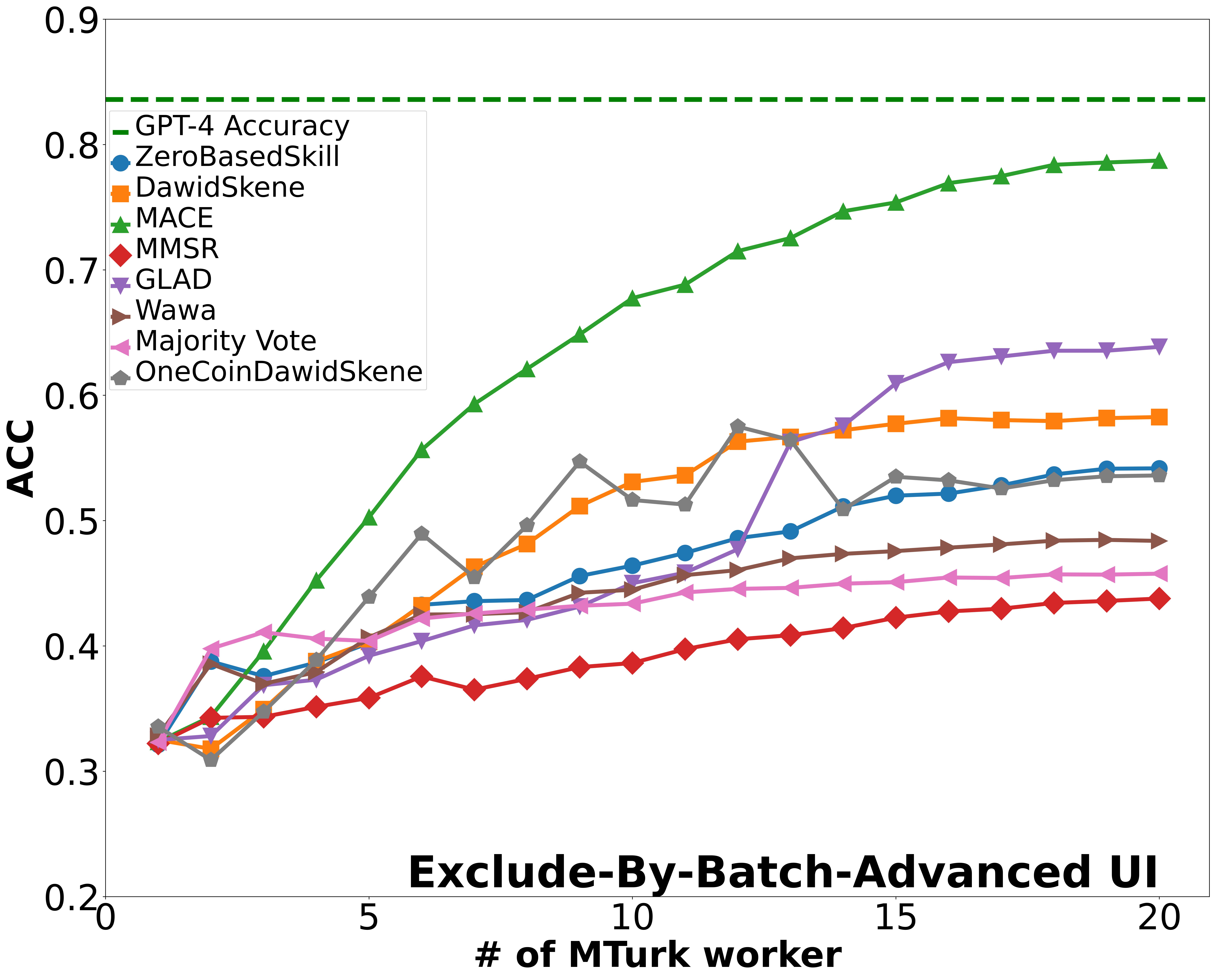}
        \caption{Exclude-By-Batch-Advanced UI}
        \label{fig:all-aggregation-together-advance-no-bad-label}
    \end{subfigure}
    
    \caption{Aggregation Methods for All Workers, Exclude-By-Worker, and Exclude-By-Batch. Among the various models and strategies employed, only the combination of the One-Coin Dawid-Skene aggregation method and workers in the Advanced Interface group using the Exclude-By-Worker approach demonstrated performance that closely approached that of GPT-4. In contrast, the other models utilizing different strategies were unable to achieve GPT-4's performance level. }
    \label{fig:all-aggregation-all-strageties}
    \Description{There are six subfigures in total for all the aggregation methods simulation on Basic/Advanced All-Workers, Basic/Advanced Exclude-By-Worker, and Basic/Advanced Exclude-By-Batch.}
\end{figure*}

To better grasp how the number of workers affects aggregated labels' accuracy, we ran simulations using varying numbers of randomly selected worker results.
The figures presented are averages from 20 rounds of these random selections. 
Figure~\ref{fig:all-aggregation-all-strageties} shows these results, with a detailed breakdown for the top-performing Exclude-By-Worker strategy in Table~\ref{table:no-bad-worker-aggregation}. 
Detailed results for the other two strategies can be found in the Appendix, specifically Table~\ref{table:all-worker-aggregation} and Table~\ref{table:no-bad-label-aggregation}.

As a result, while more crowd labels improved accuracy, even aggregating 20 could not surpass GPT-4.
We also noticed that in the simulations, accuracy improved for most aggregation methods as the number of MTurk workers increased.

As the Exclude-By-Worker label cleaning strategy systematically produced the best results, \textbf{we showcase results only using the Exclude-By-Worker cleaning strategy throughout the remainder of the paper.}

\begin{table*}[t]
\centering
\setlength{\tabcolsep}{2.8pt}
\begin{tabular}{llrrrrrrrrrrrrrrrrrr}
\toprule
\multicolumn{1}{l}{\textbf{Eval}} & \multicolumn{3}{c}{\textbf{Background}} & \multicolumn{3}{c}{\textbf{Purpose}} & \multicolumn{3}{c}{\textbf{Method}} & \multicolumn{3}{c}{\textbf{Finding}} & \multicolumn{3}{c}{\textbf{Other}} & \multicolumn{1}{c}{\multirow{2}{*}{\textbf{Acc}}} & \multicolumn{1}{c}{\multirow{2}{*}{\textbf{Kappa}}} \\ \cmidrule(lr){2-4} \cmidrule(lr){5-7} \cmidrule(lr){8-10} \cmidrule(lr){11-13} \cmidrule(lr){14-16}
\textbf{Label} & \multicolumn{1}{c}{P} & \multicolumn{1}{c}{R} & \multicolumn{1}{c}{F1} & \multicolumn{1}{c}{P} & \multicolumn{1}{c}{R} & \multicolumn{1}{c}{F1} & \multicolumn{1}{c}{P} & \multicolumn{1}{c}{R} & \multicolumn{1}{c}{F1} & \multicolumn{1}{c}{P} & \multicolumn{1}{c}{R} & \multicolumn{1}{c}{F1} & \multicolumn{1}{c}{P} & \multicolumn{1}{c}{R} & \multicolumn{1}{c}{F1} & \multicolumn{1}{c}{} & \multicolumn{1}{c}{} \\ \midrule
\multicolumn{19}{c}{\textbf{Basic UI}} \\
MV  & .722 & .342 & .465 & .166 & .548 & .254 & .408 & .606 & .488 & .794 & .564 & .660 & .857 & .286 & .429 & .522 & .337 \\
DawidSkene  & .789 & .407 & .537 & .228 & .530 & .319 & .512 & .682 & .585 & .858 & .669 & .752 & .063 & .571 & .114 & .604 & .446 & \\
OneCoin  & .817 & .461 & .590 & .361 & .599 & .451 & .506 & .726 & .597 & .820 & .757 & .787 & 1.000 & .238 & .385 & .671 & .514 & \\
GLAD  & .759 & .483 & .590 & .228 & .636 & .336 & .492 & .675 & .569 & .861 & .650 & .740 & .474 & .429 & .450 & .616 & .460 & \\
M-MSR  & .553 & .378 & .449 & .134 & .452 & .206 & .345 & .521 & .415 & .755 & .438 & .555 & .171 & .286 & .214 & .443 & .249 & \\
\textbf{MACE}  & .742 & .589 & .657 & .288 & .618 & .392 & .582 & .687 & .630 & .881 & .717 & .791 & .155 & .619 & .248 & \textbf{.675} & .536 & \\
Wawa  & .712 & .424 & .531 & .183 & .604 & .281 & .439 & .647 & .523 & .859 & .569 & .684 & .700 & .333 & .452 & .555 & .389 & \\
ZBS  & .735 & .446 & .555 & .202 & .627 & .305 & .459 & .674 & .546 & .868 & .596 & .707 & .636 & .333 & .438 & .580 & .419 & \\ \midrule
\multicolumn{19}{c}{\textbf{Advanced UI}}\\
MV  & .642 & .344 & .448 & .143 & .488 & .221 & .390 & .656 & .489 & .835 & .490 & .618 & \multicolumn{1}{c}{-} & 0 & \multicolumn{1}{c}{-} & .490 & .310 & \\ 
DawidSkene  & .688 & .411 & .515 & .187 & .618 & .288 & .720 & .725 & .722 & .936 & .705 & .804 & .054 & .476 & .097 & .637 & .501 & \\
\textbf{OneCoin}  & .874 & .768 & .818 & .500 & .618 & .553 & .703 & .853 & .771 & .909 & .853 & .880 & 1.000 & .286 & .444 & \textbf{.815} & .723 & \\
GLAD  & .807 & .610 & .695 & .254 & .645 & .365 & .582 & .765 & .661 & .927 & .707 & .802 & .615 & .381 & .471 & .692 & .564 & \\
M-MSR  & .555 & .414 & .474 & .142 & .516 & .223 & .388 & .599 & .471 & .834 & .433 & .570 & .167 & .048 & .074 & .467 & .291 & \\
MACE  & .817 & .822 & .819 & .424 & .691 & .525 & .767 & .818 & .791 & .939 & .814 & .872 & .310 & .619 & .413 & .807 & .718 & \\
Wawa  & .628 & .460 & .531 & .168 & .571 & .260 & .462 & .684 & .552 & .890 & .523 & .659 & 1.000 & .190 & .320 & .545 & .384 & \\
ZBS  & .681 & .537 & .600 & .190 & .608 & .289 & .518 & .703 & .597 & .903 & .579 & .706 & .857 & .286 & .429 & .596 & .447 \\ \midrule
GPT-4 (t=.2) & .860 & .913 & .885 & .499 & .843 & .627 & .775 & .871 & .820 & .982 & .784 & .872 & .322 & .905 & .475 & .836 & .764 \\
\bottomrule
\end{tabular}
\caption{Exclude-By-Worker Table. All models use Bio Expert as the gold standard. The baseline is the Majority Vote (MV). From Exclude-By-Worker results, the One-Coin Dawid-Skene aggregation model achieves the highest accuracy for both basic and advanced interfaces. Advanced One-Coin Dawid-Skene reaches 81.5\% and outperforms other aggregation models in every aspect. The accuracy from advanced One-Coin Dawid-Skene almost reaches the accuracy of the GPT-4 (t=.2), 83.6\%.
}
\label{table:no-bad-worker-aggregation}
\end{table*}

\subsection{GPT-4 in MTurk Pipelines\label{sec:mturk-plus-gpt}}

\begin{figure*}
    \centering
    \begin{subfigure}{0.36\textheight}
    \centering
    \includegraphics[width=\linewidth]{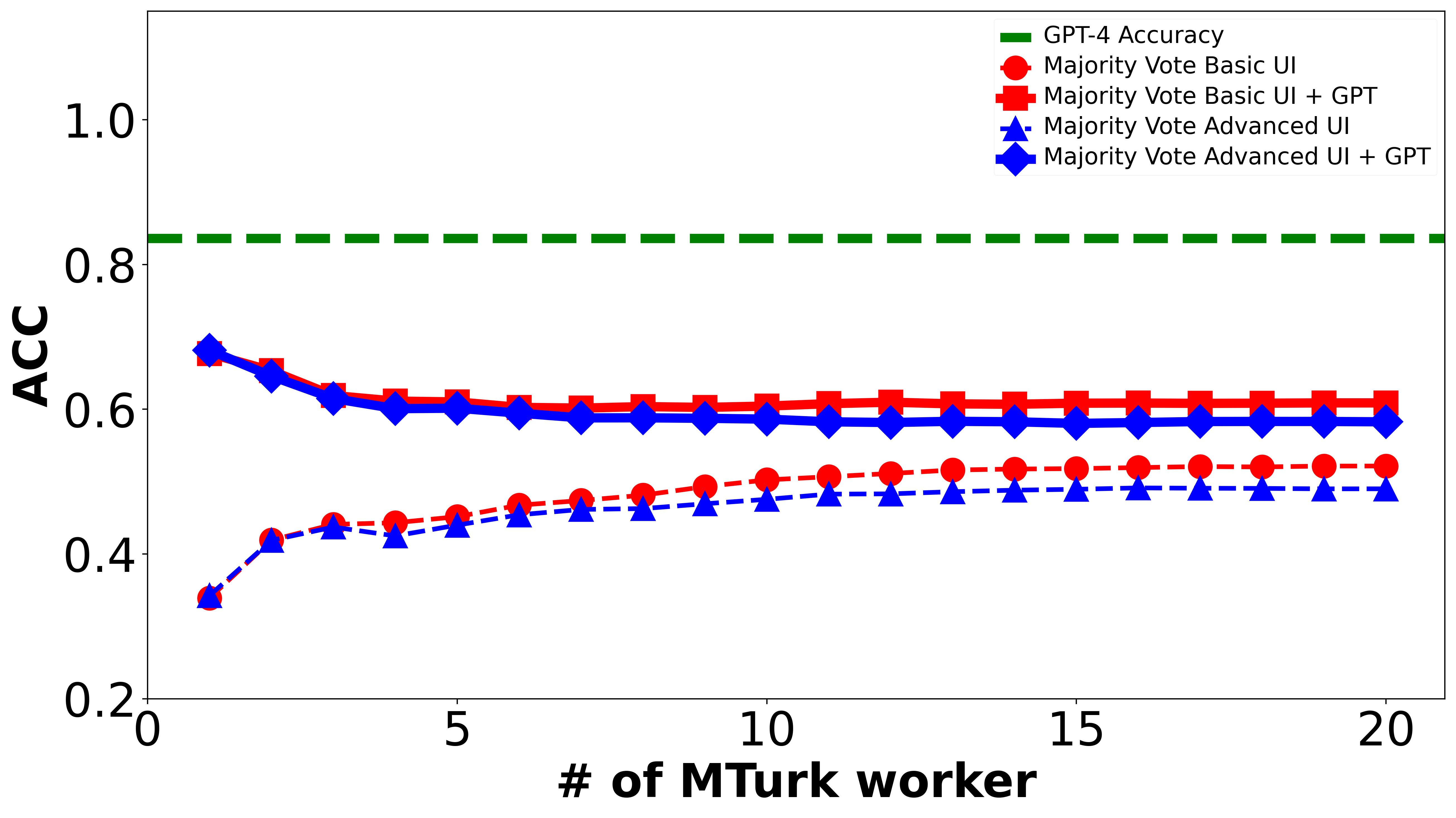}
    \caption{Majority Vote}
    \label{fig:agg-mv-baseline-no-bad-worker}
    \end{subfigure}
     \hfill
    \begin{subfigure}{0.36\textheight}
        \centering
        \includegraphics[width=\linewidth]{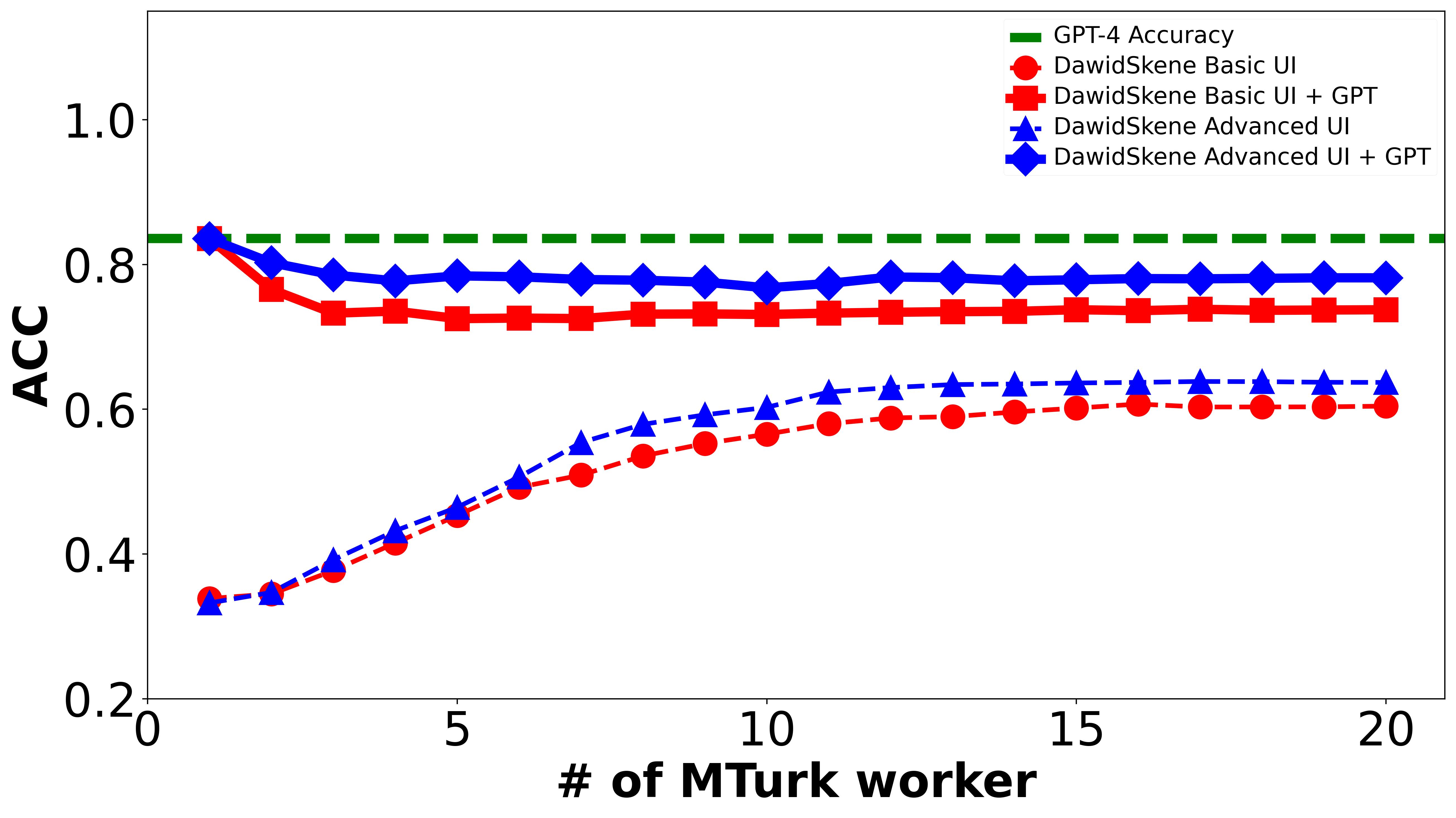}
        \caption{Dawid-Skene}
        \label{fig:agg-dawid-no-bad-worker}
    \end{subfigure}

    \begin{subfigure}{0.36\textheight}
        \centering
        \includegraphics[width=\linewidth]{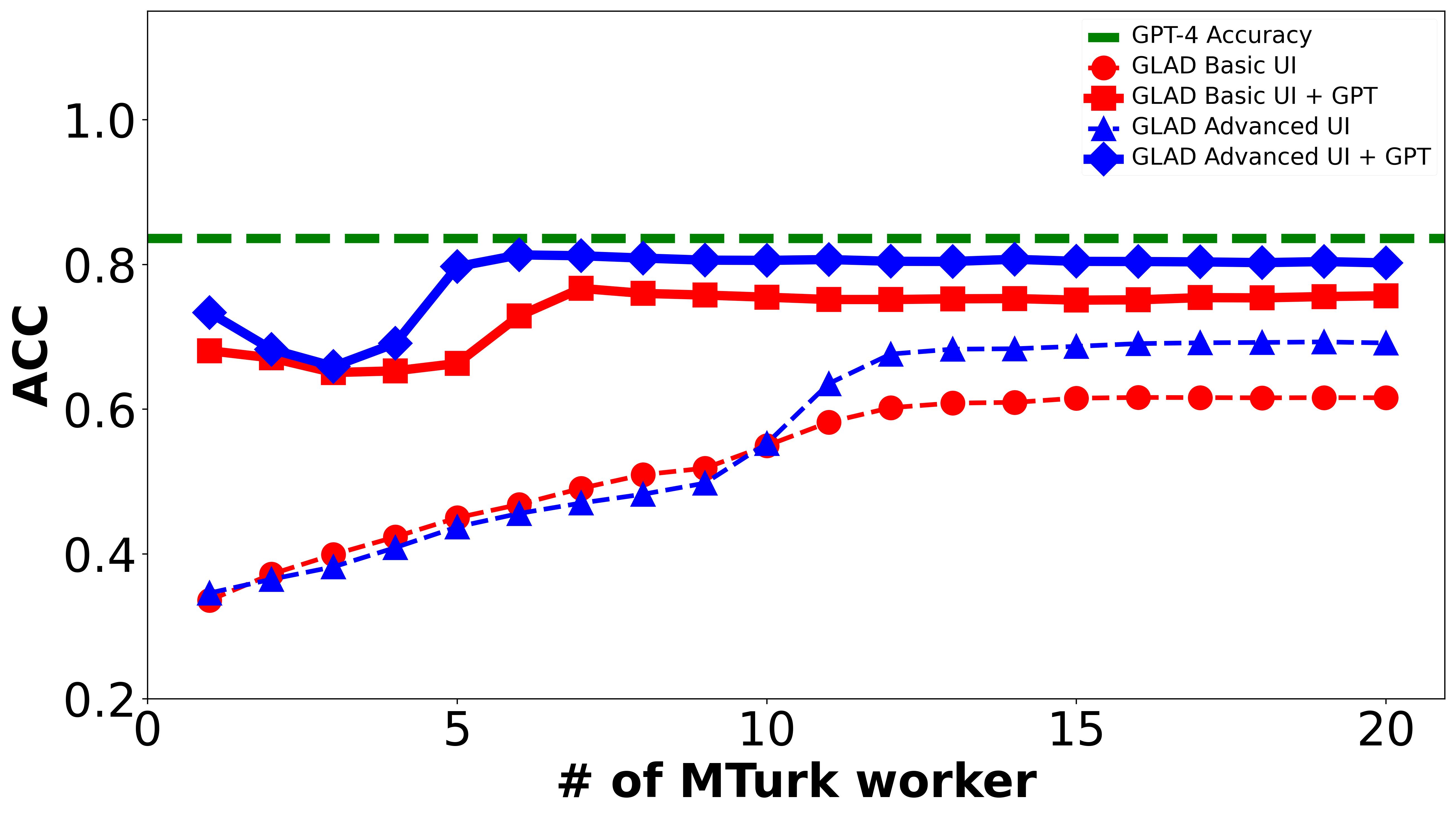}
        \caption{GLAD}
        \label{fig:agg-glad-no-bad-worker}
    \end{subfigure}
     \hfill
    \begin{subfigure}{0.36\textheight}
        \centering
        \includegraphics[width=\linewidth]{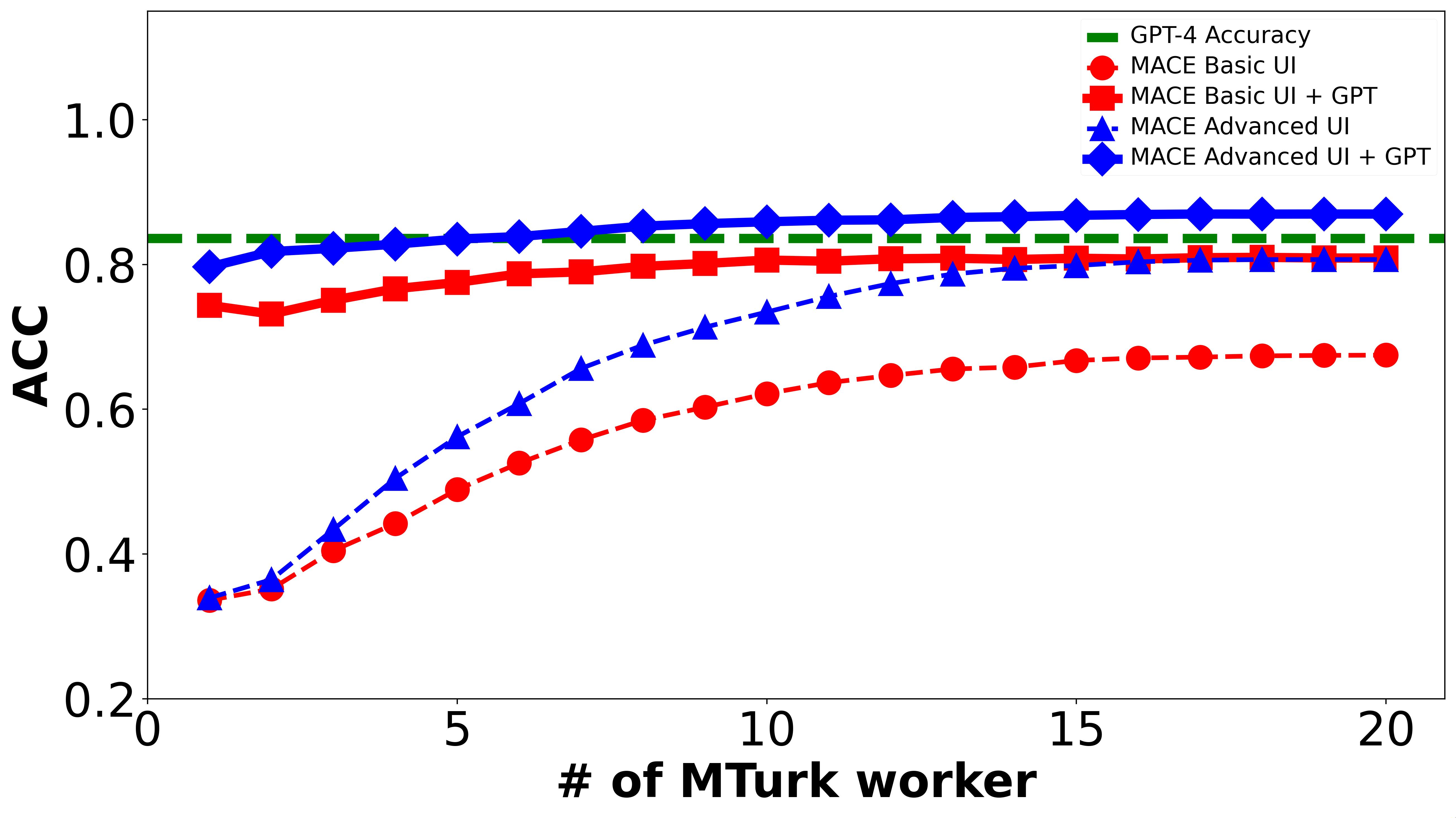}
        \caption{MACE}
        \label{fig:agg-mace-no-bad-worker}
    \end{subfigure}

    \begin{subfigure}{0.36\textheight}
        \centering
        \includegraphics[width=\linewidth]{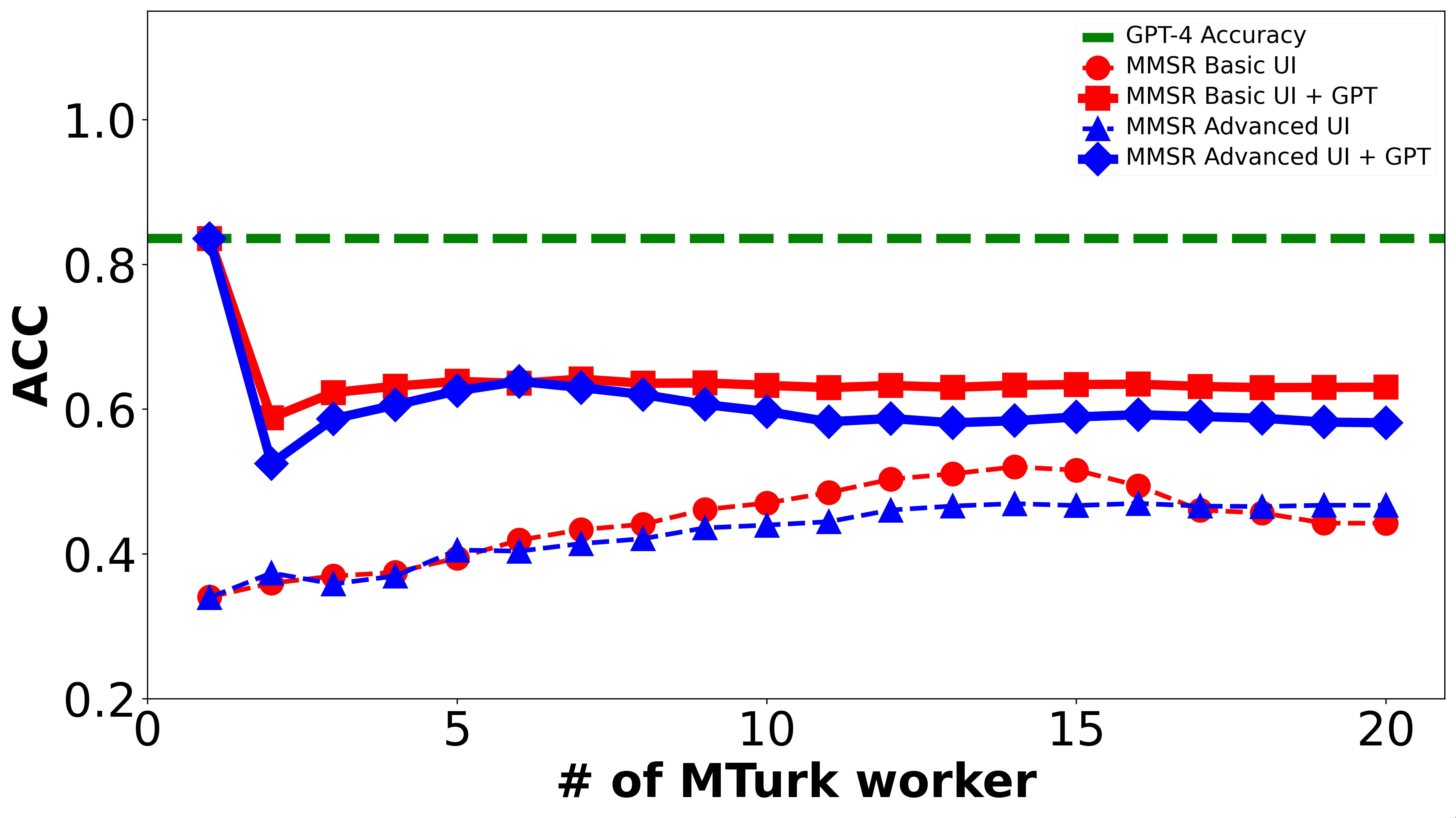}
        \caption{MMSR}
        \label{fig:agg-mmsr-no-bad-worker}
    \end{subfigure}
    \hfill
    \begin{subfigure}{0.36\textheight}
        \centering
        \includegraphics[width=\linewidth]{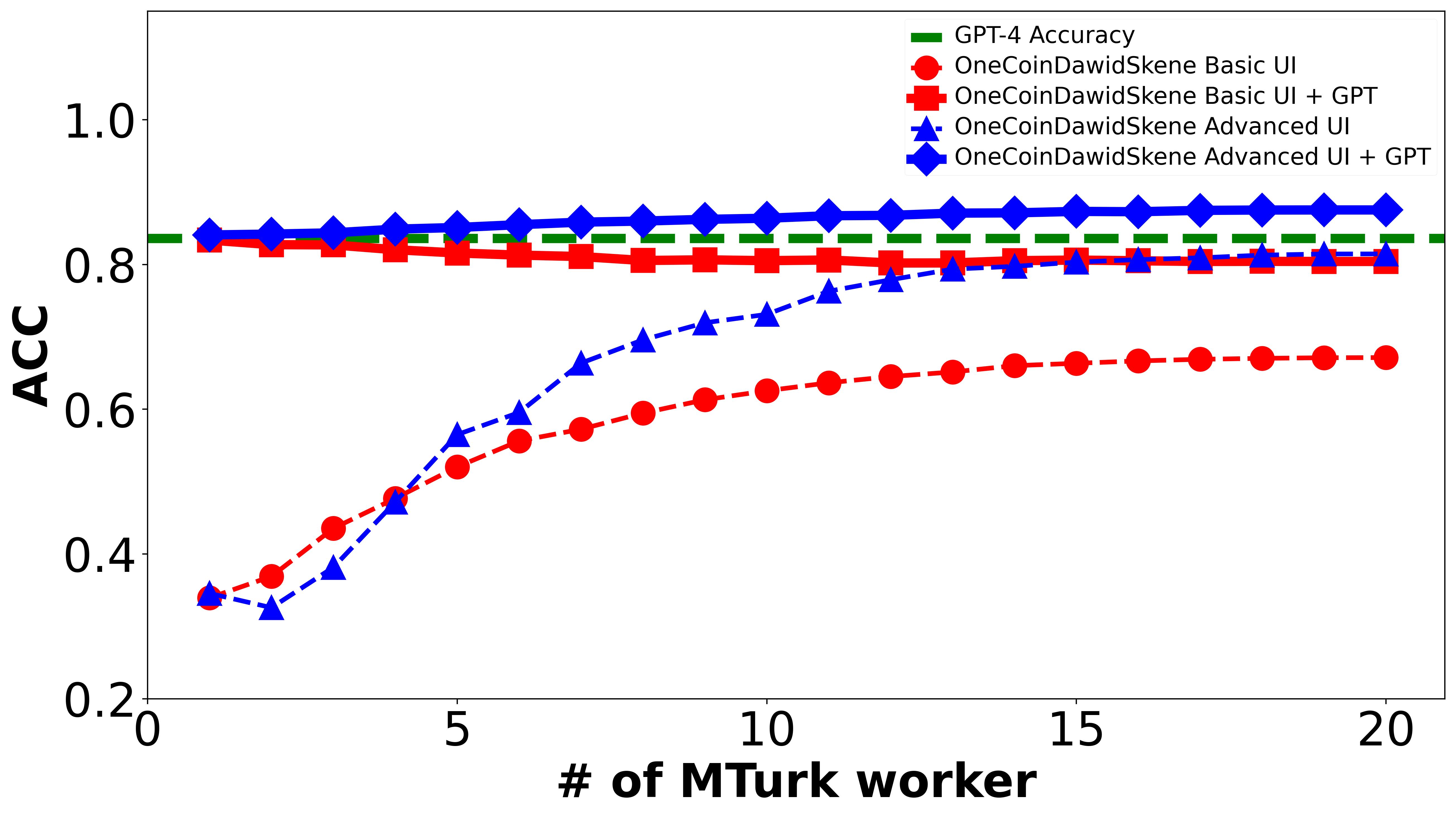}
        \caption{One-Coin Dawid-Skene}
        \label{fig:agg-onecoin-no-bad-worker}
    \end{subfigure}
    
    \begin{subfigure}{0.36\textheight}
        \centering
        \includegraphics[width=\linewidth]{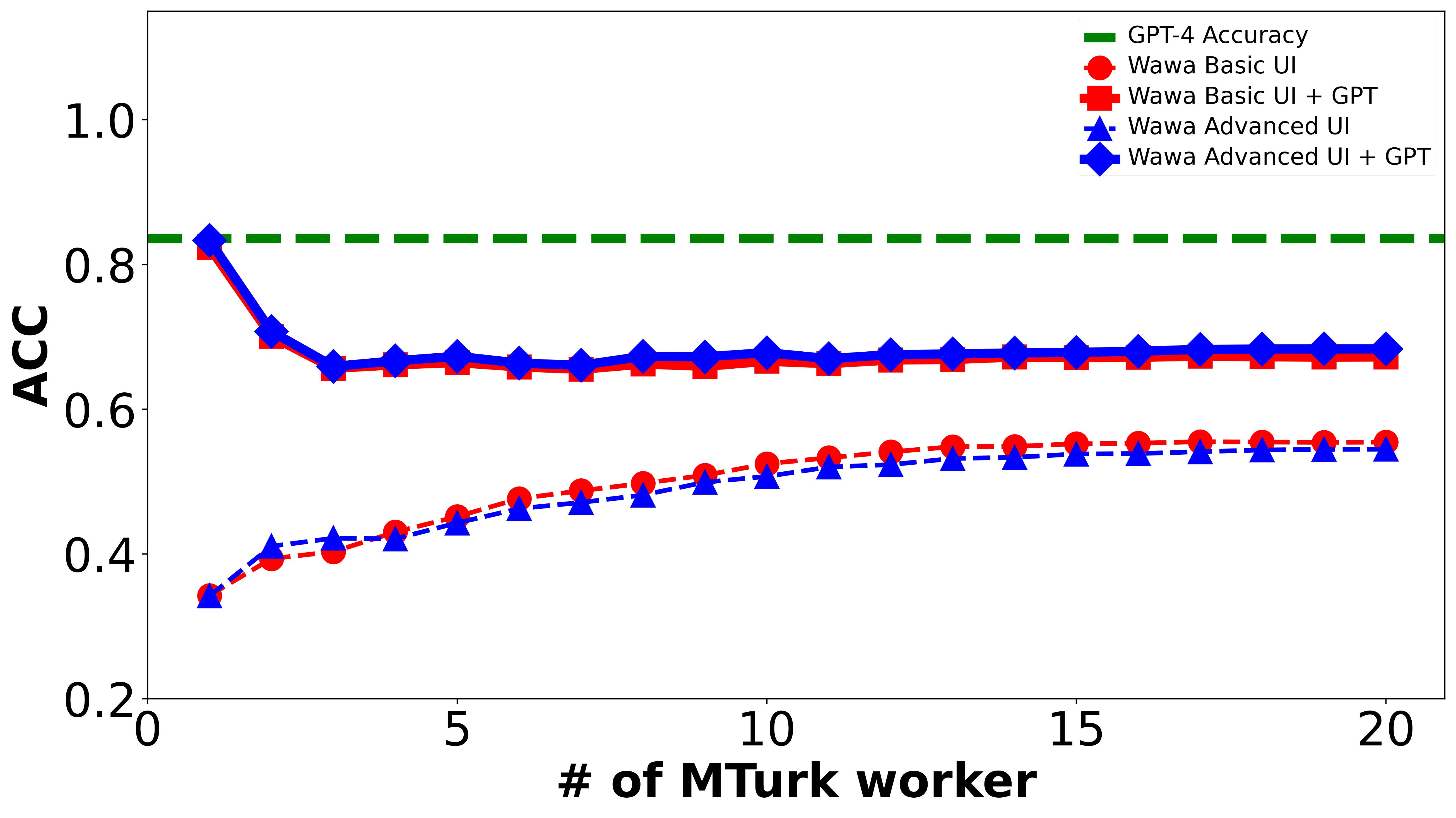}
        \caption{WAWA}
        \label{fig:agg-wawa-no-bad-worker}
    \end{subfigure}
     \hfill
    \begin{subfigure}{0.36\textheight}
        \centering
        \includegraphics[width=\linewidth]{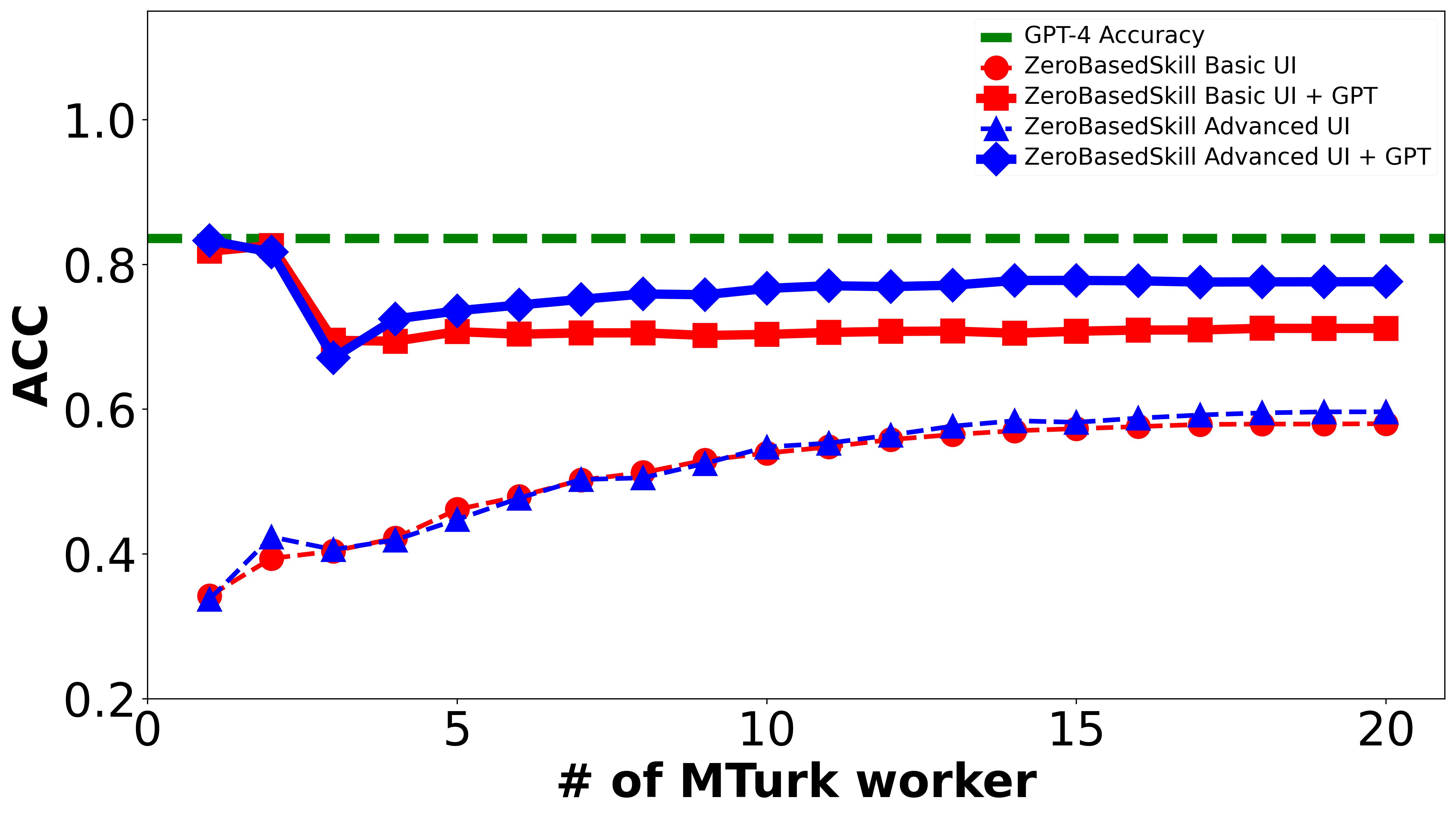}
        \caption{ZeroBasedSkill}
        \label{fig:agg-zero-no-bad-worker}
    \end{subfigure}
\caption{Exclude-By-Worker simulation results applied to different aggregation models. One-Coin Dawid-Skene and MACE algorithms for a combination of advanced interface results had the best accuracy and outperformed GPT-4 at temperature 0.2 (see (d) (f)). }
\Description{There are eight subfigures for all Exclude-By-Worker aggregation methods simulation. They are Majority Vote, Dawid-Skene, GLAD, MACE, MMSR, One-Coin Dawid-Skene, Wawa, and ZeroBased Skill methods.}
\label{figure:no-bad-worker-single-methods-gpt}
\end{figure*}

Driven by a collaborative perspective on crowdwork, this paper emphasizes the importance of aggregating results from all labels for the final output.
The previous section compared GPT-4 against MTurk pipelines as separate entities, and despite our efforts, nothing surpassed GPT-4's performance. 
It is intriguing to consider the potential impact of integrating \textbf{GPT-4 as a worker} into the label aggregation process.
This subsection presents the findings.

\subsubsection{Combining GPT-4 with crowd labels can potentially exceed GPT-4's solo performance.}
In our simulation study, we treated GPT-4 as an MTurk worker and selected it at t=0.2 due to its high accuracy.
The results are shown in Figure~\ref{figure:no-bad-worker-single-methods-gpt}.
The x-axis in each figure indicates the number of human workers in the aggregation.
A count of 5 workers, for example, refers to a mix of 5 MTurk workers and the GPT-4 model.

In aggregations that included GPT-4, the Advanced Interface results using One-Coin Dawid-Skene (Figure~\ref{figure:no-bad-worker-single-methods-gpt}f) and MACE (Figure~\ref{figure:no-bad-worker-single-methods-gpt}d) consistently \textbf{surpassed GPT-4's performance (83.6\%), peaking at an accuracy of 87.5\%}. 
All these two settings confirm the statistical significance of the improvement, where results are detailed in Table \ref{table:advanced-crowd-acc-result}.
Notably, this was \textbf{the only two settings in our experiments that bested GPT-4}, demonstrating the potential and the challenge to enhance accuracy by incorporating crowd labels.
This finding
suggests that even a handful of crowd labels can be beneficial.


We also conducted a \textbf{t-test analysis} to compare GPT-4 (t=0.2) across various settings, including different aggregation methods, cleaning strategies, and user interfaces.
The analysis included two-tailed paired t-tests at both the sentence and article levels.
For the sentence-level, each sentence's correctness was treated as a sample (N=3,177),
while for the article-level, we considered the average score within each article (N=200) as a sample.
Despite the sentence-level t-test aligning more with our other analyses,
its large sample size (N=3,177) resulted in very small p-values (p-value $\leq$ 0.001), limiting meaningful interpretation.
The only exception was observed in the One-Coin Dawid-Skene method combined with the Exclude-By-Worker strategy and Advanced Interface, yielding a p-value of 0.007.
Conversely, the article-level t-test produced marginally more interpretable results with slightly higher p-values.
Consequently, we have included the article-level t-test p-values in Tables~\ref{table:basic-acc-result} to~\ref{table:advanced-crowd-acc-result}, noting that most p-values remained quite small (p-value $\leq$ 0.001).
These t-test results suggest that the integration of GPT-4 with the Advanced Interface, particularly with the One-Coin Dawid-Skene and MACE methods, significantly outperformed the pure GPT-4 setting (as detailed in Table~\ref{table:advanced-crowd-acc-result}).

To further aid in interpreting these results, we calculated \textbf{95\% confidence intervals} for sentence-level accuracy (N=3,177). Tables \ref{table:basic-acc-result}, \ref{table:advanced-acc-result}, \ref{table:basic-crowd-gpt-acc-result}, and \ref{table:advanced-crowd-acc-result} present these accuracies alongside their respective 95\% confidence intervals, highlighting the precision and reliability of our findings.

However, in contrast to MTurk-only results (Section~\ref{sec:mturk-plus-gpt}), merging GPT-4's outputs showed varied trends across aggregation models as shown in Figure~\ref{figure:no-bad-worker-single-methods-gpt}.
Interestingly, as more MTurk workers were added, accuracy generally declined.

\begin{table*}[]
\centering
\begin{tabular}{lccccccccc}
\multicolumn{10}{c}{\textbf{Basic Interface}}\\
\toprule
\tiny Acc of GPT-4 (t=0.2) =\textbf{.836} & \multicolumn{3}{c}{\textbf{All Workers}} & \multicolumn{3}{c}{\textbf{Exclude-By-Worker}} & \multicolumn{3}{c@{}}{\textbf{Exclude-By-Batch}} \\ \cmidrule(lr){2-4} \cmidrule(lr){5-7} \cmidrule(lr){8-10}
\textbf{Method} & \textbf{Acc} & \textbf{P-Value} & \textbf{95\% CI} & \textbf{Acc} & \textbf{P-Value} & \textbf{95\% CI} & \textbf{Acc} & \textbf{P-Value} & \textbf{95\% CI} \\ \midrule
\textbf{MV} & .551 & <.001 & {[}.534, .569{]} & .609 & <.001 & {[}.592, .626{]} & .567 & <.001 & {[}.550, .584{]} \\
\textbf{DawidSkene} & .675 & <.001 & {[}.659, .691{]} & .737 & <.001 & {[}.722, .753{]} & .712 & <.001 & {[}.696, .727{]} \\
\textbf{OneCoin} & .797 & .002 & {[}.783, .811{]}  & .804 & .002 & {[}.790, .818{]} & .800 & .006 & {[}.786, .814{]}\\
\textbf{GLAD} & .734 & <.001 & {[}.719, .749{]}  & .756 & <.001 & {[}.741, .771{]} & .731 & <.001 & {[}.715, .746{]}\\
\textbf{M-MSR} & .531 & <.001 & {[}.513, .548{]}  & .630 & <.001 & {[}.614, .647{]} & .590 & <.001 & {[}.572, .607{]}\\
\textbf{MACE} & \textbf{\underline{.811}} & .001 & \textbf{\underline{{[}.797, .824{]}}}  & \textbf{.809} & .001 & \textbf{{[}.795, .823{]}} & \textbf{.809} & <.001 & \textbf{{[}.796, .823{]}}\\
\textbf{Wawa} & .633 & <.001 & {[}.617, .650{]} & .672 & <.001 & {[}.656, .689{]} & .636 & <.001 & {[}.619, .653{]} \\
\textbf{ZBS} & .677 & <.001 & {[}.661, .694{]} & .712 & <.001 & {[}.696, .727{]} & .675 & <.001 & {[}.659, .691{]} \\ \midrule
\textbf{Avg. Acc} & .676 & - & - & .716 & - & - & .690 & - & - \\ \midrule
\textbf{\#workers} & \multicolumn{3}{c}{216} & \multicolumn{3}{c}{134} & \multicolumn{3}{c@{}}{176}
\\
\bottomrule
\end{tabular}
\caption{Aggregation Accuracy Results of the Basic Interface integrated with GPT4 Group. 
\textbf{Bold} and \underline{underline} highlight the highest score within the column and across the table, respectively. P-value is obtained by comparing with GPT-4 over the article-level accuracy. 
(\textsuperscript{**}: p<0.01; \textsuperscript{***}: p<0.001. Paired t-test. N=200)
} 
\label{table:basic-crowd-gpt-acc-result}
\end{table*}

\begin{table*}[t]
\centering
\begin{tabular}{lccccccccc}
\multicolumn{10}{c}{\textbf{Advanced Interface}}\\
\toprule
\tiny Acc of GPT-4 (t=0.2) =\textbf{.836} & \multicolumn{3}{c}{\textbf{All Workers}} & \multicolumn{3}{c}{\textbf{Exclude-By-Worker}} & \multicolumn{3}{c}{\textbf{Exclude-By-Batch}} \\ \cmidrule(lr){2-4} \cmidrule(lr){5-7} \cmidrule(lr){8-10}
\textbf{Method} & \textbf{Acc} & \textbf{P-Value} & \textbf{95\% CI} & \textbf{Acc} & \textbf{P-Value} & \textbf{95\% CI} & \textbf{Acc} & \textbf{P-Value} & \textbf{95\% CI} \\ \midrule
\textbf{MV} & .517 & <.001 & {[}.500, .535{]} & .583 & <.001 & {[}.565, .600{]} & .523 & <.001 & {[}.505, .540{]} \\
\textbf{DawidSkene} & .678 & <.001 & {[}.662, .695{]} & .782 & <.001 & {[}.767, .796{]} & .718 & <.001 & {[}.702, .734{]} \\
\textbf{OneCoin} & \textbf{.873} & .002 & {[}.861, .884{]} & \textbf{\underline{.875}} &.001 & {[}.864, .887{]} & \textbf{.874} & <.001 & {[}.863, .886{]} \\
\textbf{GLAD} & .781 & <.001 & {[}.763, .792{]} & .802 & <.001 & {[}.788, .816{]} & .767 & <.001 & {[}.753, .782{]} \\
\textbf{M-MSR} & .512 & <.001 & {[}.494, .529{]} & .581 & <.001 & {[}.564, .599{]} & .522 & <.001 & {[}.504, .539{]} \\
\textbf{MACE} & .869 & .004 & {[}.857, .880{]} & .870 & .010 & {[}.858, .881{]} & .869 & .003 & {[}.858, .881{]} \\
\textbf{Wawa} & .586 & <.001 & {[}.569, .604{]} & .683 & <.001 & {[}.667, .700{]} & .597 & <.001 & {[}.580, .614{]} \\
\textbf{ZBS} & .715 & <.001 & {[}.700, .731{]} & .776 & <.001 & {[}.762, .791{]} & .711 & <.001 & {[}.695, .727{]} \\ \midrule
\textbf{Avg. Acc} & .691 & - & - & .744 & - & - & .698 & - & - \\ \midrule
\textbf{\#workers} & \multicolumn{3}{c}{199} & \multicolumn{3}{c}{129} & \multicolumn{3}{c}{162}
\\ \bottomrule
\end{tabular}
\caption{Aggregation Accuracy Results of the Advanced Interface integrated with GPT4 Group. 
\textbf{Bold} and \underline{underline} highlight the highest score within the column and across the table, respectively. OneCoin and MACE are only two aggregation methods that outperform GPT-4 and the differences are statistically significant, shown in the table. P-value is obtained by comparing with GPT-4 over the article-level accuracy.
(\textsuperscript{**}: p<0.01; \textsuperscript{***}: p<0.001. Paired t-test. N=200)
}
\label{table:advanced-crowd-acc-result}
\end{table*}


\subsection{Analyzing Cases Where Crowd Labels Enhance GPT's Label Accuracy\label{sec:mturk-gpt-analysis}}
In Section~\ref{sec:mturk-plus-gpt}'s study that combined GPT-4 and crowd labels collected via the Advanced Interface, of the 16 distinct configurations tested (spanning 2 variations in interface and 8 aggregation methods), only 2 showed enhanced accuracy beyond what was achieved by GPT-4 alone.
This subsection analyzes the potential factors contributing to these improvements.

\subsubsection{Improvement occurs as the crowd and GPT-4's labeling capabilities complement each other.}
\begin{figure*}[ht]
    \centering
    \includegraphics[width=0.97\textwidth]{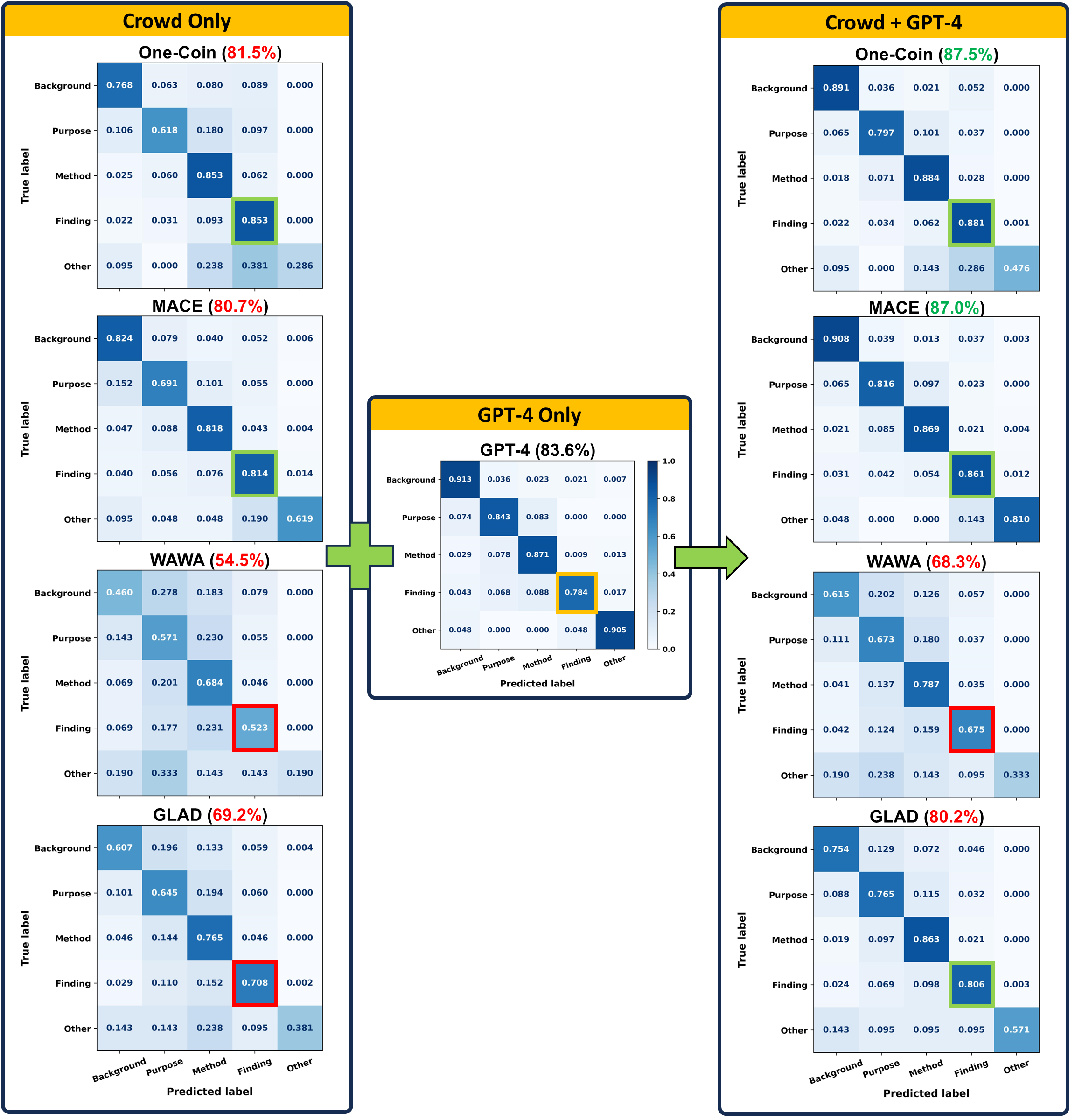}
    \caption{Comparison of the confusion matrices before and after combining with GPT-4's predictions for \textbf{Advanced} Interface when applying \textbf{Exclude-By-Worker} cleaning strategy. Integration of crowd labels with GPT-4 enhances overall performance only when the crowd-only conditions' finding accuracy surpasses that of GPT-4 alone. This is evident in the cases of both the One-Coin and MACE methods.}
    \label{fig:pretty-confusion-exclude-worker}
    \Description{There are 9 subfigures overall within three columns. The first column has 4 subfigures of normalized confusion matrix for Crowd Only results applied with One-Coin Dawid-Skene, MACE, Wawa, and GLAD aggregation methods. The second column contains 1 subfigure of the normalized confusion matrix for GPT-4 (t=0.2). The third column contains 4 subfigures of normalized confusion matrix for Crowd plus GPT results applied with One-Coin Dawid-Skene, MACE, Wawa, and GLAD aggregation methods. All figures are under Exclude-By-Worker cleaning strategy and using Advanced Interface result. The purpose of this whole figure is to show the comparison of confusion matrices before and after combining with GPT-4’s prediction for Advanced Interface when applying Exclude-By-Worker. Integration of crowd labels with GPT-4 enhances overall performance only when the crowd-only conditions' finding accuracy surpasses that of GPT-4 alone. This is evident in the cases of both the One-Coin and MACE methods.}
\end{figure*}

We first noticed that, in Table~\ref{table:no-bad-worker-aggregation}, 
OneCoin and MACE were the only two algorithms that exceeded or matched GPT-4's class-specific F1 scores in any individual class.
When aggregating crowd labels, most algorithms failed to surpass GPT-4 in F1 scores for any label class.
However, OneCoin (0.880) and MACE (0.872) achieved higher or similar F1 scores than GPT-4 (0.872) in the Finding/Contribution class (Table~\ref{table:no-bad-worker-aggregation}).
We further visualized OneCoin and MACE's confusion matrices, alongside WAWA and GLAD for comparison, in Figure~\ref{fig:pretty-confusion-exclude-worker}.
The confusion matrices show that while OneCoin and MACE did not outperform GPT-4 in all other classes, they achieved higher recalls than GPT-4, specifically in the Finding/Contribution class.
These observations suggest a hypothesis: 
when crowd labels exhibit specific strengths surpassing GPT-4's capabilities, 
effectively compensating for GPT-4's weaknesses (in our case, labeling the Finding/Contribution class),
their aggregation can lead to even higher accuracy.

\begin{figure*}[ht]
    \centering
    \begin{subfigure}[t]{0.49\textwidth}
        \includegraphics[width=\textwidth]{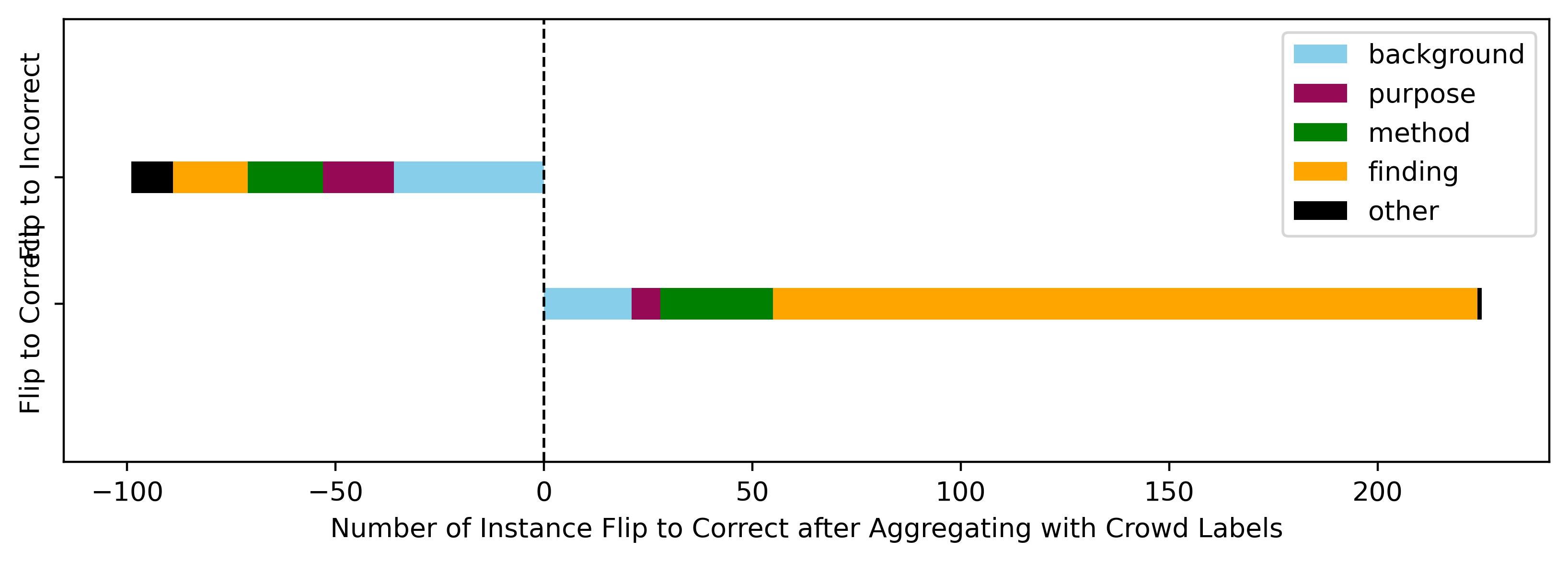}
        \caption{One-Coin Dawid-Skene on Crowd + GPT Exclude-By-Worker results in \textbf{flip to correct} and \textbf{flip to incorrect} different annotations from GPT result, under Advanced Interface condition.}
        \label{fig:improve-damaga-gpt-1}
    \end{subfigure}
    \hfill
    \begin{subfigure}[t]{0.49\textwidth}
        \includegraphics[width=\textwidth]{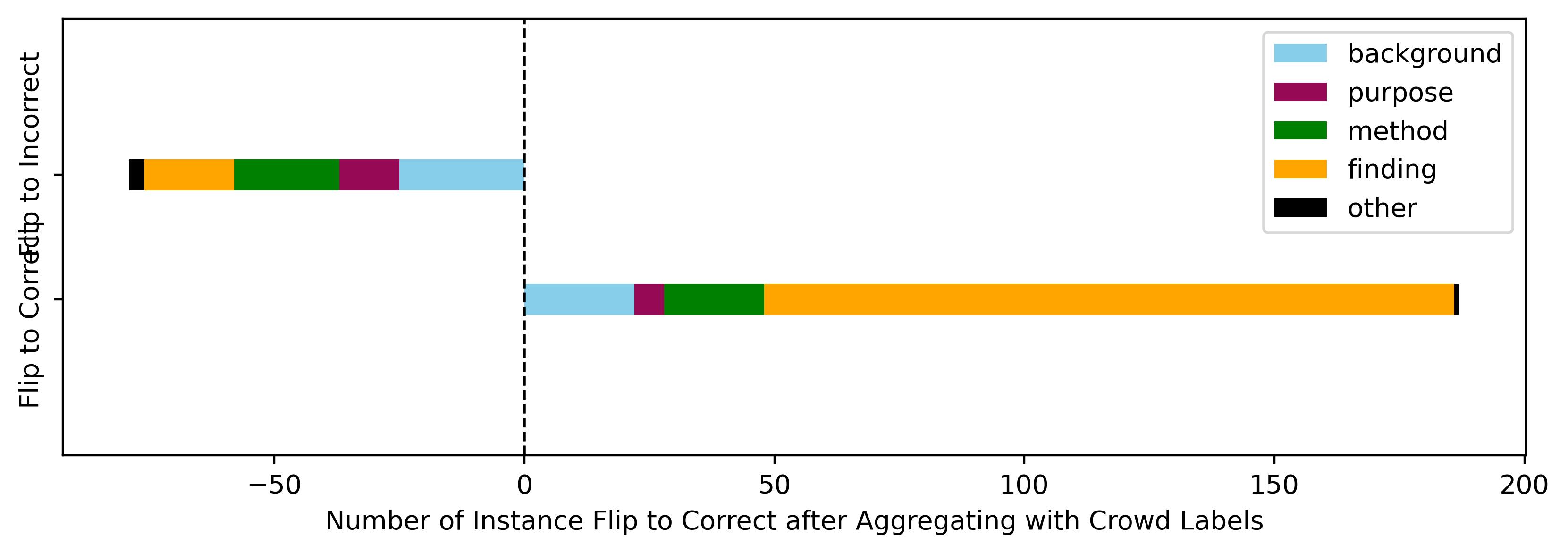}
        \caption{MACE on Crowd + GPT results \textbf{flip to correct} and \textbf{flip to incorrect} different annotations from Pure Crowd Workers result, under Advanced Interface Exclude-By-Worker condition.}
        \label{fig:improve-damaga-crowd-1}
    \end{subfigure}
    \caption{One-Coin Dawid-Skene and MACE Flip to Correct/Incorrect Stacked Bar Chart for Advanced Interface under Exclude-By-Worker, comparing to GPT-Only results, respectively. Figure~\ref{fig:improve-damaga-gpt-1} and~\ref{fig:improve-damaga-crowd-1} show that crowd workers benefit GPT results mainly in the ``Finding'' category (positive X-axis), while the decline in some of GPT's annotations across all labels at different levels (negative X-axis)}.
    \label{fig:pretty-exclude-worker-one-coin-improve-damage}
    \Description{It contains two subfigures of two stacked bar charts for One-Coin Dawid-Skene and MACE aggregation methods for Advanced Interface applying Exclude-By-Worker cleaning strategy. The stacked bar chart displayed horizontally where the x-axis represents the number of GPT results benefit or decline after combining with crowd results. There are two bars in each subfigures, the one on the top and has direction towards negative is the stacked bar displaying the number of declines in GPT’s annotations for five annotation labels “Finding”, “Purpose”, “Method”, “Finding”, and “Other”. For both subfigures, the decline after combining the crowd includes all labels without much difference. The bottom right stacked bar represents the number of improvements in GPT’s annotation for five labels. From both subfigures, GPT is benefited most on “Finding” labels than other labels.}
\end{figure*}

To test this hypothesis, we calculated the number of GPT's labels that were flipped to correct and incorrect, following aggregation with crowd labels. 
The results are shown in Figure~\ref{fig:pretty-exclude-worker-one-coin-improve-damage}.
It appears that the ``Finding'' class played the most important role in enhancing the accuracy of GPT labels when aggregating with crowd labels using MACE and OneCoin algorithms.
As Finding is the only class where crowd outperformed GPT, these results suggest that the aggregation algorithms leveraged the strengths of both the crowd workers and GPT-4 to achieve better overall accuracy.

\subsubsection{Where did the crowd and GPT-4 disagree with gold-standard labels?}
We manually coded a small set of error cases of crowd and GPT, identifying four primary sources for disagreeing with the gold-standard labels:

\begin{itemize}

    \item \textbf{[Ambiguous]:}
    First, \textbf{the text was unclear and could be categorized under either of two different labels.} 
For instance, 
this following sentence segment mentioned both ``Purpose'' and ``Method'':
\textit{``allow us to provide health strategies with more intelligent capabilities to accurately predict the outcomes of the disease in daily life and the hospital and prevent the progression of this disease and its many complications.''}
GPT-4 labeled it as ``Purpose'', while the Bio Expert marked it as ``Method''.

\item \textbf{[Keyword]:}
Second, \textbf{the sentence segment began with an annotation keyword}, yet the subsequent content does not align with the implied meaning of that keyword.
For example, this segment contained a keyword ``Conclusion'', which ideally should be categorized as ``Finding'':
\textit{``CONCLUSION: Regarding all initial variables,''}
However, GPT-4 classified it as ``Other''.

\item \textbf{[Short]:}
Third, \textbf{the text was just a short fragment of a longer sentence}, containing minimal information.
For instance, 
\textit{``Most participants were female (68\%),''} and 
\textit{``4.5 M ) in Canada,''}.

\item \textbf{[Context]:}
Finally, \textbf{some sentence segments can only be annotated correctly within the context} because it could be classified as a different label when it appears alone.
For example,
the gold-standard labels of the following two segments were both ``Finding'':
\textit{``and examine associations longitudinally to better understand causality.''}, 
\textit{``and financial strain may be mutually reinforcing and compound the health consequence of smoking.''}
When appearing alone, the former could be labeled as ``Purpose,'' and the latter could be labeled as ``Background''.

\end{itemize}


To understand how frequently each of the four main reasons for disagreement between crowd workers and the gold standard occured, 
we selected 200 cases where the top-performing model (Advanced Interface using One-Coin Dawid-Skene, excluding GPT-4 labels) and the gold standard did not match. 
The first author then classified these disagreements into five categories: 
\textbf{Ambiguous}, \textbf{Keyword}, \textbf{Short}, \textbf{Context}, and \textbf{Other}, based on the main reasons for disagreement described earlier.
The ``Other'' category included disagreements that either did not align with the first four categories or had unclear causes.
Upon reviewing 200 cases of disagreement, our analysis showed the following distribution: 
65 (32.5\%) cases were due to Ambiguous, 
9 (4.5\%) to Keyword, 
17 (8.5\%) to Short, 
27 (13.5\%) to Context, and 
82 (41\%) fell into the ``Other'' category.
Notably, ambiguity was the predominant reason for disagreement, presenting classification challenges even to our experts. 
We also found that around 40\% of the disagreements were categorized as ``Other,'' highlighting the need for more research to better align crowd workers with expert judgments.

In an additional parallel analysis, we examined 200 randomly selected instances of disagreement between GPT-4 annotations and the gold standard labels. 
The results were as follows: 82 (41\%) ``Ambiguous'', 6 (3\%) as ``Keyword'', 29 (14.5\%) as ``Short'', 53 (26.5\%) as ``Context'', and 30 (15\%) as ``Other''.
Compared to the disagreements observed between crowd workers and the gold standard, discrepancies involving GPT-4 featured significantly more instances of ``Context''.
This could suggest that GPT-4 struggled more with sentences requiring contextual understanding.


\section{Discussion}

\subsection{Why did some aggregation algorithms outperform others?}
In our study, the One-Coin Dawid-Skene and MACE methods were the top two algorithms throughout most cases with the best accuracies.
\citet{10.14778/3055540.3055547} surveyed 17 label aggregation algorithms and concluded that Dawid-Skene algorithm generally provides reliable results.
The OneCoin Dawid-Skene is a variant of the original Dawid-Skene algorithm. 
Its simplicity allows it to perform more effectively with smaller datasets.
Given the modest size of the dataset used in our study, this attribute likely contributed to its superior performance.

Another insight we could offer is that many label aggregation algorithms approach the problem of inferring final labels by dividing it into two aspects: {\em (i)} assessing the capabilities of the crowd workers and {\em (ii)} estimating the difficulty of the tasks. 
Based on our experience, we observed that in crowdsourcing data labeling tasks, particularly within the same data batch, the difficulty level of tasks tends to be relatively uniform. 
Therefore, the crucial factor for accuracy seems to be the effective estimation of the workers' performance.

\subsection{Generalizability}
Our study focused solely on the CODA-19 data labeling task, and thus, we cannot guarantee that our specific findings, such as the settings that could outperform GPT-4, will apply to other tasks. 
However, we believe the methodology we have used, as well as high-level takeaways, are broadly applicable to a range of labeling tasks.
Our confidence is based on two key observations. 
Firstly, the distribution of worker performance in our ``MTurk vs. GPT'' study (Figure~\ref{fig:crow-histogram}) aligns with previous research at the individual worker level~\cite{tornberg2023chatgpt}.
This could suggest that, at the pipeline level, MTurk annotation pipelines would behave similarly to the tasks from these prior studies, like tweet labeling. 
Secondly, the principle of combining two labeling sources with complementary strengths to enhance annotation quality is well-established in existing research~\cite{steyvers2022bayesian,bansal2021does}.
Based on these points, we are optimistic that our higher-level findings-- particularly that GPT-4 generally surpasses traditional crowdsourcing pipelines in accuracy, and that combining crowd and GPT inputs can further improve results-- could be applicable to other data labeling tasks.

\subsection{Crowdsourced Data Annotation Practices in the Era of LLMs}

Our study demonstrates that even a meticulously crafted MTurk pipeline may not outperform the zero-shot GPT-4 in labeling accuracy. 
(Note that we did not even test GPT-4 in few-shot settings.)
Despite our extensive experience in crowdsourcing, with several years using MTurk, this paper's crowdsourced data annotation effort was still demanding.
We spent weeks developing, testing, and implementing Basic and Advanced interfaces.
After finalizing these interfaces, another two weeks were dedicated to posting tasks and gathering data from MTurk workers.
This was mixed with significant effort to review and filter their submissions. 
However, even with this level of commitment, we could not match GPT-4's performance.
In contrast, the efficiency of GPT-4 was outstanding.
The design, testing, and execution of annotation tasks took two days. 
From a cost perspective, GPT-4 is also more affordable than hiring MTurk workers.
The main experiment with GPT-4 totaled only \$122.08, compared to the \$4,508 spent on MTurk (\$3,388 for annotation tasks and \$1,120 for qualification tasks).
This brings us to a pivotal question: In light of the fact that LLMs can now, in some instances, outperform human annotators, \textbf{how will the practices of data annotation evolve?}
While we cannot definitively answer it, we want to give a few thoughts based on our study:

\begin{itemize}
    \item 
Firstly, \textbf{the value of expert-level, high-quality labels will likely rise significantly.}
In our study, gold labels played a central role in several critical decisions:
refining prompts for greater efficacy,
choosing the most effective label-cleaning strategy,
and selecting the best label-aggregation algorithms.
These decisions led us to the few parameter combinations (Advanced Interface + OneCoin/MACE + incorporating GPT-4) that eventually surpassed GPT-4's performance.
Given that GPT-4 has matched crowd workers in many scenarios but has not yet consistently outdone domain experts, we anticipate a shift in data annotation towards amassing smaller, exceptionally high-quality datasets.

\item
Second, the research focus might \textbf{shift from ``using AI to support human labelers'' to ``using humans to enhance AI labeling.''}
Our study showed that by carefully adding a few crowd labels, GPT-4's accuracy can be improved.
Given the cost and difficulty of finding expert labelers, using non-expert labels to enhance LLM's performance will likely become more critical. 

\item
Finally, while it might appear as a nuanced point, we believe that \textbf{the Human-Computer Interaction (HCI) challenges in the human annotation process will become central again}.
In our study, initial observations suggest marginal differences between the Basic and Advanced interfaces: the variations in accuracy among workers from both groups are subtle, and aggregation methods do not offer an apparent advantage to those using the Advanced interface.
The majority voting even favored the Basic interface group (Table~\ref{table:no-bad-worker-aggregation}).
However, the more detailed analysis, {\em i.e.}, worker number analysis and the GPT-4 aggregation study, show the strengths of the advanced interface.
Workers using it provided more consistent labels, making it the only interface to surpass GPT-4. 
Given LLMs' high labeling accuracy, we will likely need even more reliable human labels in the future to boost their performance further.
Developing systems that allow users, especially non-expert annotators, to perform reliably and consistently is essentially an HCI problem.

\end{itemize}
\subsection{Limitations}

Our study has several limitations worth noting.
First, we focused exclusively on one annotation task, CODA-19. 
Consequently, our findings may not apply to other tasks, especially where crowd workers might surpass LLMs.
Second, our research employed only GPT-4, meaning it may not reflect the behaviors of other LLMs like GPT-3.5. 
Additionally, GPT-4 operates as a closed model, making generalizations to open models like LLama potentially inapplicable.
Third, although it appears unlikely, we cannot dismiss the possibility that crowd workers might use LLMs while completing tasks, as noted by~\citet{veselovsky2023prevalence}.
We designed our interface to enable quick completion of each sentence segment, usually within seconds.
This design discouraged workers from copying text to use ChatGPT for answers, which would take significantly longer.
%
However, admittedly, our interfaces lack specific mechanisms to prevent the use of LLMs by workers.
Finally, we invested significant resources in worker recruitment and segmentation through Qualification tasks, which was essential for our study's objectives. 
However, this approach diverges from many MTurk tasks with broader recruitment strategies.
While we stand by our conclusions, given our study's expansive scale, crowd performance might be marginally lower in more open recruitment settings due to the inclusion of newer workers and the lack of a cohesive group dynamic.


\section{Conclusion and Future Work}

This paper evaluates GPT-4's labeling capabilities in contrast to a well-executed, ethical crowdsourcing pipeline for annotating unseen data.
Utilizing the CODA-19 labeling scheme, we exhaustively tested various label-cleaning strategies, label-aggregation techniques, and interface designs on MTurk. 
Despite adhering to best crowdsourcing practices, the best-performing MTurk pipeline achieved an accuracy of 81.5\%, slightly below GPT-4's 83.6\%.
Interestingly, by optimizing the combination of label aggregation techniques and interfaces, integrating GPT-4 labels with the MTurk aggregation process boosted accuracy to 87.5\%.

Moving forward, our research will focus on generating a smaller set of high-quality labels via MTurk, aiming to further enhance the labeling performance of already sophisticated LLMs like GPT-4.
Additionally, we will delve deeper into the influence of worker interface design on label quality to further improve LLM performance.

\begin{acks}
We extend our gratitude to Dmitry Ustalov and Nikita Pavlichenko for their valuable feedback and insights on the Crowd-Kit implementations. 
Our thanks also go to Xingwei He and Jian Jiao for clarifying details about the QK dataset mentioned in their research. 
We are grateful to Man Luo for detailing the data-labeling process in her publications.
Additionally, we appreciate the contributions of the crowd workers on Amazon Mechanical Turk who participated in our studies.

\end{acks}

\bibliographystyle{Style/ACM-Reference-Format}
\bibliography{Bibtex/sample-base,Bibtex/software,Bibtex/main}

\appendix

\section{All Workers and Exclude-By-Batch Result Table}
In this section, we show all the additional aggregation results for different distinct configurations tested (2 different interfaces, and 3 different cleaning strategies). Table \ref{table:all-worker-aggregation} to \ref{table:no-bad-label-gpt-aggregation} is the aggregation accuracy results on both crowd-only and crowd+gpt results for both Basic and Advanced Interfaces under different cleaning strategies. They show detailed P, R, F1, Accuracy, and Kappa for each aggregation model under different user interfaces.
\begin{table*}[t]
\centering
\setlength{\tabcolsep}{2.8pt}
\begin{tabular}{llrrrrrrrrrrrrrrrrrr}
\toprule
\multicolumn{1}{l}{\textbf{Eval}} & \multicolumn{3}{c}{\textbf{Background}} & \multicolumn{3}{c}{\textbf{Purpose}} & \multicolumn{3}{c}{\textbf{Method}} & \multicolumn{3}{c}{\textbf{Finding}} & \multicolumn{3}{c}{\textbf{Other}} & \multicolumn{1}{c}{\multirow{2}{*}{\textbf{Acc}}} & \multicolumn{1}{c}{\multirow{2}{*}{\textbf{Kappa}}} \\ \cmidrule(lr){2-4} \cmidrule(lr){5-7} \cmidrule(lr){8-10} \cmidrule(lr){11-13} \cmidrule(lr){14-16}
\textbf{Label} & \multicolumn{1}{c}{P} & \multicolumn{1}{c}{R} & \multicolumn{1}{c}{F1} & \multicolumn{1}{c}{P} & \multicolumn{1}{c}{R} & \multicolumn{1}{c}{F1} & \multicolumn{1}{c}{P} & \multicolumn{1}{c}{R} & \multicolumn{1}{c}{F1} & \multicolumn{1}{c}{P} & \multicolumn{1}{c}{R} & \multicolumn{1}{c}{F1} & \multicolumn{1}{c}{P} & \multicolumn{1}{c}{R} & \multicolumn{1}{c}{F1} & \multicolumn{1}{c}{} & \multicolumn{1}{c}{} \\ \midrule
\multicolumn{18}{c}{\textbf{Basic UI}}\\
MV & .713 & .281 & .403 & .149 & .525 & .232 & .368 & .599 & .456 & .772 & .507 & .612 & 1.000 & .286 & .444 & .477 & .285 \\
DawidSkene & .676 & .503 & .577 & .181 & .502 & .266 & .504 & .588 & .543 & .868 & .561 & .682 & .036 & .429 & .066 & .549 & .392 \\
OneCoin & .852 & .428 & .570 & .328 & .585 & .421 & .496 & .737 & .593 & .824 & .752 & .787 & 1.000 & .238 & .385 & .663 & .504 \\
GLAD & .767 & .448 & .566 & .224 & .645 & .333 & .482 & .675 & .563 & .866 & .648 & .741 & .360 & .429 & .391 & .608 & .451 \\
M-MSR & .519 & .337 & .408 & .140 & .498 & .218 & .347 & .526 & .418 & .768 & .435 & .555 & .143 & .238 & .179 & .436 & .244 \\
\textbf{MACE} & .733 & .586 & .651 & .288 & .631 & .396 & .581 & .688 & .630 & .888 & .716 & .793 & .165 & .619 & .260 & \textbf{.675} & .537 \\
Wawa & .718 & .380 & .497 & .173 & .576 & .266 & .408 & .647 & .501 & .839 & .536 & .654 & .750 & .286 & .414 & .527 & .353 \\
ZBS & .741 & .398 & .518 & .181 & .594 & .277 & .426 & .666 & .520 & .858 & .559 & .677 & .600 & .286 & .387 & .547 & .380 \\ \midrule
\multicolumn{18}{c}{\textbf{Advanced UI}}\\
MV & .598 & .307 & .405 & .112 & .438 & .179 & .373 & .634 & .469 & .815 & .425 & .559 & \multicolumn{1}{c}{-} & 0 & \multicolumn{1}{c}{-} & .442 & .259 \\ 
DawidSkene & .565 & .362 & .442 & .134 & .475 & .209 & .599 & .657 & .627 & .931 & .609 & .736 & .046 & .429 & .084 & .555 & .401 \\
OneCoin & .686 & .404 & .509 & .144 & .530 & .226 & .433 & .691 & .533 & .881 & .498 & .636 & \multicolumn{1}{c}{-} & 0 & \multicolumn{1}{c}{-} & .517 & .352 \\
GLAD & .745 & .553 & .635 & .192 & .594 & .290 & .554 & .719 & .626 & .910 & .637 & .750 & .600 & .286 & .387 & .631 & .488 \\
M-MSR & .534 & .368 & .436 & .113 & .470 & .183 & .380 & .587 & .461 & .809 & .384 & .521 & .286 & .095 & .143 & .428 & .249 \\
\textbf{MACE} & .824 & .807 & .815 & .412 & .700 & .519 & .757 & .819 & .787 & .937 & .803 & .865 & .196 & .476 & .278 & \textbf{.798} & .707 \\
Wawa & .586 & .377 & .459 & .128 & .516 & .205 & .411 & .641 & .501 & .858 & .435 & .577 & 1.000 & .095 & .174 & .470 & .299 \\
ZBS & .648 & .438 & .523 & .142 & .525 & .224 & .457 & .671 & .544 & .883 & .510 & .647 & 1.000 & .238 & .385 & .528 & .365 \\ 
\midrule
GPT-4 (t=.2) & .860 & .913 & .885 & .499 & .843 & .627 & .775 & .871 & .820 & .982 & .784 & .872 & .322 & .905 & .475 & .836 & .764 \\\bottomrule
\end{tabular}
\caption{All Workers Table. All models use Bio Expert as the gold standard. Baseline is the Majority Vote (MV).}
\label{table:all-worker-aggregation}
\end{table*}
\begin{table*}[ht]
\centering
\setlength{\tabcolsep}{2.8pt}
\begin{tabular}{lrrrrrrrrrrrrrrrrr}
\toprule
\multicolumn{1}{l}{\textbf{Eval}} & \multicolumn{3}{c}{\textbf{Background}} & \multicolumn{3}{c}{\textbf{Purpose}} & \multicolumn{3}{c}{\textbf{Method}} & \multicolumn{3}{c}{\textbf{Finding}} & \multicolumn{3}{c}{\textbf{Other}} & \multicolumn{1}{c}{\multirow{2}{*}{\textbf{Acc}}} & \multicolumn{1}{c}{\multirow{2}{*}{\textbf{Kappa}}} \\ \cmidrule(lr){2-4} \cmidrule(lr){5-7} \cmidrule(lr){8-10} \cmidrule(lr){11-13} \cmidrule(lr){14-16}
\textbf{Label} & \multicolumn{1}{c}{P} & \multicolumn{1}{c}{R} & \multicolumn{1}{c}{F1} & \multicolumn{1}{c}{P} & \multicolumn{1}{c}{R} & \multicolumn{1}{c}{F1} & \multicolumn{1}{c}{P} & \multicolumn{1}{c}{R} & \multicolumn{1}{c}{F1} & \multicolumn{1}{c}{P} & \multicolumn{1}{c}{R} & \multicolumn{1}{c}{F1} & \multicolumn{1}{c}{P} & \multicolumn{1}{c}{R} & \multicolumn{1}{c}{F1} & \multicolumn{1}{c}{} & \multicolumn{1}{c}{} \\ \midrule
\multicolumn{18}{c}{\textbf{Basic UI}}\\
MV & .688 & .297 & .414 & .147 & .535 & .231 & .382 & .619 & .472 & .804 & .504 & .620 & 1.000 & .286 & .444 & .484 & .299 \\
DawidSkene & .773 & .463 & .579 & .197 & .539 & .289 & .518 & .616 & .563 & .861 & .647 & .739 & .060 & .524 & .107 & .592 & .435 \\
OneCoin & .839 & .426 & .565 & .304 & .594 & .402 & .479 & .741 & .582 & .842 & .723 & .778 & 1.000 & .238 & .385 & .649 & .490 \\
GLAD & .767 & .447 & .565 & .211 & .636 & .317 & .476 & .691 & .564 & .871 & .620 & .724 & .529 & .429 & .474 & .597 & .440 \\
M-MSR & .539 & .340 & .417 & .131 & .475 & .205 & .314 & .522 & .392 & .756 & .392 & .516 & .583 & .333 & .424 & .414 & .220 \\
\textbf{MACE} & .744 & .580 & .652 & .287 & .627 & .394 & .575 & .694 & .629 & .885 & .717 & .793 & .178 & .619 & .277 & \textbf{.675} & .537 \\
Wawa & .694 & .397 & .505 & .166 & .571 & .257 & .411 & .643 & .501 & .855 & .525 & .651 & .750 & .286 & .414 & .524 & .353 \\
ZBS & .723 & .420 & .531 & .178 & .585 & .273 & .431 & .663 & .522 & .878 & .564 & .687 & .750 & .286 & .414 & .553 & .389 \\ \midrule
\multicolumn{18}{c}{\textbf{Advanced UI}}\\
MV & .616 & .311 & .413 & .123 & .465 & .194 & .376 & .635 & .473 & .824 & .451 & .583 & \multicolumn{1}{c}{-} & 0 & \multicolumn{1}{c}{-} & .458 & .275 \\ 
DawidSkene & .642 & .354 & .456 & .144 & .498 & .223 & .616 & .684 & .648 & .933 & .655 & .770 & .047 & .429 & .085 & .583 & .433 \\
OneCoin & .698 & .417 & .522 & .156 & .558 & .244 & .449 & .712 & .551 & .890 & .517 & .654 & \multicolumn{1}{c}{-} & 0 & \multicolumn{1}{c}{-} & .536 & .374 \\
GLAD & .729 & .572 & .641 & .202 & .618 & .305 & .567 & .726 & .637 & .918 & .639 & .753 & .667 & .286 & .400 & .639 & .499 \\
M-MSR & .481 & .394 & .433 & .128 & .493 & .203 & .383 & .565 & .456 & .822 & .400 & .538 & .143 & .048 & .071 & .438 & .258 \\
\textbf{MACE} & .825 & .792 & .808 & .404 & .696 & .511 & .739 & .807 & .772 & .931 & .793 & .856 & .167 & .476 & .247 & \textbf{.787} & .692 \\
Wawa & .608 & .383 & .470 & .135 & .530 & .215 & .416 & .650 & .507 & .866 & .455 & .596 & 1.000 & .143 & .250 & .484 & .314 \\
ZBS & .655 & .451 & .534 & .154 & .558 & .242 & .468 & .676 & .554 & .887 & .525 & .659 & 1.000 & .286 & .444 & .542 & .381 \\ 
\midrule
GPT-4 (t=.2) & .860 & .913 & .885 & .499 & .843 & .627 & .775 & .871 & .820 & .982 & .784 & .872 & .322 & .905 & .475 & .836 & .764 \\
\bottomrule
\end{tabular}
\caption{Exclude-By-Batch Table. All models use Bio Expert as the gold standard. Baseline is the Majority Vote (MV).}
\label{table:no-bad-label-aggregation}
\end{table*}

\begin{table*}[ht]
\centering
\setlength{\tabcolsep}{2.8pt}
\begin{tabular}{lrrrrrrrrrrrrrrrrr}
\toprule
\multicolumn{1}{l}{\textbf{Eval}} & \multicolumn{3}{c}{\textbf{Background}} & \multicolumn{3}{c}{\textbf{Purpose}} & \multicolumn{3}{c}{\textbf{Method}} & \multicolumn{3}{c}{\textbf{Finding}} & \multicolumn{3}{c}{\textbf{Other}} & \multicolumn{1}{c}{\multirow{2}{*}{\textbf{Acc}}} & \multicolumn{1}{c}{\multirow{2}{*}{\textbf{Kappa}}} \\ \cmidrule(lr){2-4} \cmidrule(lr){5-7} \cmidrule(lr){8-10} \cmidrule(lr){11-13} \cmidrule(lr){14-16}
\textbf{Label} & \multicolumn{1}{c}{P} & \multicolumn{1}{c}{R} & \multicolumn{1}{c}{F1} & \multicolumn{1}{c}{P} & \multicolumn{1}{c}{R} & \multicolumn{1}{c}{F1} & \multicolumn{1}{c}{P} & \multicolumn{1}{c}{R} & \multicolumn{1}{c}{F1} & \multicolumn{1}{c}{P} & \multicolumn{1}{c}{R} & \multicolumn{1}{c}{F1} & \multicolumn{1}{c}{P} & \multicolumn{1}{c}{R} & \multicolumn{1}{c}{F1} & \multicolumn{1}{c}{} & \multicolumn{1}{c}{} \\ \midrule
\multicolumn{18}{c}{\textbf{Basic UI}}\\
MV & .798 & .380 & .515 & .187 & .604 & .286 & .435 & .685 & .532 & .828 & .566 & .672 & 1.000 & .286 & .444 & .551 & .381 \\
DawidSkene & .830 & .595 & .693 & .250 & .659 & .363 & .693 & .762 & .726 & .945 & .678 & .789 & .046 & .524 & .085 & .675 & .554 \\
OneCoin & .892 & .695 & .781 & .510 & .696 & .589 & .673 & .831 & .744 & .888 & .848 & .867 & .875 & .333 & .483 & .797 & .696 \\
GLAD & .887 & .630 & .737 & .329 & .760 & .460 & .609 & .797 & .690 & .922 & .751 & .828 & .632 & .571 & .600 & .734 & .619 \\
M-MSR & .652 & .440 & .525 & .184 & .594 & .281 & .419 & .628 & .503 & .851 & .523 & .648 & .222 & .286 & .250 & .531 & .362 \\
\textbf{MACE} & .872 & .807 & .838 & .445 & .802 & .572 & .720 & .819 & .766 & .961 & .810 & .879 & .353 & .857 & .500 & \textbf{.811} & .726 \\
Wawa & .823 & .511 & .631 & .235 & .696 & .352 & .516 & .757 & .614 & .897 & .628 & .739 & .889 & .381 & .533 & .633 & .491 \\
ZBS & .845 & .554 & .670 & .278 & .737 & .404 & .548 & .779 & .643 & .916 & .683 & .782 & .750 & .429 & .545 & .677 & .546 \\ \midrule
\multicolumn{18}{c}{\textbf{Advanced UI}}\\
MV & .665 & .364 & .470 & .150 & .544 & .235 & .437 & .701 & .539 & .866 & .509 & .641 & 1.000 & .048 & .091 & .517 & .349 \\ 
DawidSkene & .819 & .473 & .599 & .206 & .668 & .315 & .794 & .781 & .787 & .968 & .728 & .831 & .053 & .571 & .097 & .678 & .559 \\
\textbf{OneCoin} & .911 & .877 & .893 & .571 & .779 & .659 & .805 & .894 & .847 & .951 & .880 & .914 & .909 & .476 & .625 & \textbf{.873} & .811 \\
GLAD & .867 & .712 & .782 & .344 & .724 & .466 & .695 & .860 & .769 & .953 & .790 & .864 & .818 & .429 & .563 & .781 & .684 \\
M-MSR & .633 & .433 & .514 & .146 & .535 & .229 & .440 & .663 & .529 & .865 & .483 & .620 & .500 & .143 & .222 & .512 & .345 \\
MACE & .889 & .910 & .899 & .538 & .816 & .648 & .835 & .871 & .852 & .967 & .858 & .909 & .415 & .810 & .548 & .869 & .807 \\
Wawa & .706 & .501 & .586 & .181 & .608 & .279 & .504 & .722 & .593 & .911 & .567 & .699 & 1.000 & .238 & .385 & .586 & .437 \\
ZBS & .830 & .650 & .729 & .266 & .705 & .386 & .633 & .813 & .712 & .942 & .708 & .808 & .889 & .381 & .533 & .715 & .599 \\ 
\midrule
GPT-4 (t=.2) & .860 & .913 & .885 & .499 & .843 & .627 & .775 & .871 & .820 & .982 & .784 & .872 & .322 & .905 & .475 & .836 & .764 \\\bottomrule
\end{tabular}
\caption{All Workers integrated with GPT-4 Table. All models use Bio Expert as the gold standard. Baseline is the Majority Vote (MV).}
\label{table:all-worker-gpt-aggregation}
\end{table*}
\begin{table*}[t]
\centering
\setlength{\tabcolsep}{2.8pt}
\begin{tabular}{llrrrrrrrrrrrrrrrrrr}
\toprule
\multicolumn{1}{l}{\textbf{Eval}} & \multicolumn{3}{c}{\textbf{Background}} & \multicolumn{3}{c}{\textbf{Purpose}} & \multicolumn{3}{c}{\textbf{Method}} & \multicolumn{3}{c}{\textbf{Finding}} & \multicolumn{3}{c}{\textbf{Other}} & \multicolumn{1}{c}{\multirow{2}{*}{\textbf{Acc}}} & \multicolumn{1}{c}{\multirow{2}{*}{\textbf{Kappa}}} \\ \cmidrule(lr){2-4} \cmidrule(lr){5-7} \cmidrule(lr){8-10} \cmidrule(lr){11-13} \cmidrule(lr){14-16}
\textbf{Label} & \multicolumn{1}{c}{P} & \multicolumn{1}{c}{R} & \multicolumn{1}{c}{F1} & \multicolumn{1}{c}{P} & \multicolumn{1}{c}{R} & \multicolumn{1}{c}{F1} & \multicolumn{1}{c}{P} & \multicolumn{1}{c}{R} & \multicolumn{1}{c}{F1} & \multicolumn{1}{c}{P} & \multicolumn{1}{c}{R} & \multicolumn{1}{c}{F1} & \multicolumn{1}{c}{P} & \multicolumn{1}{c}{R} & \multicolumn{1}{c}{F1} & \multicolumn{1}{c}{} & \multicolumn{1}{c}{} \\ \midrule 
\multicolumn{18}{c}{\textbf{Basic UI}} \\
MV  & .814 & .426 & .559 & .223 & .654 & .333 & .488 & .712 & .579 & .853 & .644 & .734 & 1.000 & .286 & .444 & .609 & .451 \\
DawidSkene  & .872 & .519 & .650 & .280 & .737 & .406 & .746 & .803 & .773 & .936 & .808 & .867 & .125 & .667 & .211 & .737 & .626 & \\
OneCoin  & .888 & .725 & .798 & .507 & .714 & .593 & .689 & .826 & .751 & .897 & .849 & .872 & .875 & .333 & .483 & .804 & .708 & \\
GLAD  & .876 & .669 & .759 & .354 & .760 & .483 & .644 & .806 & .716 & .928 & .775 & .845 & .609 & .667 & .636 & .757 & .649 & \\
M-MSR  & .756 & .536 & .627 & .225 & .627 & .331 & .511 & .697 & .590 & .886 & .647 & .748 & .900 & .429 & .581 & .630 & .482 & \\
\textbf{MACE} & .873 & .801 & .836 & .442 & .797 & .569 & .719 & .819 & .766 & .960 & .809 & .878 & .327 & .857 & .474 & \textbf{.809} & .724 & \\
Wawa  & .833 & .577 & .682 & .264 & .728 & .388 & .553 & .759 & .640 & .912 & .673 & .774 & .818 & .429 & .563 & .672 & .540 & \\
ZBS  & .849 & .613 & .712 & .305 & .747 & .433 & .592 & .788 & .676 & .921 & .721 & .809 & .667 & .476 & .556 & .712 & .590 & \\ \midrule
\multicolumn{18}{c}{\textbf{Advanced UI}}\\
MV  & .740 & .417 & .533 & .189 & .590 & .286 & .475 & .738 & .578 & .885 & .595 & .712 & 1.000 & .048 & .091 & .583 & .424 & \\ 
DawidSkene  & .874 & .663 & .754 & .311 & .756 & .441 & .833 & .860 & .847 & .972 & .804 & .880 & .126 & .762 & .216 & .782 & .690 & \\
\textbf{OneCoin}  & .908 & .891 & .899 & .579 & .797 & .671 & .814 & .884 & .848 & .952 & .881 & .915 & .909 & .476 & .625 & \textbf{.875} & .815 & \\
GLAD  & .880 & .754 & .812 & .384 & .765 & .512 & .718 & .863 & .784 & .958 & .806 & .875 & .750 & .571 & .649 & .802 & .714 & \\
M-MSR  & .663 & .539 & .594 & .186 & .581 & .282 & .497 & .696 & .580 & .897 & .555 & .686 & .333 & .238 & .278 & .581 & .429 & \\
MACE & .890 & .908 & .899 & .540 & .816 & .650 & .838 & .869 & .853 & .966 & .861 & .910 & .425 & .810 & .557 & .870 & .809 & \\
Wawa  & .779 & .615 & .687 & .253 & .673 & .367 & .586 & .787 & .672 & .934 & .675 & .784 & 1.000 & .333 & .500 & .683 & .556 & \\
ZBS  & .852 & .732 & .787 & .351 & .737 & .475 & .685 & .850 & .759 & .953 & .773 & .854 & .909 & .476 & .625 & .776 & .678 \\
\midrule
GPT-4 (t=.2) & .860 & .913 & .885 & .499 & .843 & .627 & .775 & .871 & .820 & .982 & .784 & .872 & .322 & .905 & .475 & .836 & .764 \\ \bottomrule
\end{tabular}
\caption{Exclude-By-Worker integrated with GPT-4 Table. All models use Bio Expert as the gold standard. Baseline is the Majority Vote (MV). From Exclude-By-Worker results, One-Coin Dawid-Skene aggregation model achieves the highest accuracy for both basic and advanced interface. Advanced One-Coin Dawid-Skene reaches 86.6\% and outperforms other aggregation models in every aspects. The accuracy from advanced One-Coin Dawid-Skene almost reaches the accuracy of the GPT-4 (t=.2), 82.7\%. 
}
\label{table:no-bad-worker-gpt-aggregation}
\end{table*}
\begin{table*}[ht]
\centering
\setlength{\tabcolsep}{2.8pt}
\begin{tabular}{lrrrrrrrrrrrrrrrrr}
\toprule
\multicolumn{1}{l}{\textbf{Eval}} & \multicolumn{3}{c}{\textbf{Background}} & \multicolumn{3}{c}{\textbf{Purpose}} & \multicolumn{3}{c}{\textbf{Method}} & \multicolumn{3}{c}{\textbf{Finding}} & \multicolumn{3}{c}{\textbf{Other}} & \multicolumn{1}{c}{\multirow{2}{*}{\textbf{Acc}}} & \multicolumn{1}{c}{\multirow{2}{*}{\textbf{Kappa}}} \\ \cmidrule(lr){2-4} \cmidrule(lr){5-7} \cmidrule(lr){8-10} \cmidrule(lr){11-13} \cmidrule(lr){14-16}
\textbf{Label} & \multicolumn{1}{c}{P} & \multicolumn{1}{c}{R} & \multicolumn{1}{c}{F1} & \multicolumn{1}{c}{P} & \multicolumn{1}{c}{R} & \multicolumn{1}{c}{F1} & \multicolumn{1}{c}{P} & \multicolumn{1}{c}{R} & \multicolumn{1}{c}{F1} & \multicolumn{1}{c}{P} & \multicolumn{1}{c}{R} & \multicolumn{1}{c}{F1} & \multicolumn{1}{c}{P} & \multicolumn{1}{c}{R} & \multicolumn{1}{c}{F1} & \multicolumn{1}{c}{} & \multicolumn{1}{c}{} \\ \midrule
\multicolumn{18}{c}{\textbf{Basic UI}}\\
MV & .780 & .387 & .517 & .194 & .613 & .294 & .450 & .710 & .551 & .854 & .582 & .692 & 1.000 & .286 & .444 & .567 & .402 \\
DawidSkene & .864 & .484 & .621 & .263 & .724 & .386 & .721 & .784 & .751 & .936 & .782 & .852 & .088 & .619 & .154 & .712 & .593 \\
OneCoin & .888 & .712 & .790 & .513 & .710 & .596 & .675 & .834 & .746 & .896 & .843 & .869 & .875 & .333 & .483 & .800 & .702 \\
GLAD & .873 & .629 & .731 & .328 & .765 & .459 & .608 & .797 & .690 & .924 & .744 & .825 & .684 & .619 & .650 & .731 & .615 \\
M-MSR & .689 & .491 & .574 & .202 & .581 & .299 & .479 & .656 & .553 & .858 & .609 & .713 & .500 & .333 & .400 & .590 & .428 \\
\textbf{MACE} & .871 & .804 & .836 & .444 & .797 & .570 & .718 & .818 & .765 & .960 & .810 & .878 & .346 & .857 & .493 & \textbf{.810} & .724 \\
Wawa & .807 & .526 & .637 & .239 & .700 & .356 & .513 & .757 & .612 & .912 & .627 & .743 & 1.000 & .381 & .552 & .636 & .495 \\
ZBS & .834 & .569 & .676 & .276 & .747 & .403 & .548 & .778 & .643 & .919 & .671 & .776 & .818 & .429 & .563 & .675 & .544 \\ \midrule
\multicolumn{18}{c}{\textbf{Advances UI}}\\
MV & .694 & .370 & .482 & .149 & .525 & .232 & .434 & .701 & .536 & .864 & .520 & .649 & 1.000 & .048 & .091 & .523 & .354 \\ 
DawidSkene & .838 & .556 & .668 & .242 & .700 & .360 & .792 & .784 & .788 & .969 & .765 & .855 & .077 & .667 & .139 & .718 & .607 \\
\textbf{OneCoin} & .909 & .884 & .896 & .581 & .793 & .671 & .808 & .893 & .848 & .953 & .879 & .914 & .909 & .476 & .625 & \textbf{.874} & .813 \\
GLAD & .865 & .698 & .772 & .325 & .700 & .444 & .671 & .857 & .753 & .951 & .773 & .853 & .889 & .381 & .533 & .767 & .665 \\
M-MSR & .537 & .450 & .489 & .169 & .544 & .257 & .451 & .650 & .532 & .869 & .498 & .634 & .313 & .238 & .270 & .522 & .353 \\
MACE & .884 & .910 & .897 & .546 & .816 & .654 & .835 & .871 & .852 & .968 & .859 & .910 & .425 & .810 & .557 & .869 & .808 \\
Wawa & .723 & .511 & .599 & .186 & .613 & .286 & .509 & .726 & .598 & .913 & .580 & .710 & 1.000 & .286 & .444 & .597 & .449 \\
ZBS & .817 & .653 & .726 & .260 & .691 & .377 & .631 & .810 & .710 & .942 & .700 & .803 & 1.000 & .381 & .552 & .711 & .593 \\ 
\midrule
GPT-4 (t=.2) & .860 & .913 & .885 & .499 & .843 & .627 & .775 & .871 & .820 & .982 & .784 & .872 & .322 & .905 & .475 & .836 & .764 \\\bottomrule
\end{tabular}
\caption{Exclude-By-Batch integrated with GPT-4 Table. All models use Bio Expert as the gold standard. Baseline is the Majority Vote (MV).}
\label{table:no-bad-label-gpt-aggregation}
\end{table*}

\section{Worker Result Visual Comparison Interfaces}
\begin{figure*}
    \centering
    \includegraphics[width=14cm]{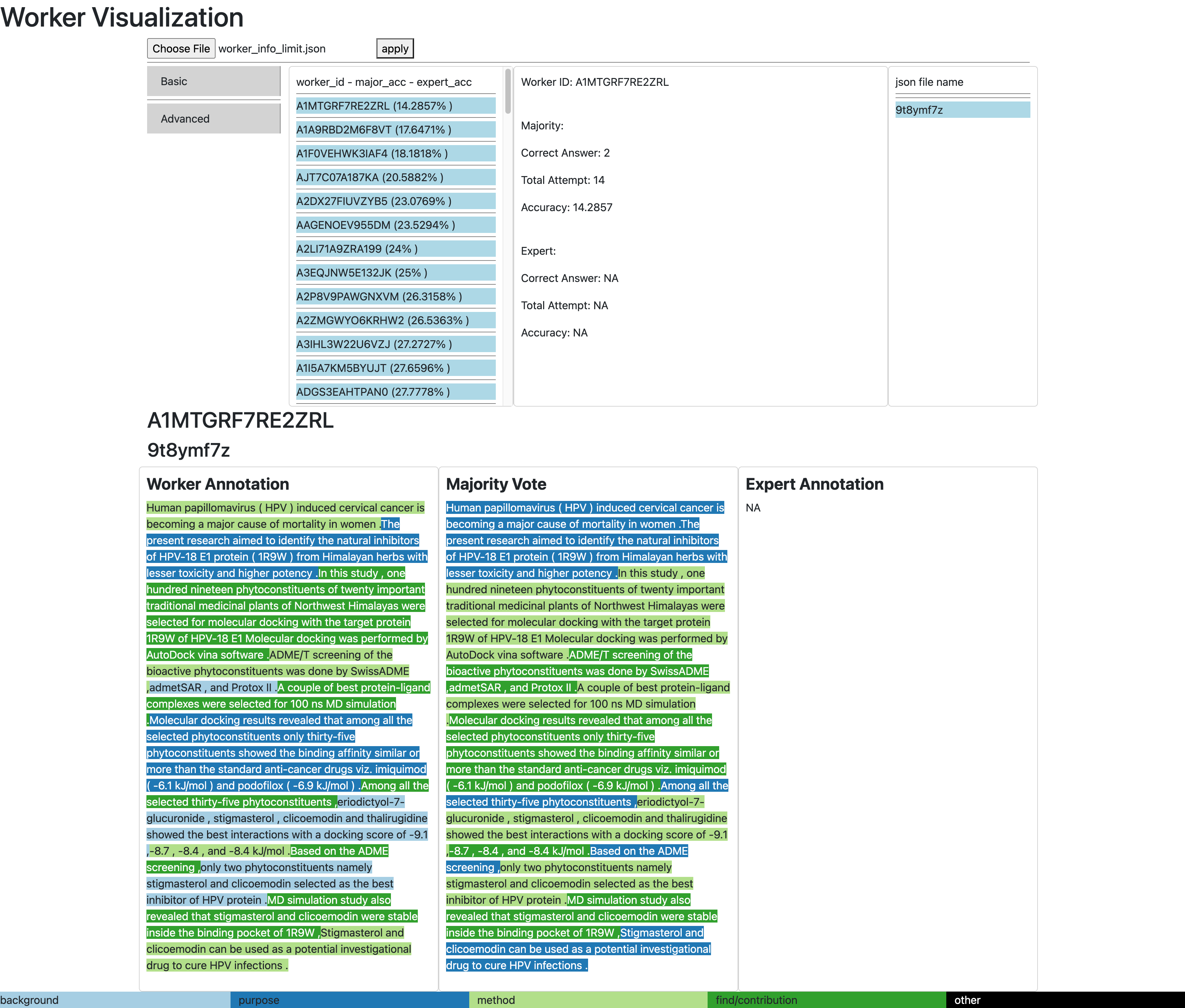}
    \caption{Worker result color-coded visualization interface for rapid label inspection. It displayed worker annotation, majority vote, and expert annotation (if applied), where different labels had different colors. To simulate the practical scenario, we pretended that we had limited expert annotations during the experiment. As a result, some abstracts did not have expert annotation when we did the comparison. }
    \label{fig:worker-result-comparison-visualization}
    \Description{It is a worker result color-coded visualization interface for rapid label inspection. It displayed worker annotation, majority vote, and expert annotation for MTurk workers from Basic and Advance UI groups by their worker ID.}
\end{figure*}

Figure \ref{fig:worker-result-comparison-visualization} displays an internal worker accuracy visualization, allowing for the monitoring of their performance against the majority vote and, if used, expert annotations. This visualization also includes a central worker panel, which shows detailed statistics for each worker, like the number of correct answers, total attempts, and their accuracy as compared to the majority vote and expert annotations.

\section{Prompt}
Table \ref{table:zero-shot-table} shows the zero-shot prompt we used for querying LLMs.
\begin{table*}
\centering
    \begin{tabular}{@{}p{\textwidth}@{}}
        \hline
        \textbf{Zero-shot Prompt} \\
        Classify the given text into one of the following labels.\\ \\

        [Background]: Text segments answer one or more of these questions: Why is this problem important?,
        What relevant works have been created before?, What is still missing in the previous works?, 
        What are the high-level research questions?, How might this help other research or researchers?\\
        
        [Purpose]: Text segments answer one or more of these questions: What specific things do the researchers want to do?, What specific knowledge do the researchers want to gain?, What specific hypothesis do the researchers want to test?\\
        
        [Method]: Text segments answer one or more of these questions: How did the researchers do the work or find what they sought?, What are the procedures and steps of the research?\\
        
        [Finding]: Text segments answer one or more of these questions: What did the researchers find out?, Did the proposed methods work?, Did the thing behave as the researchers expected?
        
        [Other]: Text fragments that do NOT fit into any of the four categories above. Text fragments that are NOT part of the article. Text fragments that are NOT in English. Text fragments that contains ONLY reference marks (e.g., "[1,2,3,4,5") or ONLY dates (e.g., "April 20, 2008"). Captions for figures and tables (e.g. "Figure 1: Experimental Result of ...", or "Table 1: The Typical Symptoms of ...") Formatting errors. I really don't know or I'm not sure.\\ \\
        
        FAQs\\
        
        1. This text fragment has terms that I don't understand. What should I do? Please use the context in the article to figure out the focus. You can look up terms you don't know if you feel like you need to understand them.\\
        2. This text fragment is too short to mean anything. What should I do? If the text fragment is too short to have significant meanings, you could consider the entire sentence and answer based on the entire sentence.\\
        3. This text fragment is NOT in English. What should I do? If the whole fragment (or the majority of words in the fragment) is in Non-English, please label it as "Other". If the majority of the words in this fragment are in English with a few non-English words, please judge the label normally.\\
        4. I'm not sure if this should be a "background" or a "finding." How do I tell? When a sentence occurs in the earlier part of an article, and it is presented as a known fact or looks authoritative, it is often a "background" information.\\
        5. Do "potential applications of the proposed work" count as "background" or "purpose"? It should be "background." The "purpose" refers to specific things the paper wants to achieve.\\
        6. If the article says it's a "literature review" (e.g., "We reviewed the literature" / "In this article, we review.." etc), would we classify those as finding/contribution or purpose? Most parts of a literature review paper should still be "background" or "purpose", and only the "insight" drew from a set of prior works can be viewed as a "finding/contribution".\\
        7. What should I do with the case study on a patient? Typically, it has a patient come in with a set of signs and symptoms in the ER, and then the patient gets assessed and diagnosed. The patient is admitted to the hospital ICU and tests are done and they may be diagnosed with something else. In such cases, please label the interventions done by the medical staff (e.g., CT scans, X-rays, and medications given) as "Method", and the patient's final result (e.g. the patient's pneumonia resolved and he was released from the hospital) as "Finding/Contribution".\\ \\
        
        Classify the following sentence into one of the label: Background, Purpose, Method, Finding, and Other. \\Provide answer in format of ```fragment-i []''' \\

        fragment-1 Text: ```\{\textbf{Sentence-1}\}'''\\
        Label: []\\

        fragment-2 Text: ```\{\textbf{Sentence-2}\}'''\\
        Label: []\\

        fragment-3 Text: ```\{\textbf{Sentence-3}\}'''\\
        Label: []\\
        ......\\
        \hline
    \end{tabular}
    \caption{Zero-shot prompt used when calling GPT-4. The \{\textbf{Sentence-n}\} will be replaced by the following sentence in the abstract we would like to predict.}
    \label{table:zero-shot-table}
\end{table*}

\section{All Aggregation Methods for All Workers and Exclude-By-Batch}
\begin{figure*}
    \centering
    \begin{subfigure}{0.36\textheight}
    \centering
    \includegraphics[width=\linewidth]{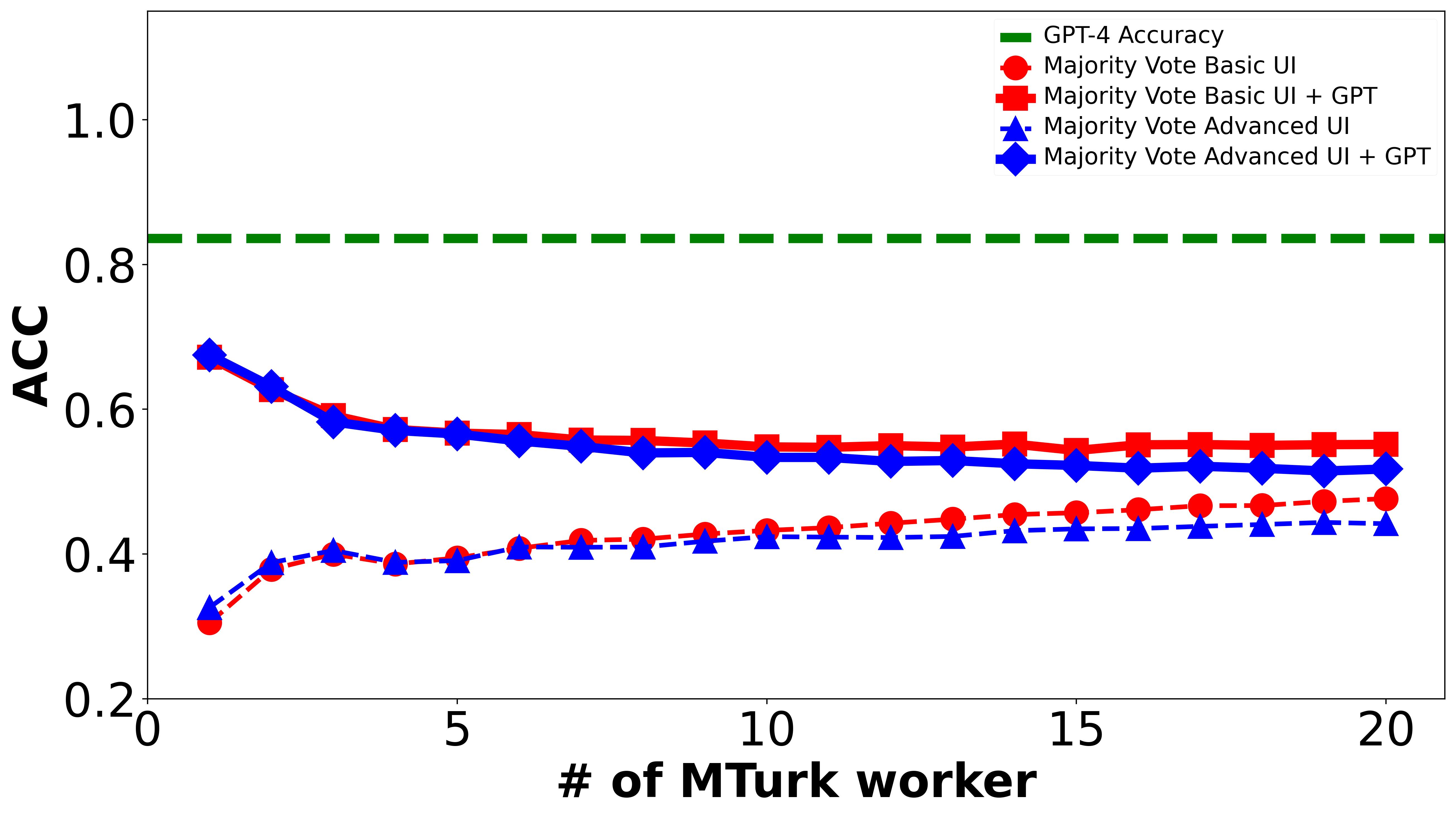}
    \caption{Majority Vote}
    \label{fig:agg-mv-baseline-all-worker}
    \end{subfigure}
     \hfill
    \begin{subfigure}{0.36\textheight}
        \centering
        \includegraphics[width=\linewidth]{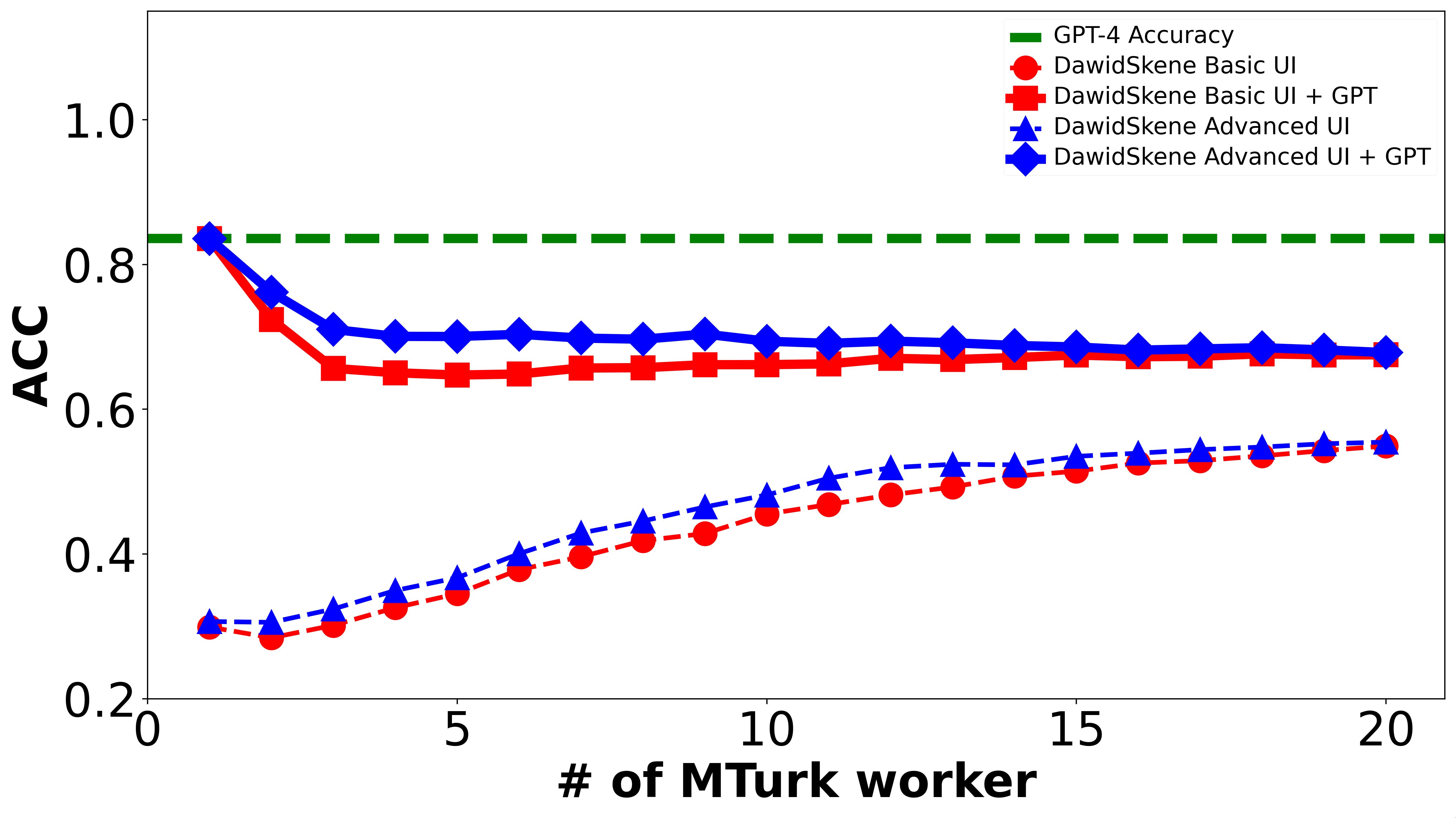}
        \caption{Dawid-Skene}
        \label{fig:agg-dawid-all-worker}
    \end{subfigure}

    \begin{subfigure}{0.36\textheight}
        \centering
        \includegraphics[width=\linewidth]{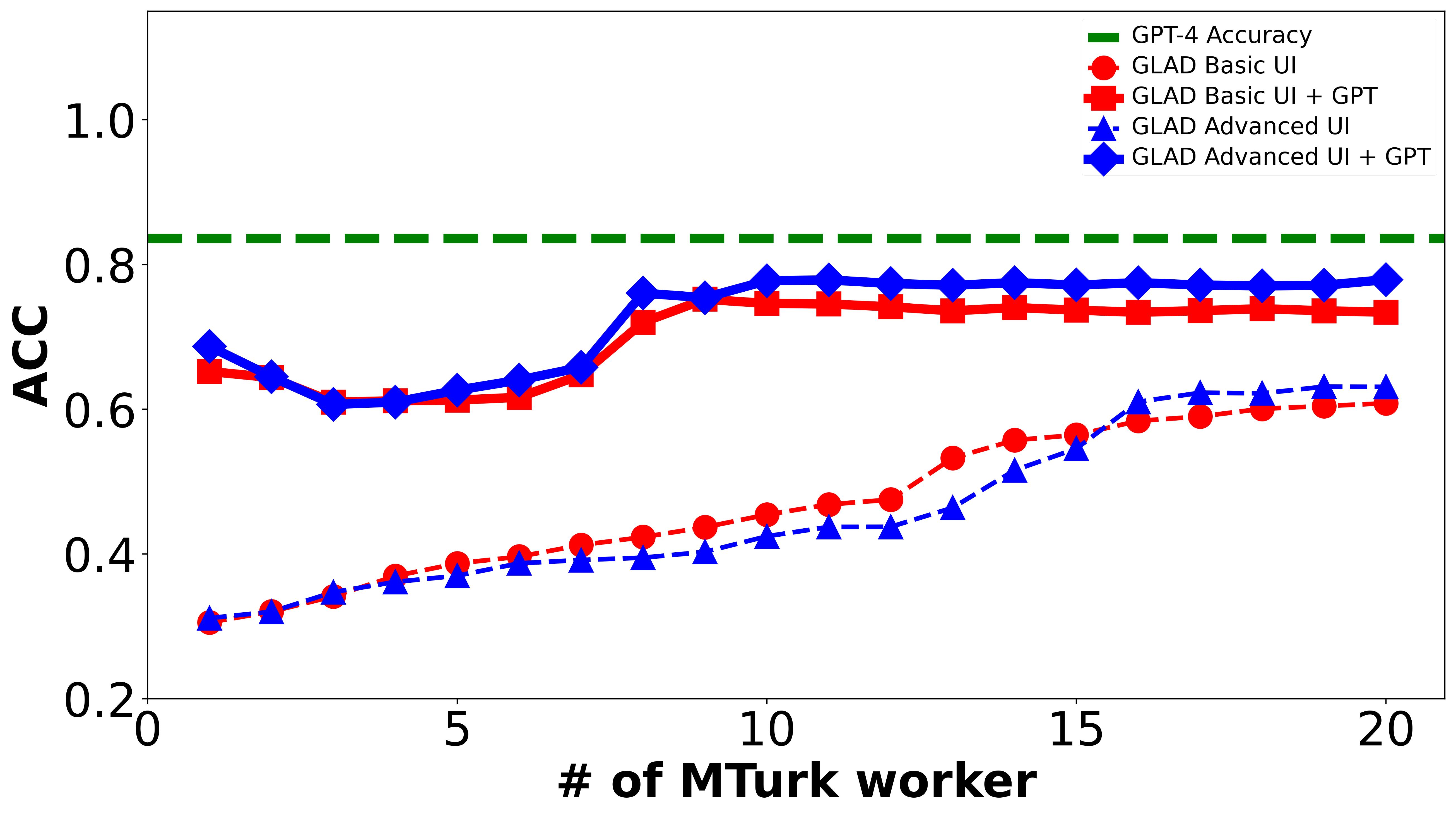}
        \caption{GLAD}
        \label{fig:agg-glad-all-worker}
    \end{subfigure}
     \hfill
    \begin{subfigure}{0.36\textheight}
        \centering
        \includegraphics[width=\linewidth]{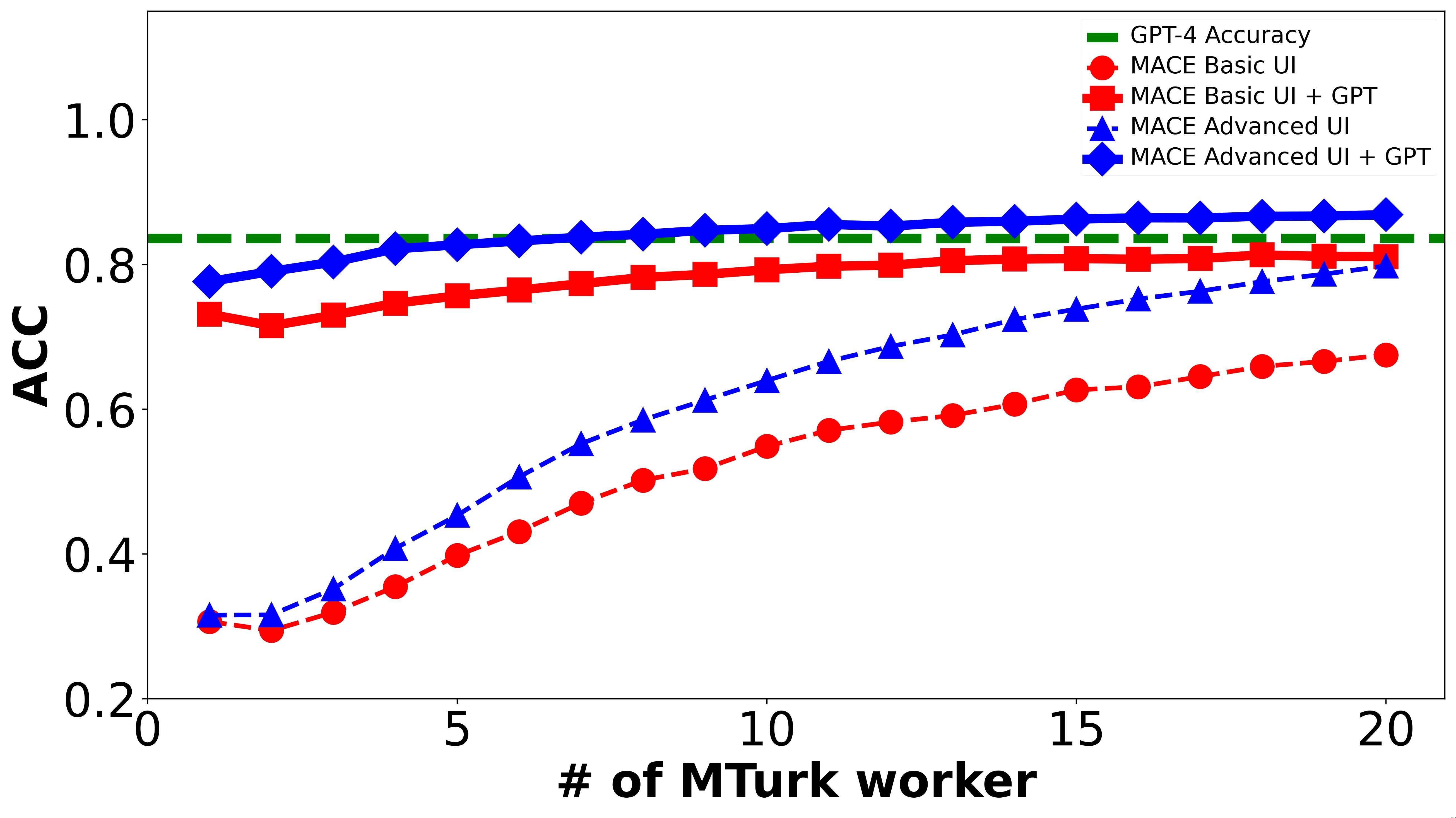}
        \caption{MACE}
        \label{fig:agg-mace-all-worker}
    \end{subfigure}

    \begin{subfigure}{0.36\textheight}
        \centering
        \includegraphics[width=\linewidth]{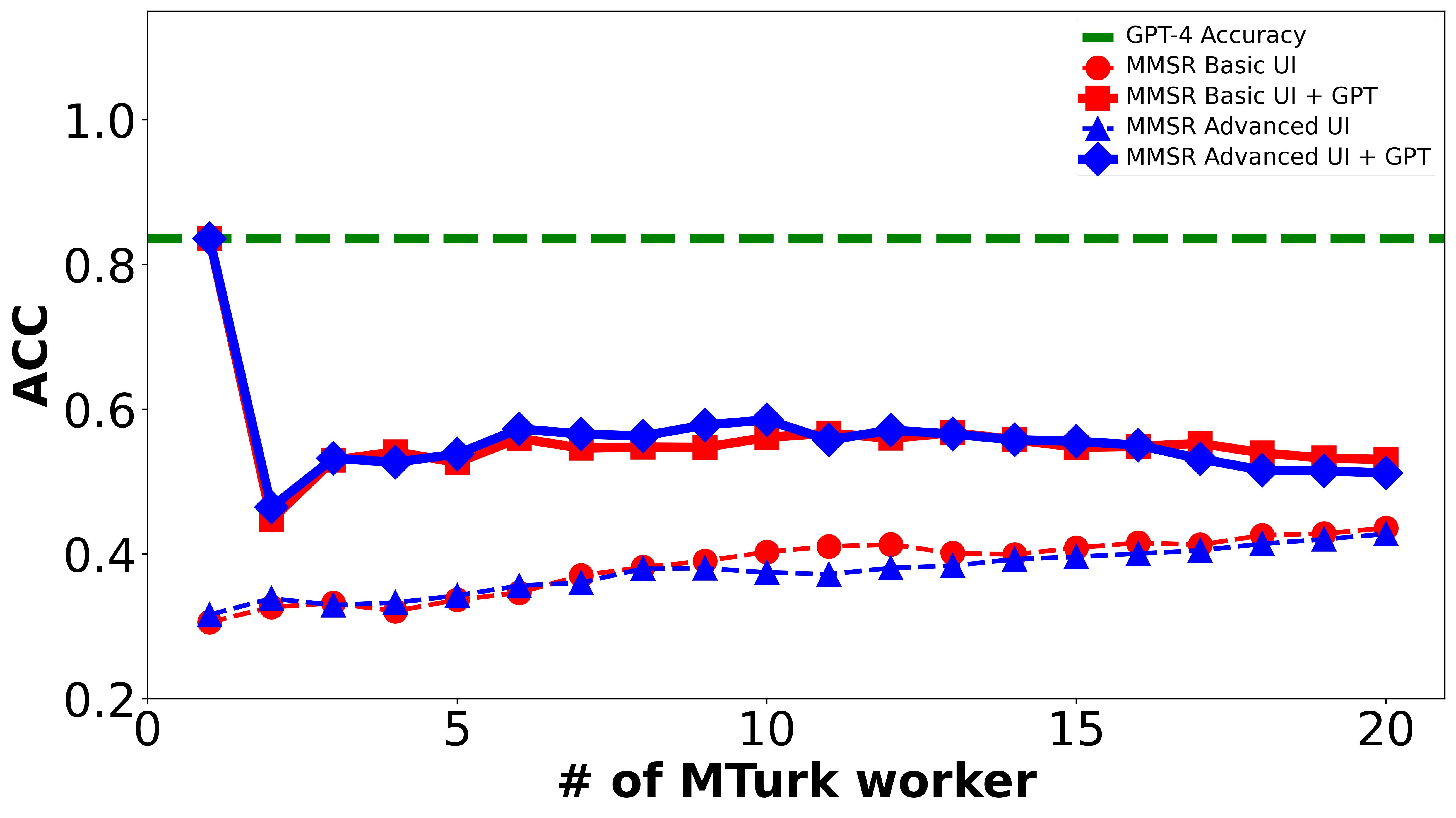}
        \caption{MMSR}
        \label{fig:agg-mmsr-all-worker}
    \end{subfigure}
    \hfill
    \begin{subfigure}{0.36\textheight}
        \centering
        \includegraphics[width=\linewidth]{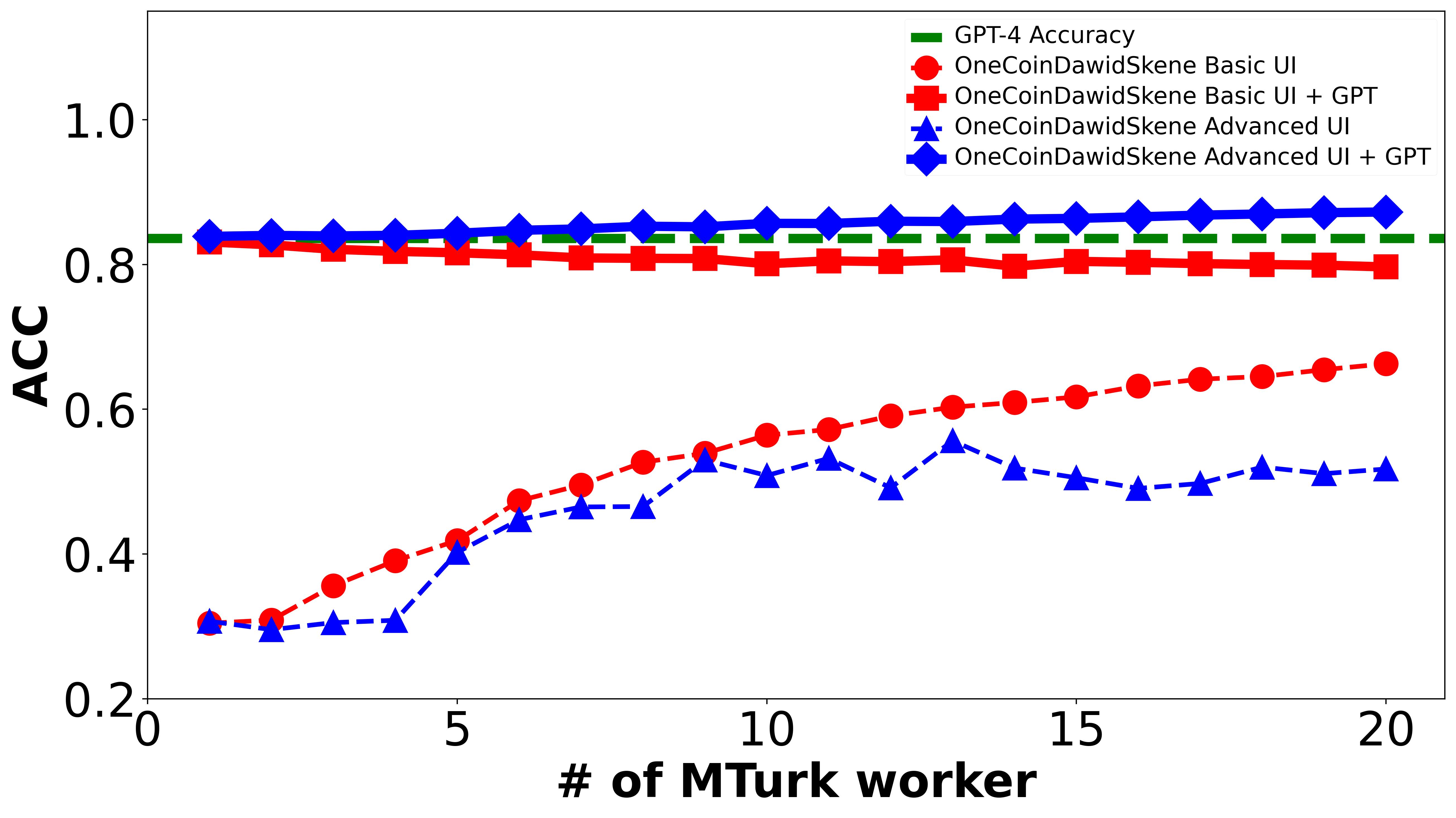}
        \caption{One-Coin Dawid-Skene}
        \label{fig:agg-onecoin-all-worker}
    \end{subfigure}
    
    \begin{subfigure}{0.36\textheight}
        \centering
        \includegraphics[width=\linewidth]{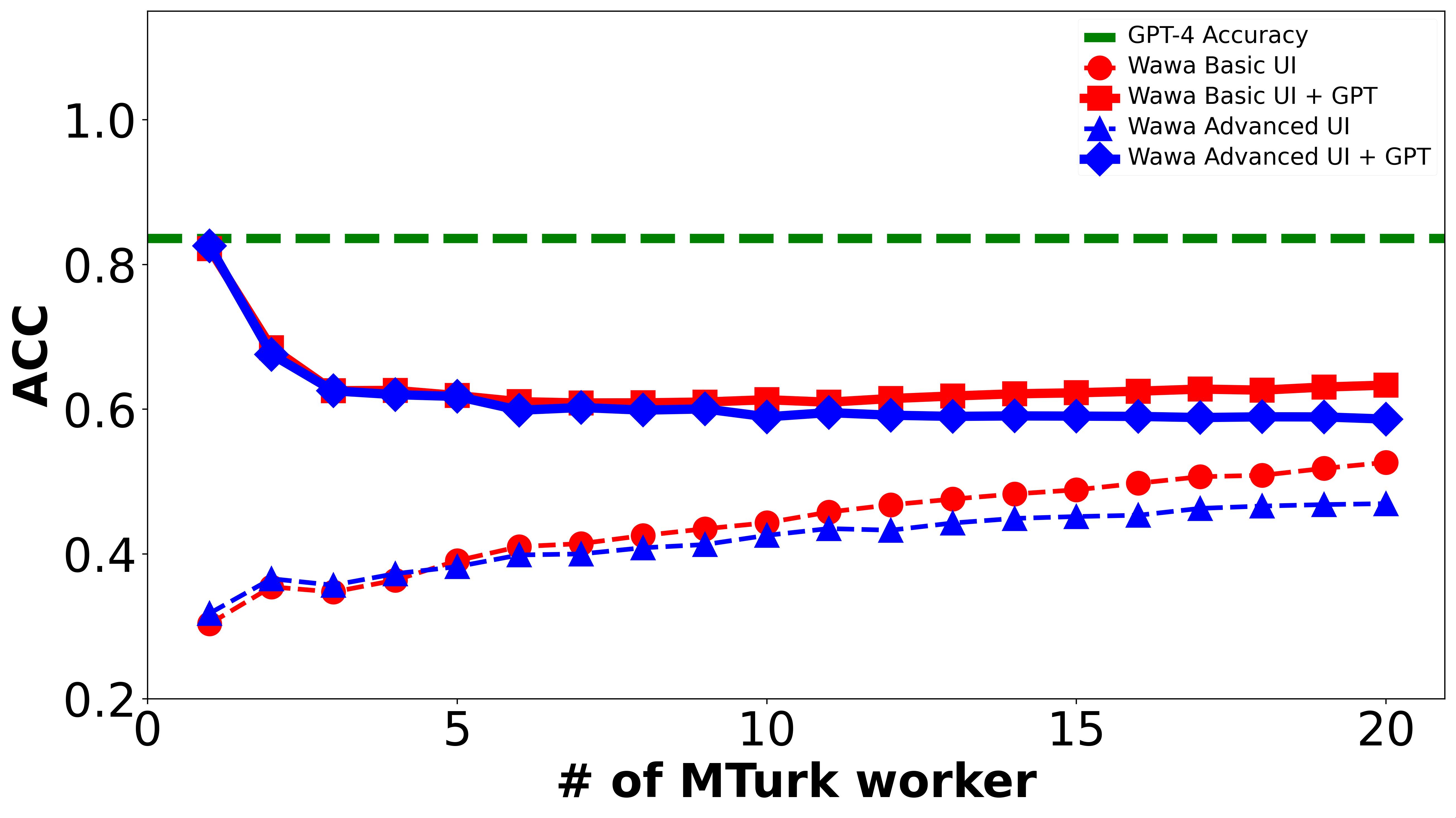}
        \caption{WAWA}
        \label{fig:agg-wawa-all-worker}
    \end{subfigure}
     \hfill
    \begin{subfigure}{0.36\textheight}
        \centering
        \includegraphics[width=\linewidth]{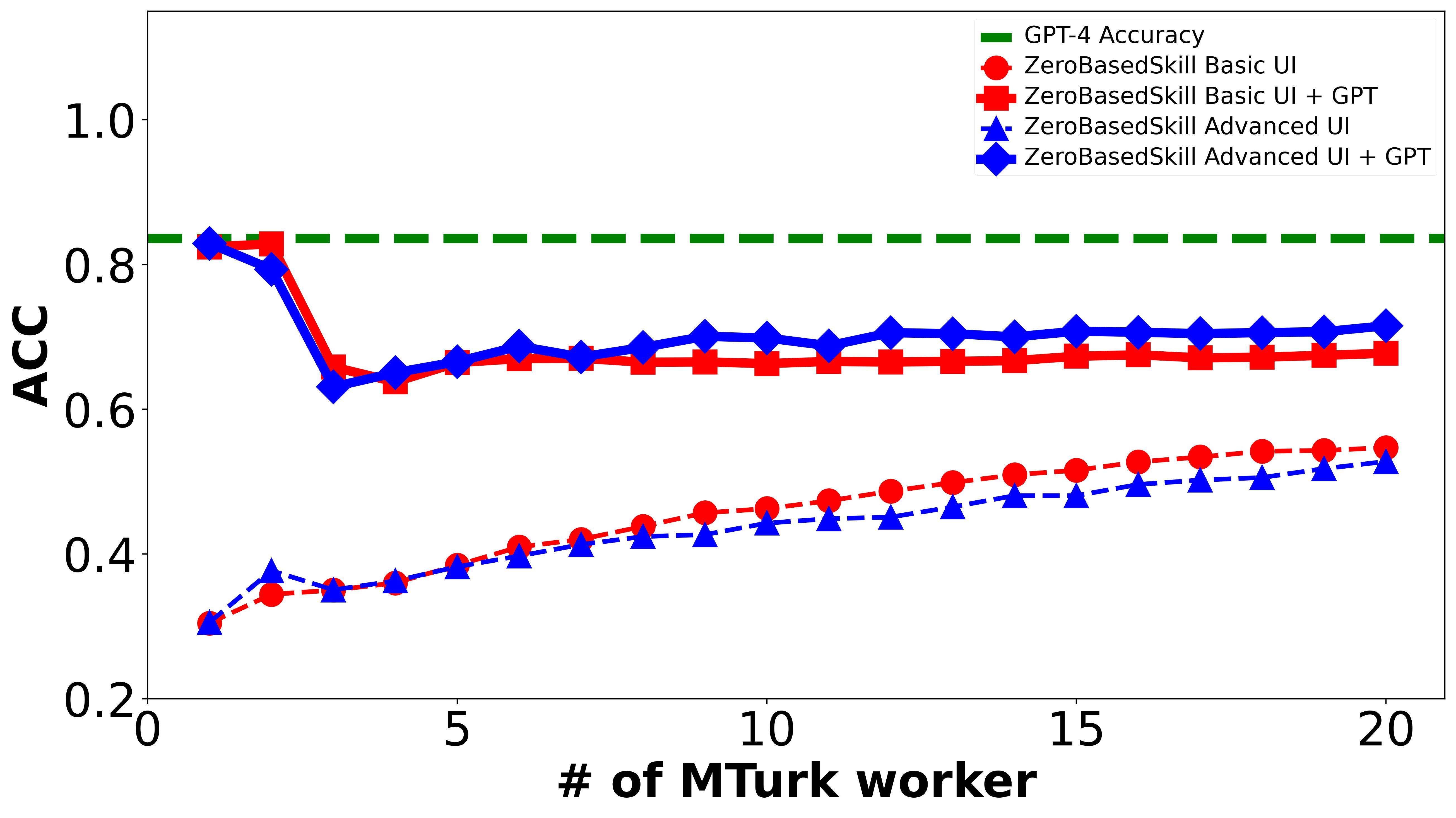}
        \caption{ZeroBasedSkill}
        \label{fig:agg-zero-all-worker}
    \end{subfigure}
\caption{All Workers simulation results applied to different aggregation models.}
\label{figure:all-worker-single-methods-gpt}
\Description{There are eight subfigures for all All-Workers aggregation methods simulation. They are Majority Vote, Dawid-Skene, GLAD, MACE, MMSR, One-Coin Dawid-Skene, Wawa, and ZeroBased Skill methods.}
\end{figure*}

\begin{figure*}
    \centering
    \begin{subfigure}{0.36\textheight}
    \centering
    \includegraphics[width=\linewidth]{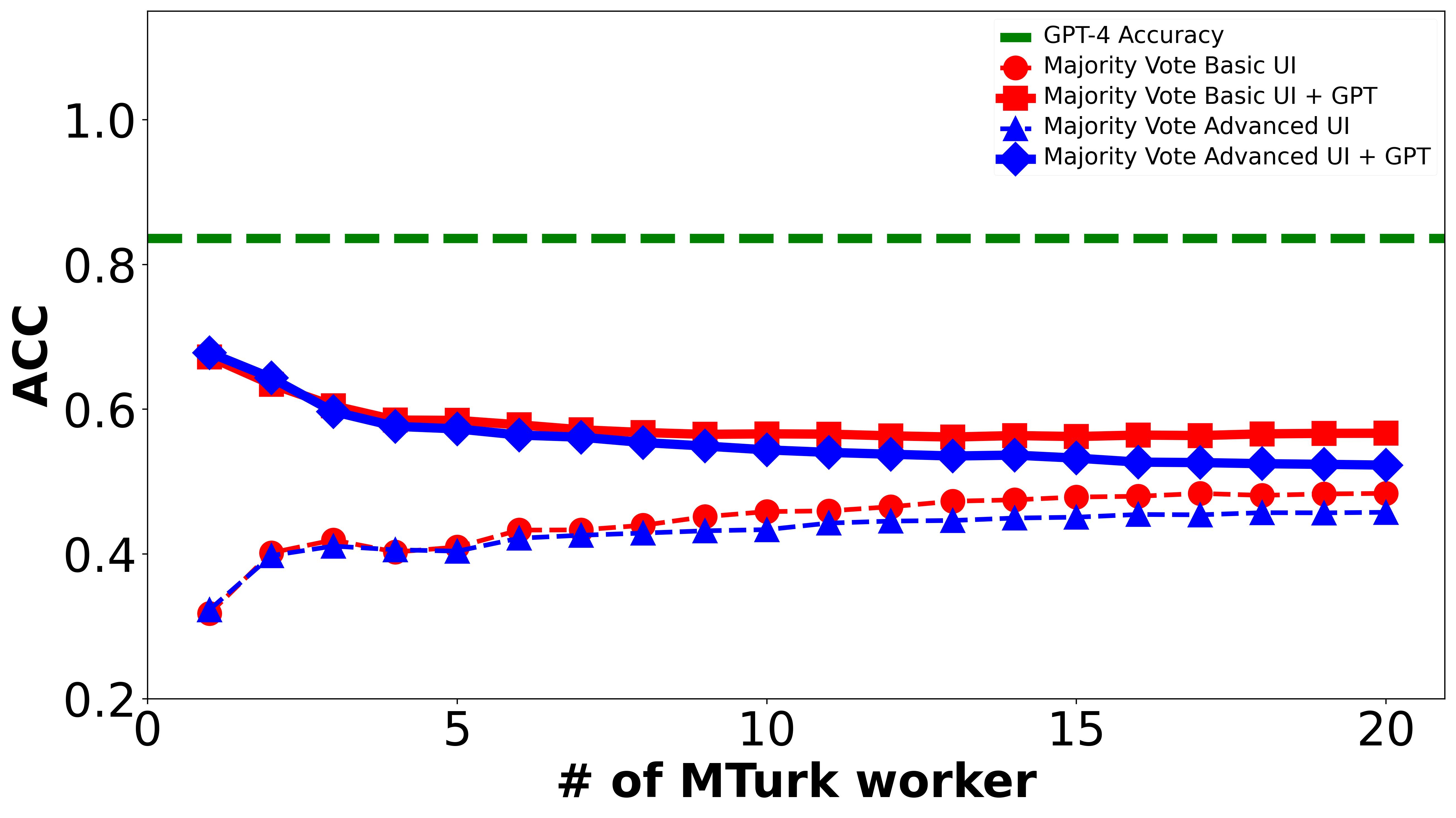}
    \caption{Majority Vote}
    \label{fig:agg-mv-baseline-no-bad-label}
    \end{subfigure}
     \hfill
    \begin{subfigure}{0.36\textheight}
        \centering
        \includegraphics[width=\linewidth]{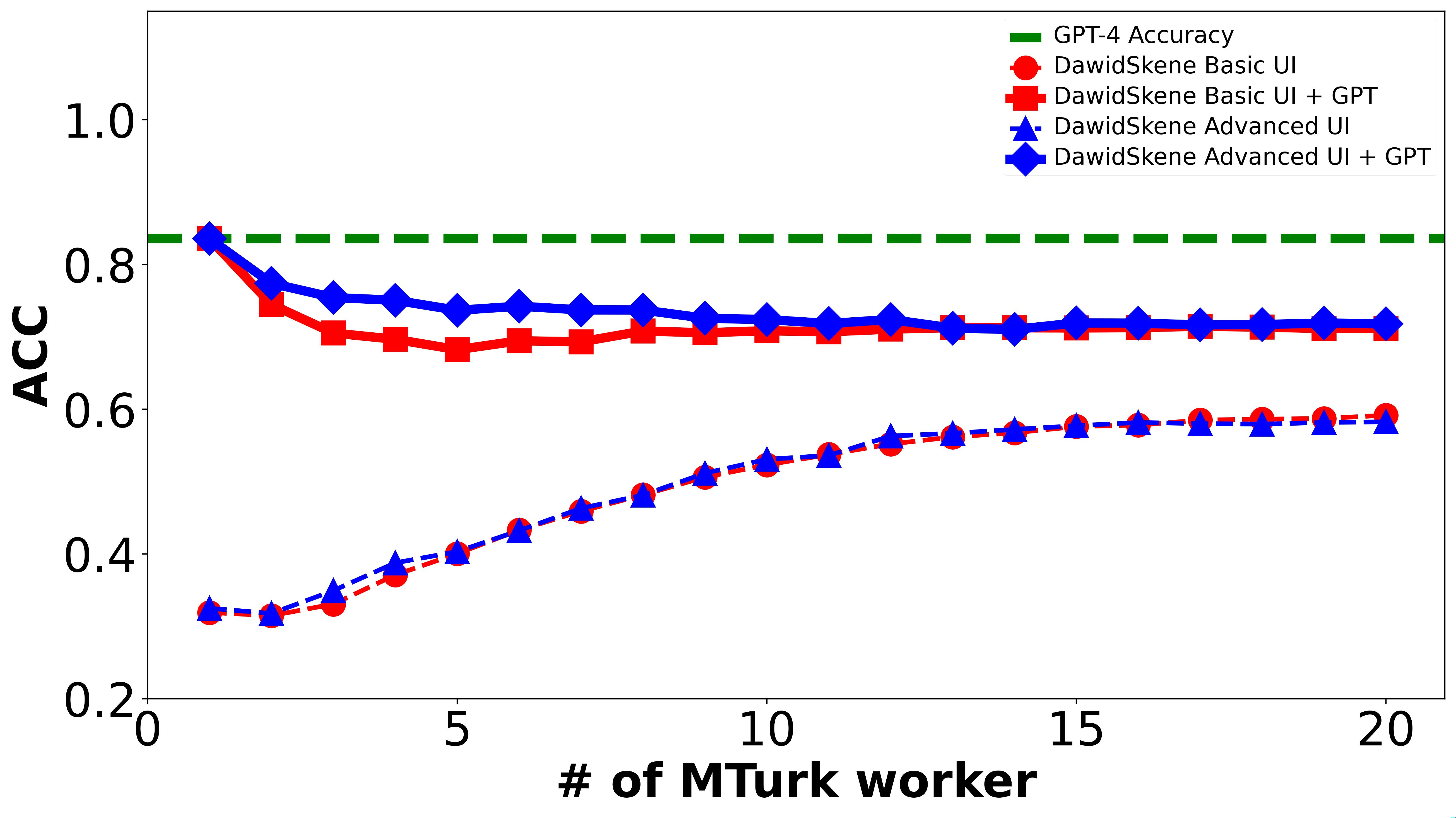}
        \caption{Dawid-Skene}
        \label{fig:agg-dawid-no-bad-label}
    \end{subfigure}

    \begin{subfigure}{0.36\textheight}
        \centering
        \includegraphics[width=\linewidth]{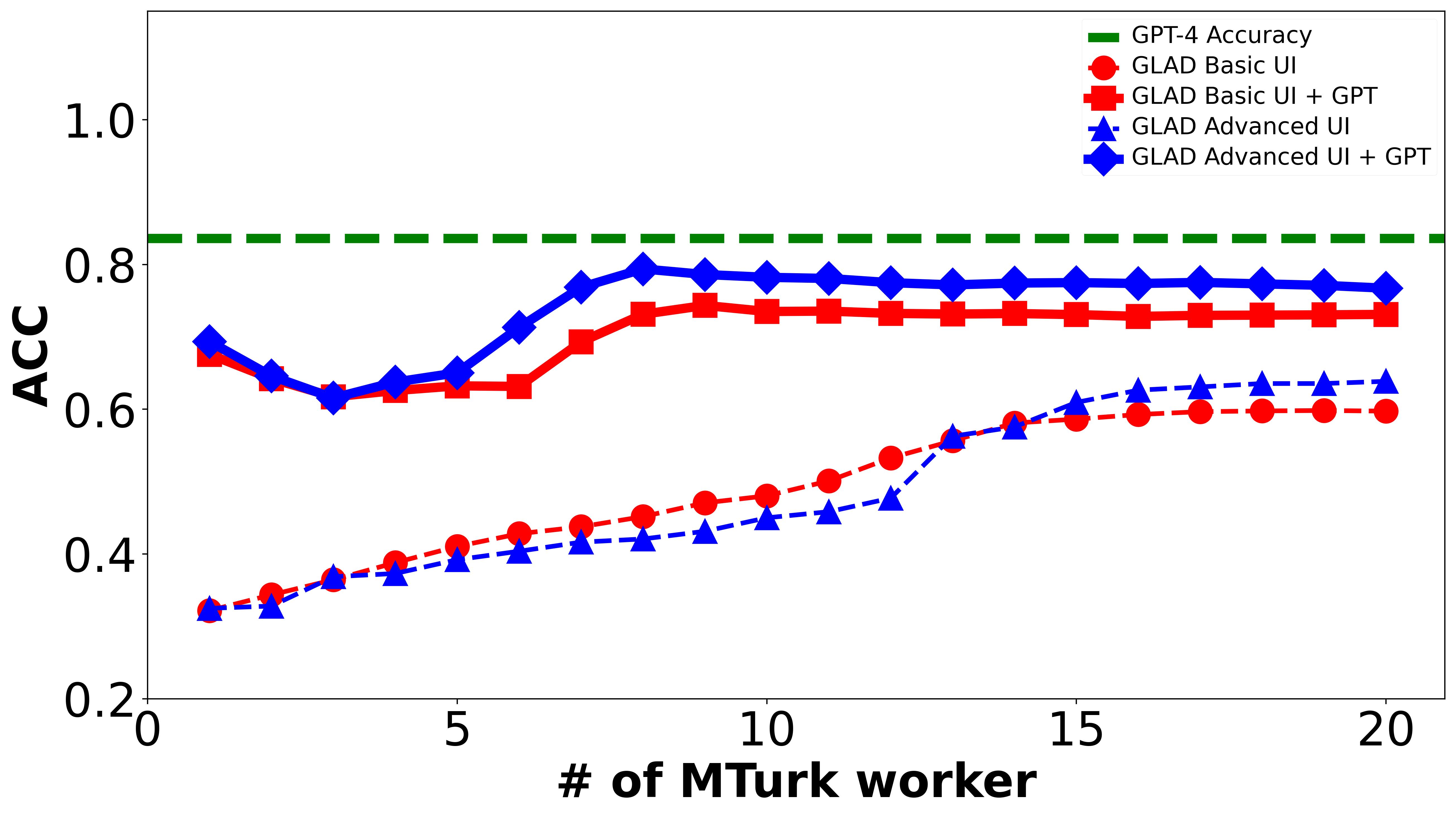}
        \caption{GLAD}
        \label{fig:agg-glad-no-bad-label}
    \end{subfigure}
     \hfill
    \begin{subfigure}{0.36\textheight}
        \centering
        \includegraphics[width=\linewidth]{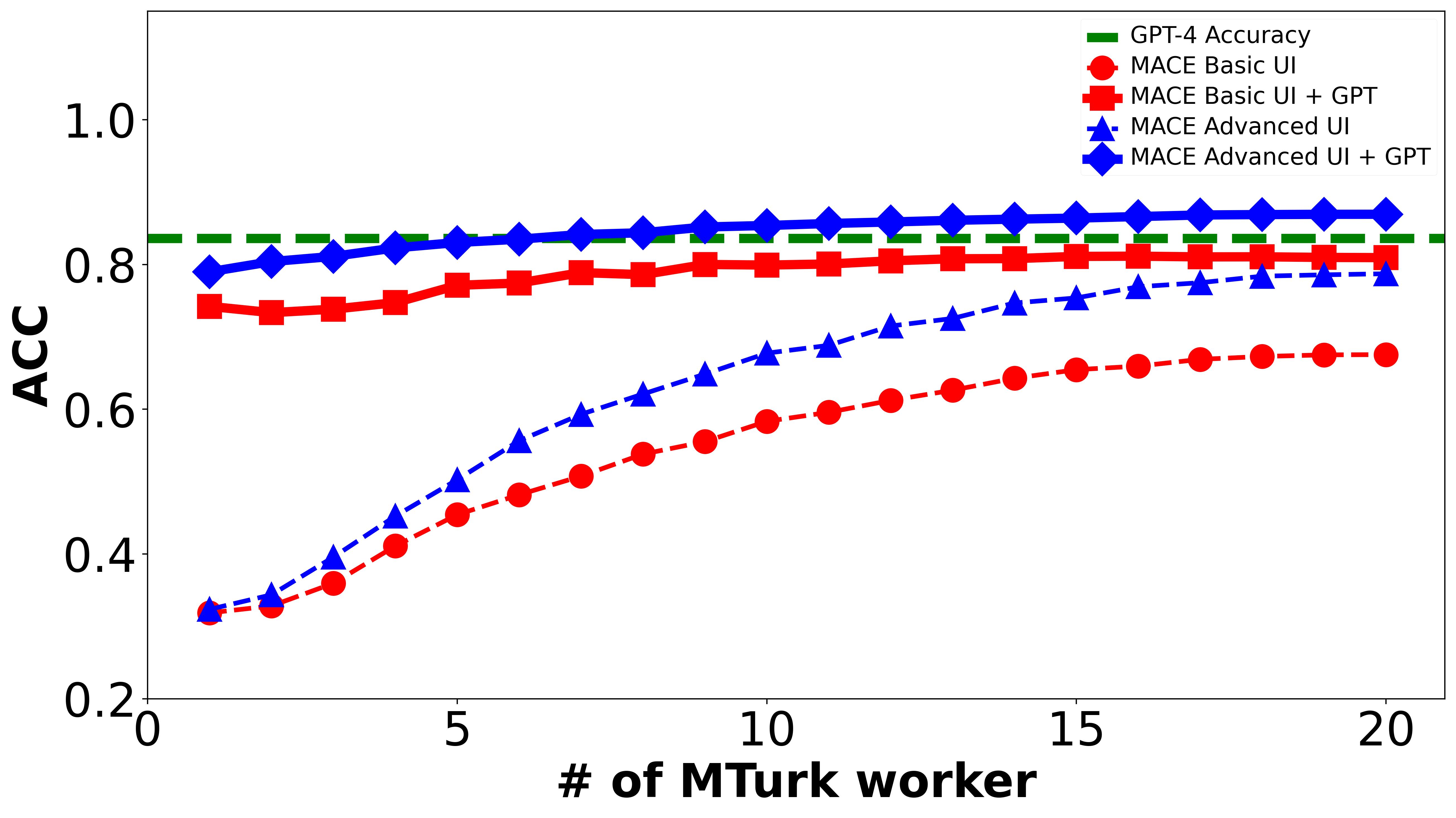}
        \caption{MACE}
        \label{fig:agg-mace-no-bad-label}
    \end{subfigure}

    \begin{subfigure}{0.36\textheight}
        \centering
        \includegraphics[width=\linewidth]{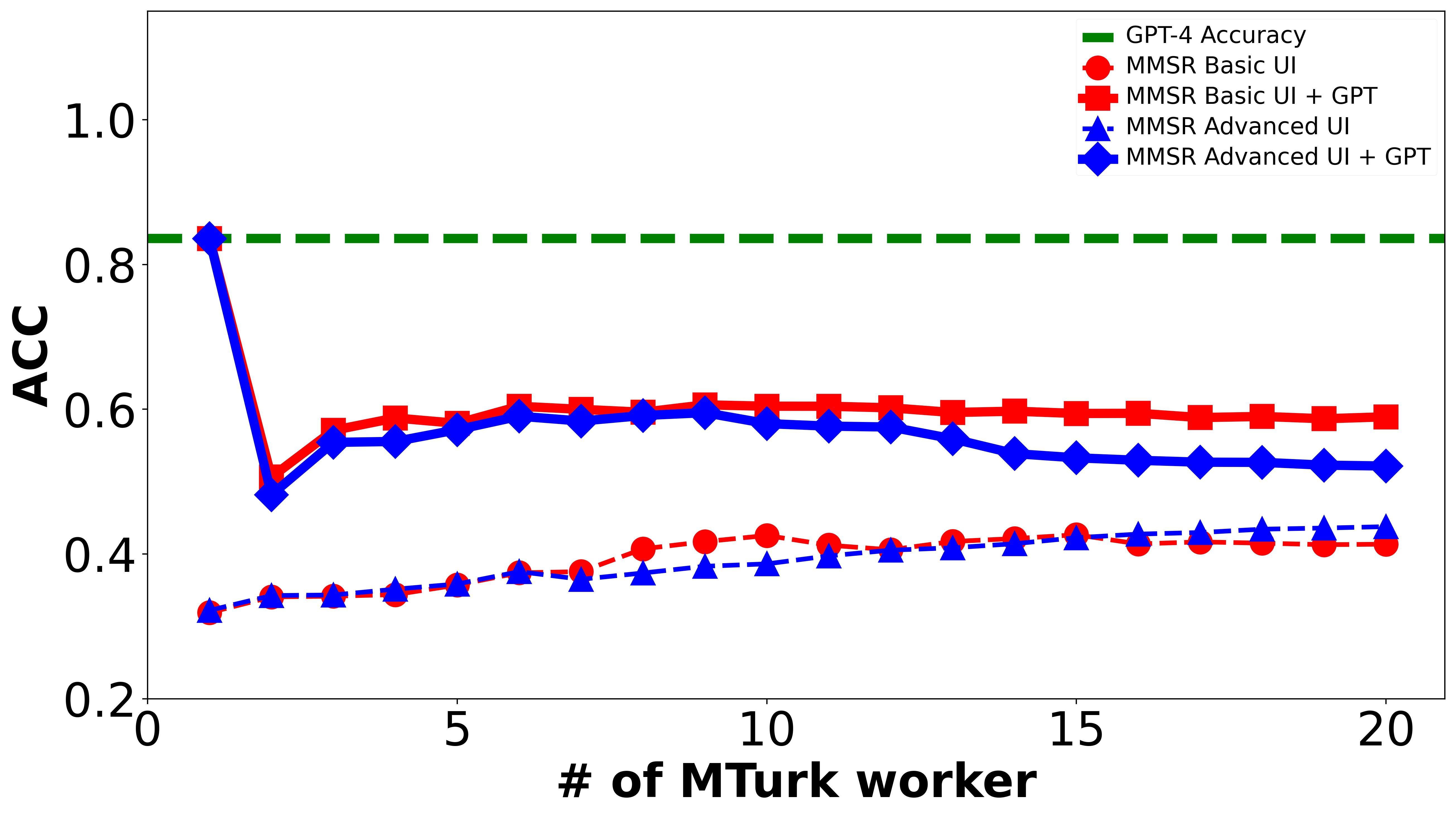}
        \caption{MMSR}
        \label{fig:agg-mmsr-no-bad-label}
    \end{subfigure}
    \hfill
    \begin{subfigure}{0.36\textheight}
        \centering
        \includegraphics[width=\linewidth]{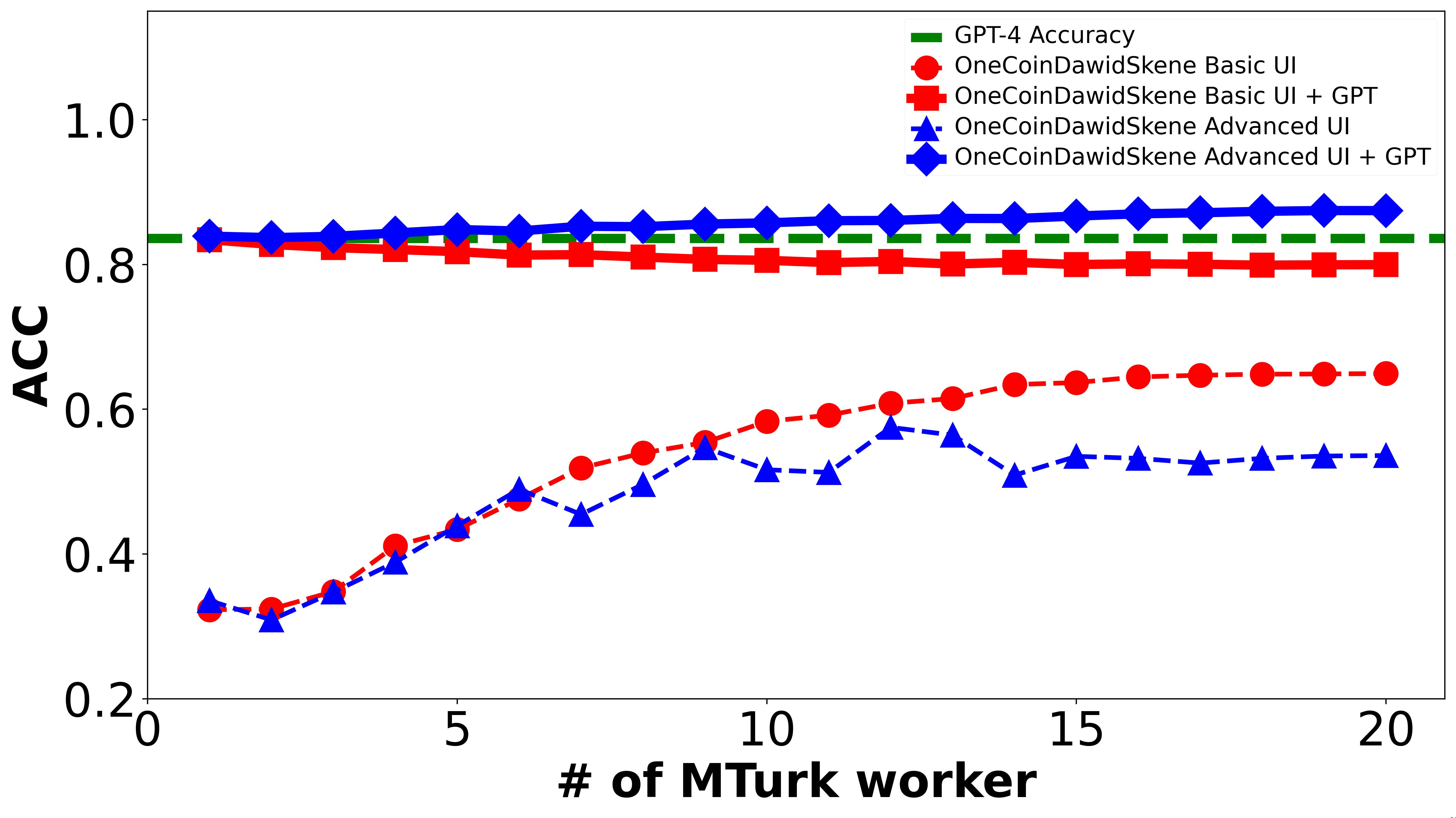}
        \caption{One-Coin Dawid-Skene}
        \label{fig:agg-onecoin-no-bad-label}
    \end{subfigure}
    
    \begin{subfigure}{0.36\textheight}
        \centering
        \includegraphics[width=\linewidth]{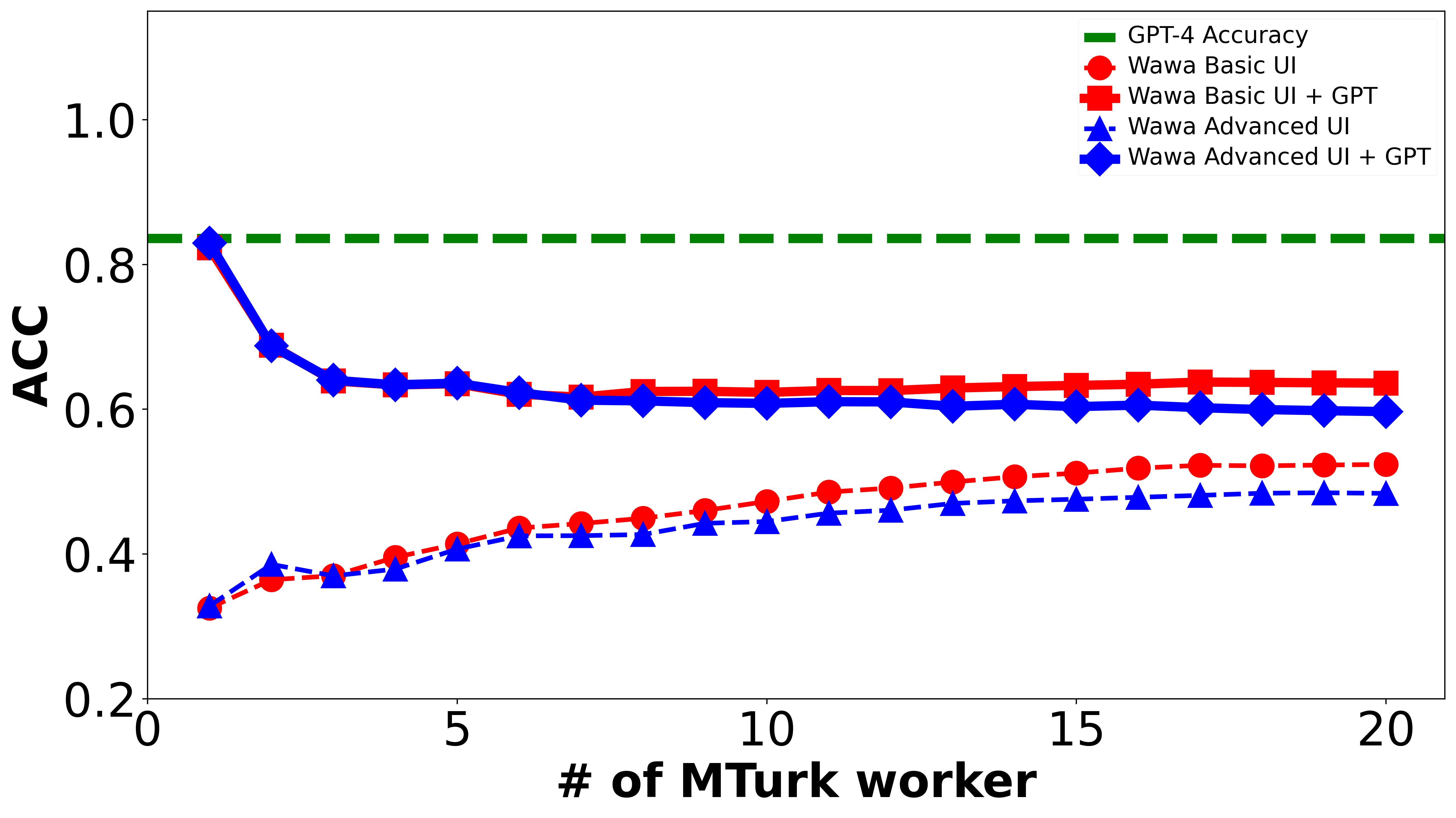}
        \caption{WAWA}
        \label{fig:agg-wawa-no-bad-label}
    \end{subfigure}
     \hfill
    \begin{subfigure}{0.36\textheight}
        \centering
        \includegraphics[width=\linewidth]{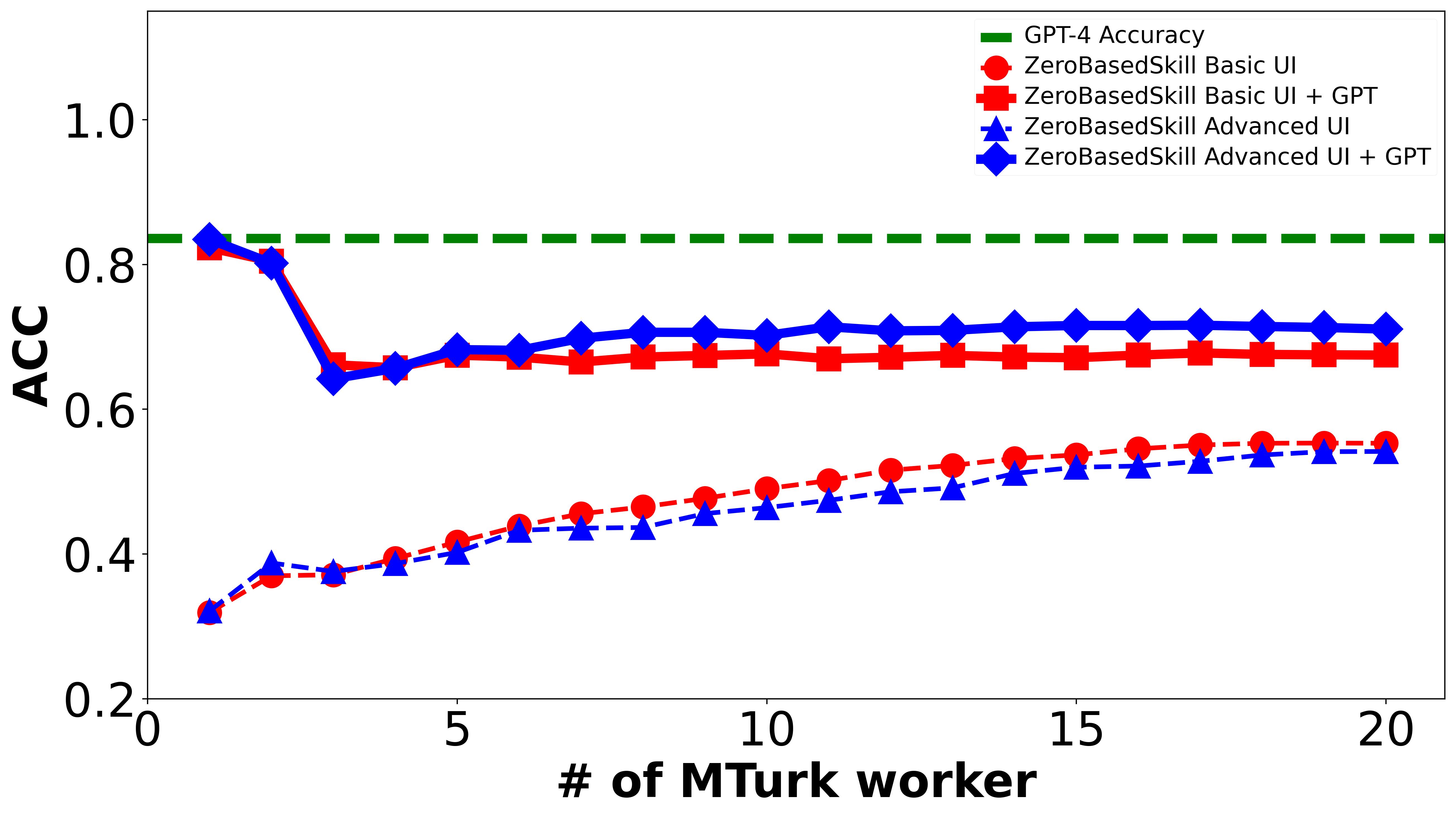}
        \caption{ZeroBasedSkill}
        \label{fig:agg-zero-no-bad-label}
    \end{subfigure}
\caption{Exclude-By-Batch simulation results applied to different aggregation models.}
\label{figure:no-bad-label-single-methods-gpt}
\Description{There are eight subfigures for all Exclude-By-Batch aggregation methods simulation. They are Majority Vote, Dawid-Skene, GLAD, MACE, MMSR, One-Coin Dawid-Skene, Wawa, and ZeroBased Skill methods.}
\end{figure*}

Figure \ref{figure:all-worker-single-methods-gpt} and \ref{figure:no-bad-label-single-methods-gpt} are simulation results applied to different aggregation models for All-Workers and Exclude-By-Batch strategies.


\clearpage

\newgeometry{top=2cm, bottom=2cm, left=2cm, right=2cm}
\newpage
\section{Corrigendum}\label{sec:corrigendum}
\noindent Minor errors were corrected on June 26, 2024. These errors did not impact the conclusion of the paper. The information in this arXiv paper has been corrected. We present the corrigendum in the following:

\noindent \textbf{Corrigendum to ``If in a Crowdsourced Data Annotation Pipeline, A GPT-4'', by He et al., CHI '24: Proceedings of the CHI Conference on Human Factors in Computing Systems, May 2024, Article No.: 1040, Pages 1-25}

\vspace{1em}

\noindent \rule{\textwidth}{1pt}

This corrigendum concerns our paper in CHI '24: Proceedings of the CHI Conference on Human Factors in Computing Systems, May 2024, Article No.: 1040, Pages 1-25 (He et al., If in a Crowdsourced Data Annotation Pipeline, A GPT-4). 

\begin{enumerate}
    \item 
    The results presented in Table 6 in the original paper should be replaced with the results in Table~\ref{table:basic-crowd-gpt-acc-result} in this document.

    In Table 6 in the original paper, the data for \textbf{the eight aggregation methods} listed under the '\textbf{Exclude-By-Worker}' and '\textbf{Exclude-By-Batch}' categories were \textbf{mistakenly swapped} during the editing process. The `Avg. Acc' and `\#workers' rows were correct.

    \item 
    The results presented in Table 12 in the original paper should be replaced by the results in Table~\ref{table:no-bad-label-gpt-aggregation} in this document.

    In \textbf{Table 12} (in the Appendix) in the original paper, within ``\textbf{Basic UI}'' section, the data in the \textbf{Majority Vote (MV)} row was pasted incorrectly. The data presented in the original paper were mistakenly copied from the \textbf{ZBS} row under \textbf{Advanced UI} section in \textbf{Table 11} (in the Appendix). This error did not affect our conclusion.

    \item Additionally, the sentence on page 8 of the original paper, 
    
    ``We also noticed that in the simulations, accuracy improved for most aggregation methods as the number of MTurk workers increased\textit{, but MACE's performance was notably inconsistent}.''
    
    should be corrected as follows: 
    
    ``We also noticed that in the simulations, accuracy improved for most aggregation methods as the number of MTurk workers increased.''

    The last part in the original sentence referred to the unstable performance of the older tool used in early experiments, the Crowd-Kit implemented MACE method, as detailed in the sections 3.6.1 and 3.6.3 of the original paper. 
    In the final version of the paper, we instead used Hovy's MACE implementation, which we also described in the sections 3.6 of the original paper.
    Therefore, the last part of the sentence needs to be removed. 
    This error did not impact the main conclusion or affect any other results presented in the original paper.

\end{enumerate}

These corrections do not affect our overall conclusions. Our core findings remain valid.

\begin{table*}[h]
\centering
\resizebox{\textwidth}{!}{%
\begin{tabular}{lccccccccc}
\multicolumn{10}{c}{\textbf{Basic Interface}}\\
\toprule
\tiny Acc of GPT-4 (t=0.2) =\textbf{.836} & \multicolumn{3}{c}{\textbf{All Workers}} & \multicolumn{3}{c}{\textbf{Exclude-By-Worker}} & \multicolumn{3}{c@{}}{\textbf{Exclude-By-Batch}} \\ \cmidrule(lr){2-4} \cmidrule(lr){5-7} \cmidrule(lr){8-10}
\textbf{Method} & \textbf{Acc} & \textbf{P-Value} & \textbf{95\% CI} & \textbf{Acc} & \textbf{P-Value} & \textbf{95\% CI} & \textbf{Acc} & \textbf{P-Value} & \textbf{95\% CI} \\ \midrule
\textbf{MV} & .551 & \textless.001 & {[}.534, .569{]} & .609 & \textless.001 & {[}.592, .626{]} & .567 & \textless.001 & {[}.550, .584{]} \\
\textbf{DawidSkene} & .675 & \textless.001 & {[}.659, .691{]} & .737 & \textless.001 & {[}.722, .753{]} & .712 & \textless.001 & {[}.696, .727{]} \\
\textbf{OneCoin} & .797 & .002 & {[}.783, .811{]}  & .804 & .002 & {[}.790, .818{]} & .800 & .006 & {[}.786, .814{]}\\
\textbf{GLAD} & .734 & \textless.001 & {[}.719, .749{]}  & .756 & \textless.001 & {[}.741, .771{]} & .731 & \textless.001 & {[}.715, .746{]}\\
\textbf{M-MSR} & .531 & \textless.001 & {[}.513, .548{]}  & .630 & \textless.001 & {[}.614, .647{]} & .590 & \textless.001 & {[}.572, .607{]}\\
\textbf{MACE} & \textbf{\underline{.811}} & .001 & \textbf{\underline{{[}.797, .824{]}}}  & \textbf{.809} & .001 & \textbf{{[}.795, .823{]}} & \textbf{.809} & \textless.001 & \textbf{{[}.796, .823{]}}\\
\textbf{Wawa} & .633 & \textless.001 & {[}.617, .650{]} & .672 & \textless.001 & {[}.656, .689{]} & .636 & \textless.001 & {[}.619, .653{]} \\
\textbf{ZBS} & .677 & \textless.001 & {[}.661, .694{]} & .712 & \textless.001 & {[}.696, .727{]} & .675 & \textless.001 & {[}.659, .691{]} \\ \midrule
\textbf{Avg. Acc} & .676 & - & - & .716 & - & - & .690 & - & - \\ \midrule
\textbf{\#workers} & \multicolumn{3}{c}{216} & \multicolumn{3}{c}{134} & \multicolumn{3}{c@{}}{176}
\\
\bottomrule
\end{tabular}
}
\caption{Aggregation Accuracy Results of the Basic Interface integrated with GPT4 Group. 
\textbf{Bold} and \underline{underline} highlight the highest score within the column and across the table, respectively. P-value is obtained by comparing with GPT-4 over the article-level accuracy. 
(\textsuperscript{**}: p\textless0.01; \textsuperscript{***}: p\textless0.001. Paired t-test. N=200)
} 
\label{table:basic-crowd-gpt-acc-result}
\end{table*}

\begin{table*}
\centering
\resizebox{\textwidth}{!}{%
\setlength{\tabcolsep}{2.8pt}
\begin{tabular}{lrrrrrrrrrrrrrrrrr}
\toprule
\multicolumn{1}{l}{\textbf{Eval}} & \multicolumn{3}{c}{\textbf{Background}} & \multicolumn{3}{c}{\textbf{Purpose}} & \multicolumn{3}{c}{\textbf{Method}} & \multicolumn{3}{c}{\textbf{Finding}} & \multicolumn{3}{c}{\textbf{Other}} & \multicolumn{1}{c}{\multirow{2}{*}{\textbf{Acc}}} & \multicolumn{1}{c}{\multirow{2}{*}{\textbf{Kappa}}} \\ \cmidrule(lr){2-4} \cmidrule(lr){5-7} \cmidrule(lr){8-10} \cmidrule(lr){11-13} \cmidrule(lr){14-16}
\textbf{Label} & \multicolumn{1}{c}{P} & \multicolumn{1}{c}{R} & \multicolumn{1}{c}{F1} & \multicolumn{1}{c}{P} & \multicolumn{1}{c}{R} & \multicolumn{1}{c}{F1} & \multicolumn{1}{c}{P} & \multicolumn{1}{c}{R} & \multicolumn{1}{c}{F1} & \multicolumn{1}{c}{P} & \multicolumn{1}{c}{R} & \multicolumn{1}{c}{F1} & \multicolumn{1}{c}{P} & \multicolumn{1}{c}{R} & \multicolumn{1}{c}{F1} & \multicolumn{1}{c}{} & \multicolumn{1}{c}{} \\ \midrule
\multicolumn{18}{c}{\textbf{Basic UI}}\\
MV & .780 & .387 & .517 & .194 & .613 & .294 & .450 & .710 & .551 & .854 & .582 & .692 & 1.000 & .286 & .444 & .567 & .402 \\
DawidSkene & .864 & .484 & .621 & .263 & .724 & .386 & .721 & .784 & .751 & .936 & .782 & .852 & .088 & .619 & .154 & .712 & .593 \\
OneCoin & .888 & .712 & .790 & .513 & .710 & .596 & .675 & .834 & .746 & .896 & .843 & .869 & .875 & .333 & .483 & .800 & .702 \\
GLAD & .873 & .629 & .731 & .328 & .765 & .459 & .608 & .797 & .690 & .924 & .744 & .825 & .684 & .619 & .650 & .731 & .615 \\
M-MSR & .689 & .491 & .574 & .202 & .581 & .299 & .479 & .656 & .553 & .858 & .609 & .713 & .500 & .333 & .400 & .590 & .428 \\
\textbf{MACE} & .871 & .804 & .836 & .444 & .797 & .570 & .718 & .818 & .765 & .960 & .810 & .878 & .346 & .857 & .493 & \textbf{.810} & .724 \\
Wawa & .807 & .526 & .637 & .239 & .700 & .356 & .513 & .757 & .612 & .912 & .627 & .743 & 1.000 & .381 & .552 & .636 & .495 \\
ZBS & .834 & .569 & .676 & .276 & .747 & .403 & .548 & .778 & .643 & .919 & .671 & .776 & .818 & .429 & .563 & .675 & .544 \\ \midrule
\multicolumn{18}{c}{\textbf{Advances UI}}\\
MV & .694 & .370 & .482 & .149 & .525 & .232 & .434 & .701 & .536 & .864 & .520 & .649 & 1.000 & .048 & .091 & .523 & .354 \\ 
DawidSkene & .838 & .556 & .668 & .242 & .700 & .360 & .792 & .784 & .788 & .969 & .765 & .855 & .077 & .667 & .139 & .718 & .607 \\
\textbf{OneCoin} & .909 & .884 & .896 & .581 & .793 & .671 & .808 & .893 & .848 & .953 & .879 & .914 & .909 & .476 & .625 & \textbf{.874} & .813 \\
GLAD & .865 & .698 & .772 & .325 & .700 & .444 & .671 & .857 & .753 & .951 & .773 & .853 & .889 & .381 & .533 & .767 & .665 \\
M-MSR & .537 & .450 & .489 & .169 & .544 & .257 & .451 & .650 & .532 & .869 & .498 & .634 & .313 & .238 & .270 & .522 & .353 \\
MACE & .884 & .910 & .897 & .546 & .816 & .654 & .835 & .871 & .852 & .968 & .859 & .910 & .425 & .810 & .557 & .869 & .808 \\
Wawa & .723 & .511 & .599 & .186 & .613 & .286 & .509 & .726 & .598 & .913 & .580 & .710 & 1.000 & .286 & .444 & .597 & .449 \\
ZBS & .817 & .653 & .726 & .260 & .691 & .377 & .631 & .810 & .710 & .942 & .700 & .803 & 1.000 & .381 & .552 & .711 & .593 \\ 
\midrule
GPT-4 (t=.2) & .860 & .913 & .885 & .499 & .843 & .627 & .775 & .871 & .820 & .982 & .784 & .872 & .322 & .905 & .475 & .836 & .764 \\\bottomrule
\end{tabular}
}
\caption{Exclude-By-Batch integrated with GPT-4 Table. All models use Bio Expert as the gold standard. Baseline is the Majority Vote (MV).}
\label{table:no-bad-label-gpt-aggregation}
\end{table*}



\vspace{1em}

\noindent (See: DOI:\href{https://dl.acm.org/doi/10.1145/3613904.3642834}{https://dl.acm.org/doi/10.1145/3613904.3642834})

\restoregeometry

\end{document}